\documentclass[%
 reprint,
 amsmath,amssymb,
 aps,
]{revtex4-1}

\usepackage{graphicx}
\usepackage{dcolumn}
\usepackage{bm}
\usepackage{color}
\usepackage{hyperref}
\usepackage{ulem}
\usepackage{graphicx}
\usepackage{float}

\usepackage[mathlines]{lineno}

\begin{document}

\preprint{APS/123-QED}

\title{Earth Tomography with ICAL at INO}

\author{Deepak Raikwal}
\email{deepakraikwal@hri.res.in}
 \affiliation{Harish-Chandra Research Institute,  A CI of Homi Bhabha National Institute, Chhatnag Road, Jhunsi, Prayagraj - 211019}
\affiliation{Homi Bhabha National Institute, Anushakti Nagar, Mumbai 400094, India}
\author{Sandhya Choubey,}%
 \email{choubey@kth.se}
\affiliation{Department of Physics, School of Engineering Sciences, KTH Royal Institute of Technology,\\ AlbaNova University Center, Roslagstullsbacken 21, SE--106 91 Stockholm, Sweden }
\affiliation{The Oskar Klein Centre, AlbaNova University Center, Roslagstullsbacken 21,\\ SE--106 91 Stockholm, Sweden}%


\begin{abstract}
Observing matter effects in atmospheric neutrinos travelling through the entire mantle and core of the Earth is a promising way of enhancing our understanding of Earth's density structure. In that context we study the prospects of Earth tomography with the ICAL detector at the India-based Neutrino Observatory. While this experiment is smaller in size in comparison to some of the other bigger detectors being proposed, it is the only neutrino experiment with charge-identification sensitivity. In particular, ICAL can see matter effects separately in neutrinos and antineutrinos. This has been seen to enhance ICAL's sensitivity to earth matter effects and hence the mass ordering sensitivity for both normal and inverted mass orderings. It is therefore, pertinent to see if the ICAL sensitivity to earth tomography is competitive or better with respect to other experiments, especially for the inverted mass ordering, where other experiments suffer reduced sensitivity. We present the sensitivity of ICAL to earth tomography by taking into consideration both the Earth's mass constraint as well as the hydrostatic equilibrium constraints. 
\end{abstract}

\maketitle


\section{\label{intro}Introduction}

Knowing and understanding our own planet remains one of mankind's biggest challenges. In particular, there is a lot of on-going effort to understand the density structure of the Earth. Today the best understanding in this area comes from seismology -- the study of earthquakes and the corresponding seismic waves that they create. The motion of the seismic waves depends crucially on the density structure of the Earth, hence, seismology has managed to provide us with a reasonable understanding of Earth's interior. Data is collected at several seismographic stations accross the surface of the Earth. Since the speed of the seismic waves depends on the distance between these stations as well as the density of matter that they cross, a careful comparative analysis of this collective data can be directly used to obtain Earth's density profile. The best density model of the Earth obtained till-date using seismological data is called the PREM (Preliminary Reference Earth Model) density profile \cite{PREM}. This model divides the Earth into various regions or zones. Broadly, Earth can be divided into the core, mantle and crust, with finer layers in each of these divisions. The density in each of these layers is given by the PREM profile. However, there remains uncertainties in these density estimates, with the uncertainty in density in some layers being significantly higher than some others. In general, the density in the deeper layers has a larger uncertainty than the outer layers. 

While efforts are on to get a better understanding of Earth's density profile via seismology, it is pertinent to ask if one could cross-check Earth's density profile using other complementary methods. Neutrino physics can provide such a complementary approach. Neutrinos are a probe for the density of matter through which they travel in two different ways. Neutrinos interact with matter via $W$ and $Z$ boson mediated weak interactions. The neutrino interaction cross-section increases with neutrino energy, becoming sizeable at the TeV energy scale. Interactions of high energy neutrinos with the ambient Earth matter results in the attenuation of the neutrino flux. Since the resultant  attenuation depends on the density of matter, measurement of this attenuation can be used to measure the average Earth matter density along the neutrino path length. This method can be used at neutrino telescopes such as IceCube \cite{pingucollaboration2017letter} and KM3NeT ARCA \cite{Adri_n_Mart_nez_2016} to determine Earth's density profile. The second way neutrinos can probe the Earth density is via neutrino oscillations. This is the subject of study in this work.

Neutrino oscillations have been observed and confirmed in a myriad of neutrino experiments. The neutrino oscillation parameters have been measured to a reasonable precision. Only the few last remaining pieces of this puzzle remain to be discovered and/or confirmed. Existence of CP violation in the lepton sector is one of the most important missing puzzle pieces. The other missing piece is the octant of the mixing angle $\theta_{23}$, {\it ie.} whether $\theta_{23} < \pi/4$ (called lower octant (LO) solution) or $\theta_{23}  > \pi/4$ (called upper octant (UO) solution). Finally, the last puzzle piece is the neutrino mass ordering (MO), {\it ie.} whether the atmospheric mass squared difference $\Delta m_{31}^2 >0 $ (called normal ordering (NO)) or $\Delta m_{31}^2 < 0 $ (called inverted ordering (IO)). Bigger and better experiments are being proposed and built to find these remaining pieces of the puzzle. Amongst these are the next generation atmospheric neutrino experiments such as IceCube (PINGU) \cite{IceCube:2016xxt}, ORCA \cite{KM3NeT:2014wix, Adri_n_Mart_nez_2016} and ICAL@INO \cite{ICAL:2015st}. These atmospheric neutrino experiments are particularly suited to measure the neutrino MO via their ability to observe Earth matter effects in neutrino oscillations. Matter effects in neutrino oscillations depend on the MO as well as density of matter. Hence, the flavor oscillations of atmospheric neutrinos while traveling inside the Earth become sensitive to the density of matter. Therefore, precise measurement of matter effects in neutrino oscillations can be used to verify the Earth matter density profile. 

Prospects of Earth tomography using atmospheric neutrinos in multi-megaton class detectors has been performed earlier for IceCube (PINGU) \cite{pingu-tomo, winter-tomo} and ORCA \cite{Capozzi2022, winter-tomo} detectors. In \cite{winter-tomo,winter-tomo-1} the author studies the prospects of earth tomography in PINGU and ORCA and concludes that the density measurements in the lower mantle region can be performed to a few percent level. Ref. \cite{Capozzi2022} further improves this analysis in the context of the ORCA detector and makes a thorough sensitivity study of Earth tomography. They include Earth mass constraints and hydrostatic equilibrium and show that the density in the outer core (mantle) can be measured to -18\%/+15\% (-6\%/+8\%) level. They also look at the impact of systematic uncertainties on the sensitivity of ORCA. In Ref. \cite{Kumar_2021, Upadhyay_2023} the authors look at the prospects of confirming the existence of the core using atmospheric neutrinos in ICAL@INO \cite{ICAL:2015st}. In this work we look at the prospect of performing Earth tomography using the atmospheric neutrino experiment ICAL@INO. We take the PREM profile as the reference density model of the Earth and study the sensitivity of ICAL@INO to deviations from PREM. We present our results as a function of the percentage change that ICAL@INO can confirm with respect to the PREM profile. We consider three different cases in our studies. We start with a simple approach where we calculate the sensitivity of ICAL@INO to Earth mass density in the mantle and core regions without any other constraint imposed. We next study the case where the Earth mass constraint is imposed by compensating an increase (decrease) in a given layer of Earth by a corresponding decrease (increase) in all the other layers such that the total mass of the Earth does not change. Finally, we take into account the fact that not all layers are equally uncertain and constrain the compensation accordingly. In all the cases we study the impact of systematic uncertainties on the sensitivity of ICAL@INO. We also take into account the condition for hydrostatic equilibrium of the Earth.

The papers is organised as follows. We begin in section \ref{sec:model} by discussing the Earth density model and the three different cases that we consider for analysing the Earth density model. In this section we also present the impact of the variations to the PREM profile on the relevant neutrino oscillation probabilities. In section \ref{sec:analysis} we present in details our analysis methodology, where we provide details of the ICAL@INO experiment, the atmospheric neutrino fluxes, the simulation tools, the oscillation framework and the statistical analysis method. In section \ref{sec:results} we present our results. Finally, we summarize our results and conclude in section \ref{sec:conclusions} along with a comparison with other experiments. 

\section{Earth Density Model \label{sec:model}}
The neutrino oscillation probabilities are calculated by using the Hamiltonian
\begin{equation}
  {\cal H} = UMU^{\dagger}+{\cal V}_e  \,, 
\end{equation}
where $U$ is the PMNS mixing matrix \cite{Giganti_2018}, $M$ is 3$\times$3 neutrino mass squared matrix $M = diag(0,\Delta m^2_{21},\Delta m_{31}^2)$, and ${\cal V}_e $ is the 3$\times$3  matrix containing the effective matter potential coming from coherent forward scattering of neutrinos with electrons in the ambient matter, with ${\cal V}_e = diag(\pm\sqrt{2}G_{F}N_A \, \rho , 0,0)$, 
where $N_A$ is the Avogadro's number and $\rho$ is the density of matter. For neutrinos the matter potential takes the positive sign, while for antineutrinos the matter potential comes with the negative sign. In this work we use the convention $\Delta m_{ij}^2=m_i^2 -m_j^2$.  As a result, neutrino oscillations are modified when neutrinos travel in matter. The effective potential depends on the density of matter in which the neutrinos are traveling. Neutrinos (and antineutrinos) with energy around 5-10 GeV, traveling through the Earth undergo large matter effects which change their oscillation probabilities. The size of these changes depends on the density of matter through which the neutrinos travel and energy of the neutrinos, and are opposite for neutrinos and antineutrinos. Atmospheric neutrinos and antineutrinos span an energy range of 100 MeV to 10 TeV and come from all directions {\it aka}, all zenith angles. Neutrinos from different zenith angles traverse different density layers of the Earth and hence experience different matter effects. ICAL@INO can separately see $\nu_\mu$ ($\mu^-$) and $\bar\nu_\mu$ ($\mu^+$), as a function of muon energy and muon angle. In addition, it can observe the corresponding hadrons produced in the charged current event. This enables ICAL@INO to observe earth matter effects rather efficiently. Thus, measurement of the earth matter effects as a function of energy and zenith angle at ICAL@INO can be effectively used to study the density structure of our Earth.

For our reference density structure of the Earth we use the PREM (Preliminary Reference Earth Model) \cite{PREM} profile. In the PREM model, Earth is broadly divided in seven major parts: 1) crust $d<30$, (2) lower lithosphere $30<d<400$, (3) upper mesosphere $400<d<600$, (4) transition zone $600<d<800$, (5) lower mesosphere $800<d<2890$, (6) outer core $2890<d<5150$ (7) and inner core $5150<d<6371$, where $d$ is the depth inside Earth, measured in km. The density varies within each of these layers and in our analysis we have taken that density variation into account by subdividing Earth into 26 layers, each with a fixed density given by the PREM profile. This gives an excellent simulation of the full PREM profile. It is known \cite{winter-tomo, Capozzi2022} and we have checked that the sensitivity of atmospheric neutrinos to the density of the crust, lower lithosphere, upper mesosphere, transition zone and inner core is rather poor. Therefore, throughout this work we discuss density changes in two major Earth zones - the {\it outer core} ($2890<d<5150$) and the lower mesosphere which we will call broadly {\it mantle} ($800<d < 2890$), where $d$ is the depth inside Earth measured in km. 

We use our simulation code, as described in the next section, to simulate the ``data" in ICAL corresponding to the reference PREM profile. This simulated ``data" is then fitted by changing the Earth's density in the theory. We work with three different scenarios for the theory, which we label as Case I, Case II and Case III, described below. For Cases II and III we also impose the condition of hydrostatic equilibrium of the Earth by demanding that the following inequalities should always hold.
\begin{equation}
    \rho_{man}^{max}\leq \rho_{OC}^{min} ~{\rm and}~ \rho_{OC}^{max} \leq \rho_{IC}^{min}\,,
    \label{equ:hydrostat}
\end{equation}
where $\rho_{man}^{max}$ and $\rho_{OC}^{max}$ are the maximum density inside the mantle (man) and outer core (OC), respectively, while $\rho_{OC}^{min}$, $\rho_{IC}^{min}$ are the minimum density inside the outer core and inner core (IC), respectively.

\subsection{Case I}

\begin{figure}
    \centering
    \includegraphics[width=0.3\textwidth]{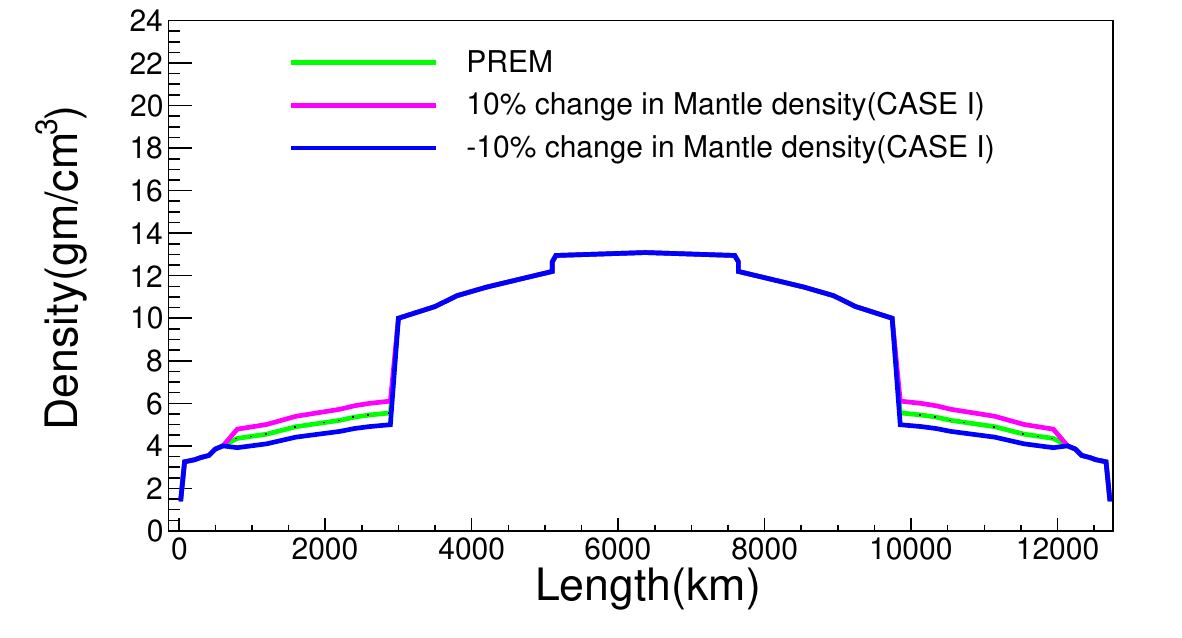} 
    \includegraphics[width=0.3\textwidth]{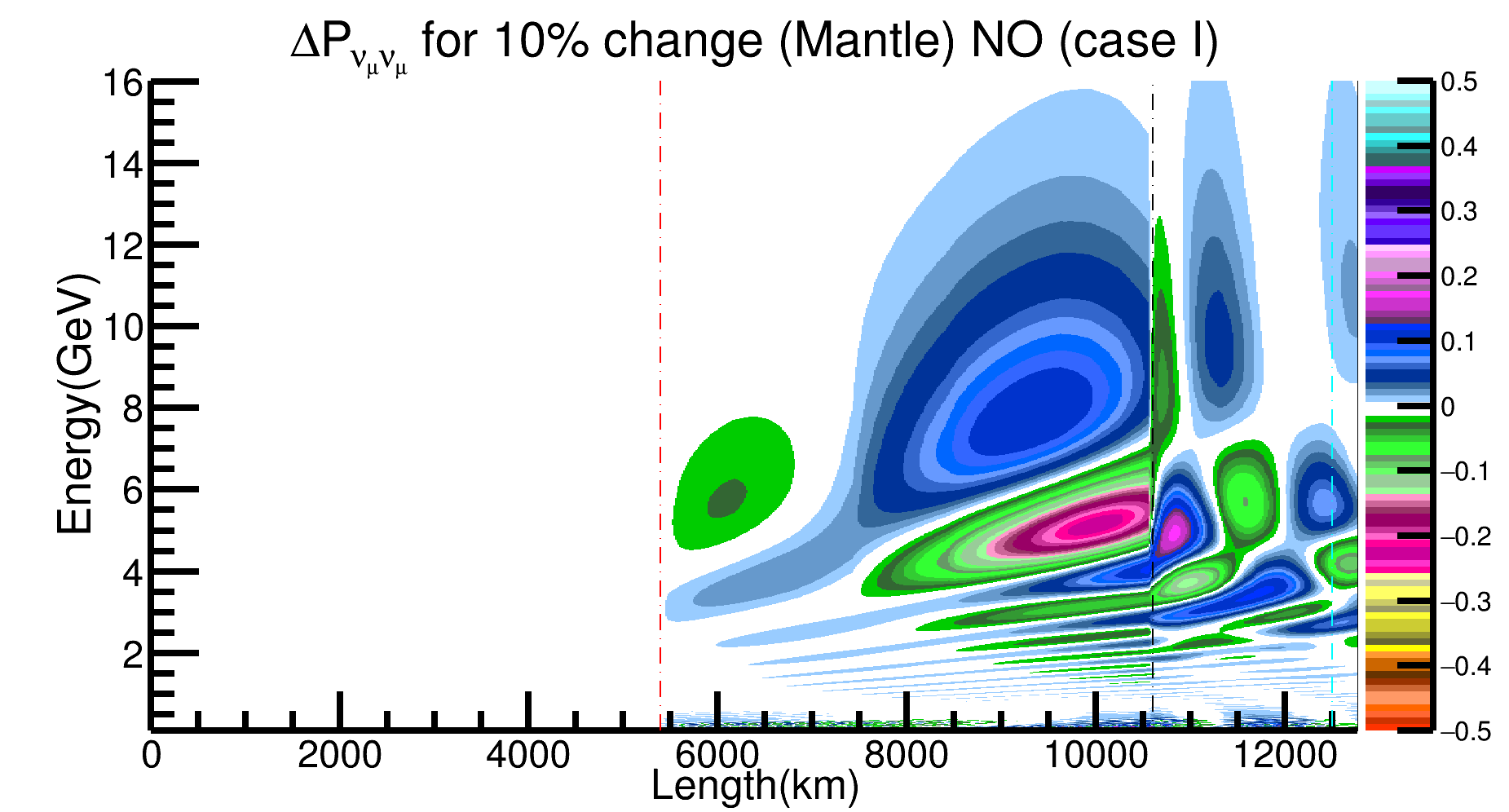} 
    \includegraphics[width=0.3\textwidth]{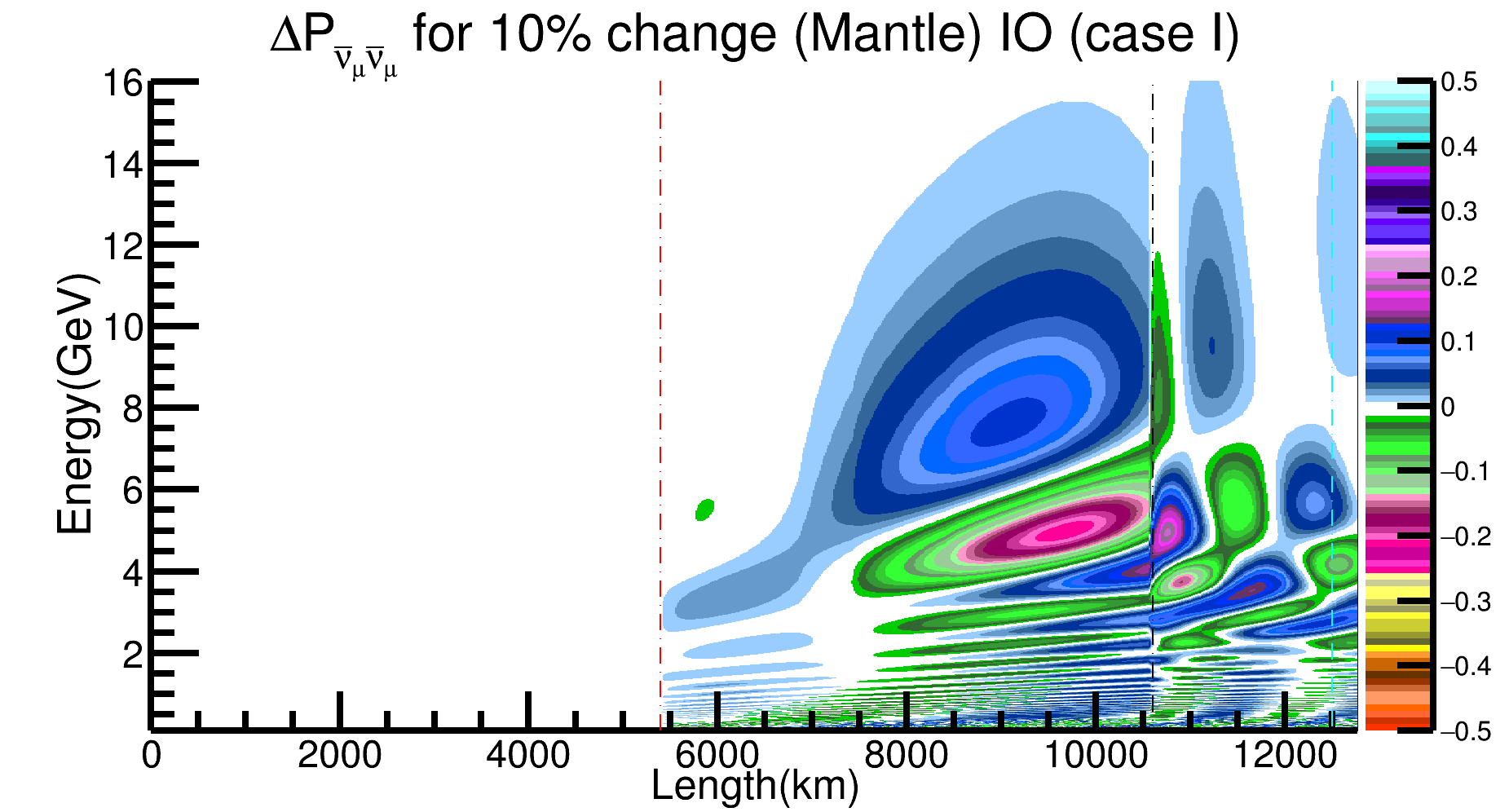}
    \caption{Top panel shows the change in density of mantle according to Case I. Middle panel shows the effect of density change on the survival probability via $\Delta P_{\nu_{\mu}\nu_{\mu}}=P_{\nu_{\mu}\nu_{\mu}}^{PREM}-P_{\nu_{\mu}\nu_{\mu}}^{newPREM}$ for NO and neutrinos, where $newPREM$ corresponds to density modified case. Lower panel is the same as panel (b) but for IO and antineutrinos.}
    \label{fig:mnprob}
\end{figure}

\begin{figure}
    \centering
    \includegraphics[width=0.3\textwidth]{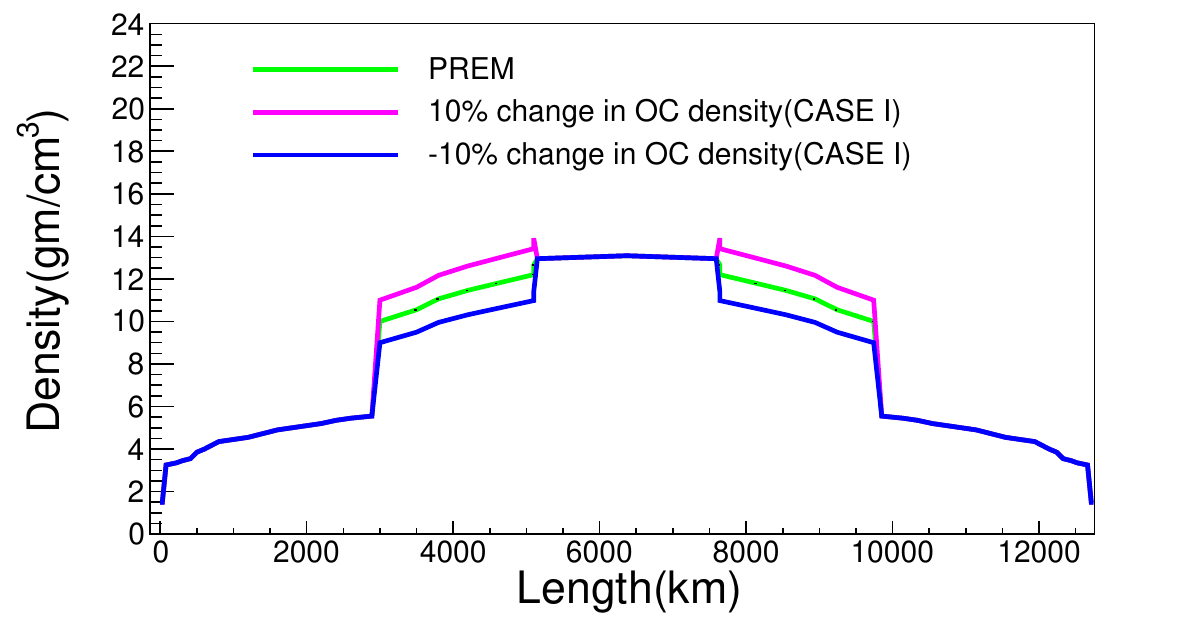}
    \includegraphics[width=0.3\textwidth]{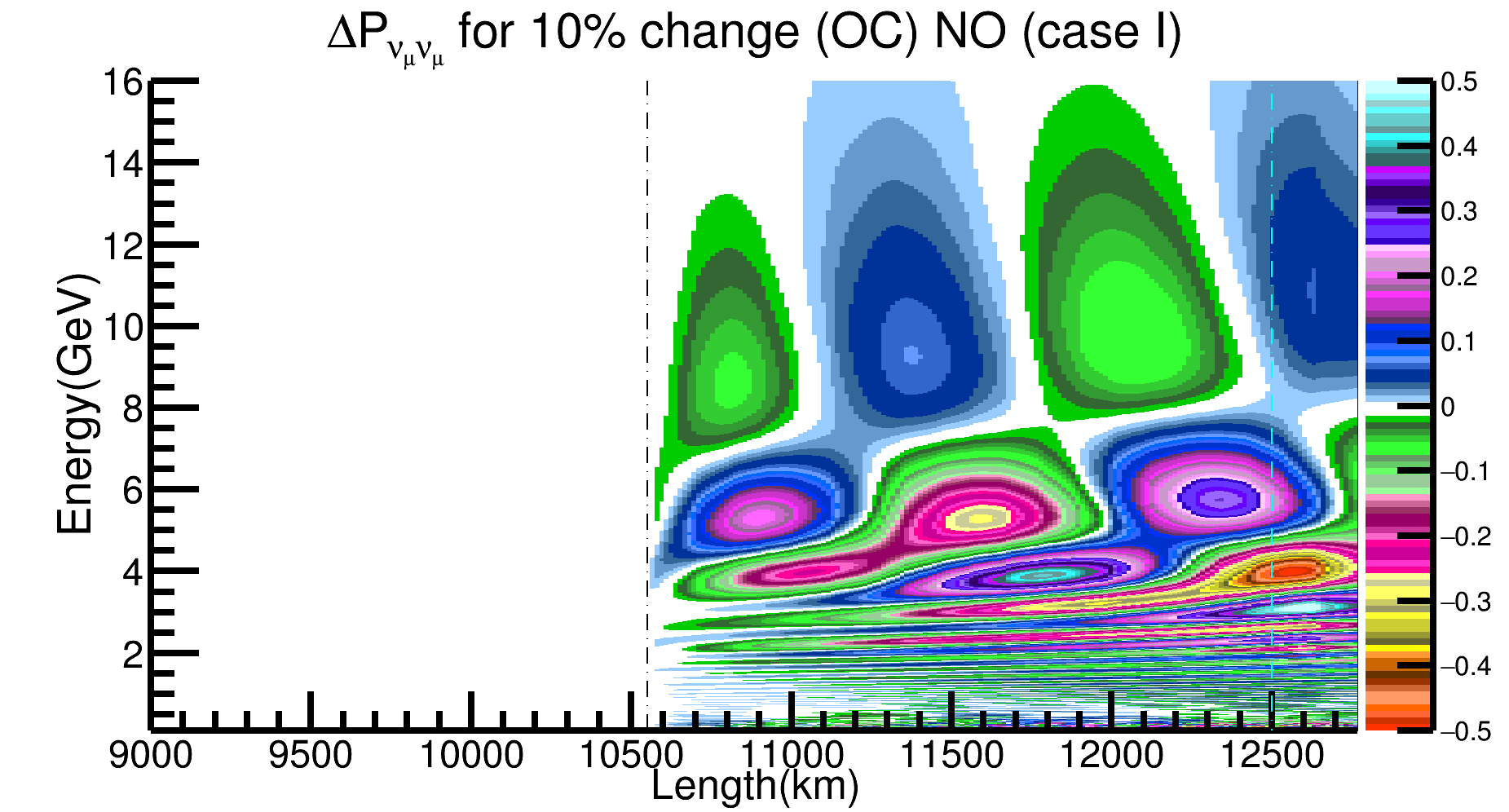}
    \includegraphics[width=0.3\textwidth]{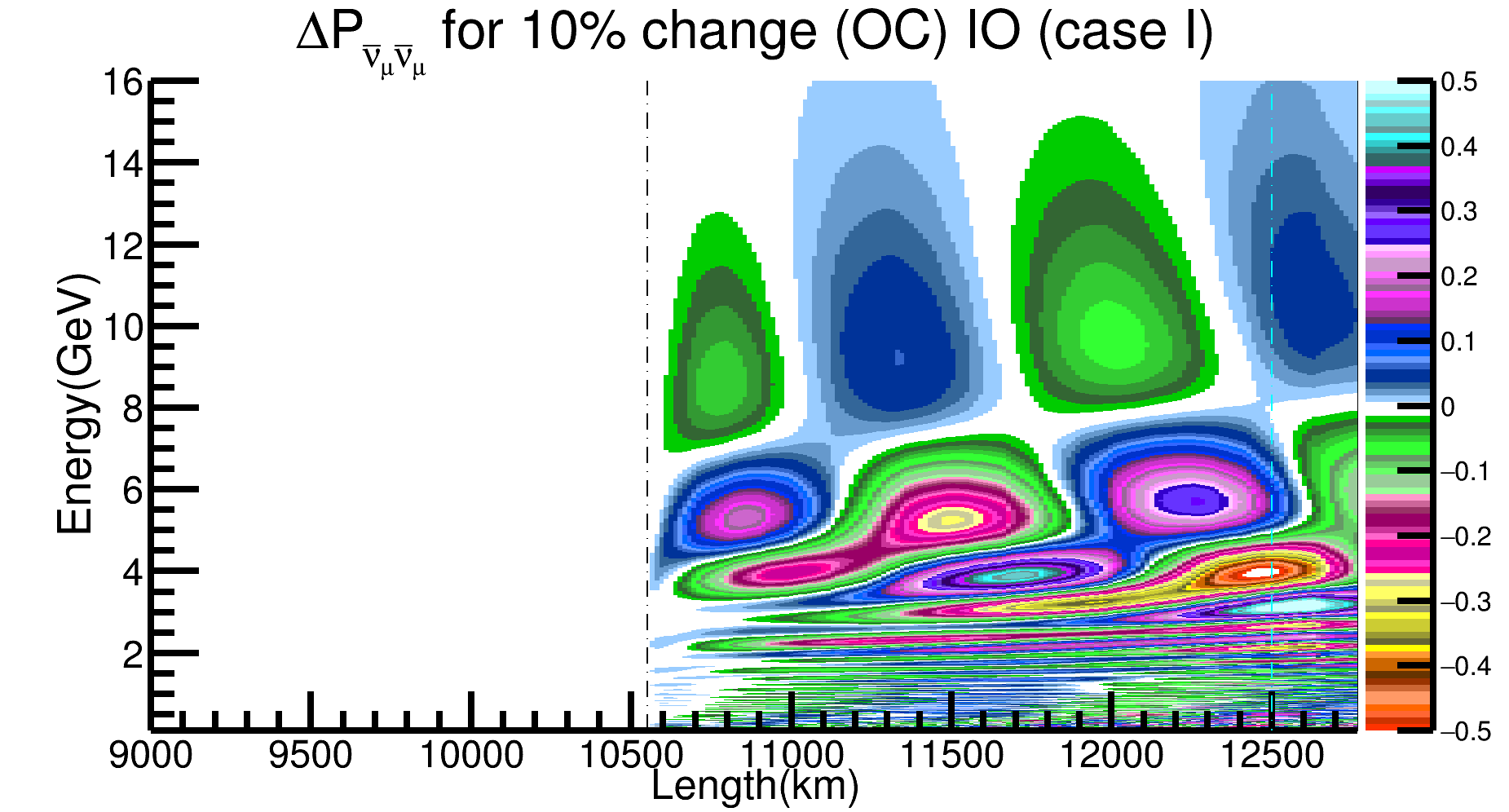}
    \caption{The panels shown here are the same as in Fig. \ref{fig:mnprob} but for density change in OC for Case I.}
    \label{fig:ocnprob}
\end{figure}

We begin with the simplest case where we change the density in a given region, outer core or mantle by a constant factor $x\%$ without any other consideration. In particular, we do not take into account the fixed mass of the Earth and/or conditions needed for its hydrostatic stability. This naive case, even though not absolutely correct, is meant to give us an understanding of how density changes in a given individual layer independently affects the neutrino oscillation probabilities. 

In left panel of Fig. \ref{fig:mnprob} we show the density profile of the Earth for this case as a function of the the radial depth $d$ in km. The green line is for the reference PREM profile while the blue and the red lines are the density profiles with $-10$\% and $+10$\% change of density in the mantle, while the density of the inner and outer core are kept fixed at their reference PREM values. The middle (for neutrinos and normal ordering (NO)) and right (for antineutrinos and inverted ordering (IO)) panels  of Fig. \ref{fig:mnprob} show how the survival probability $P_{\mu\mu}$ changes when we change the density in the mantle by $+10$\%. The change $\Delta P_{\mu\mu}$ in these panels are shown as a function of the neutrino path length $L$ in km and neutrino energy $E$ in GeV. $L$ is related to the zenith angle $\theta_z$ by the following relation

\begin{equation}
    L= \sqrt{(R+L_{0})^{2}-(R\sin\theta_{z})^{2}}-R\cos\theta_{z}\,,
\end{equation}
where $R$ is the radius of earth and $L_{0}$ is the height of atmosphere.

Note that the largest changes in the probability occur for neutrinos and antineutrinos with longer trajectories, {\it ie.}, those that cross the core as well as the mantle. We find that these changes occur for all energies less than 15 GeV, but the most prominent change happens around 2-10 GeV. The black dashed vertical line shows the $L$ corresponding to the neutrino trajectory that touches the boundary between the outer core and the mantle, while the red dashed vertical line shows the $L$ corresponding to the neutrino trajectory that crosses the boundary between the lower mesosphere and transition region. 

The three panels of Fig. \ref{fig:ocnprob} show similar results, but this time for reference PREM and $ \pm 10\%$ change of density in the outer core, keeping the density in all other layers fixed at their reference PREM value. We note that the color map shows significantly more ``islands" in the middle and right panels for this case. This implies that  better energy and angle resolutions and finer binning of data would be more crucial for this case as compared to the case for the mantle.

\subsection{Case II}

\begin{figure}
    \centering
    \includegraphics[width=0.3\textwidth]{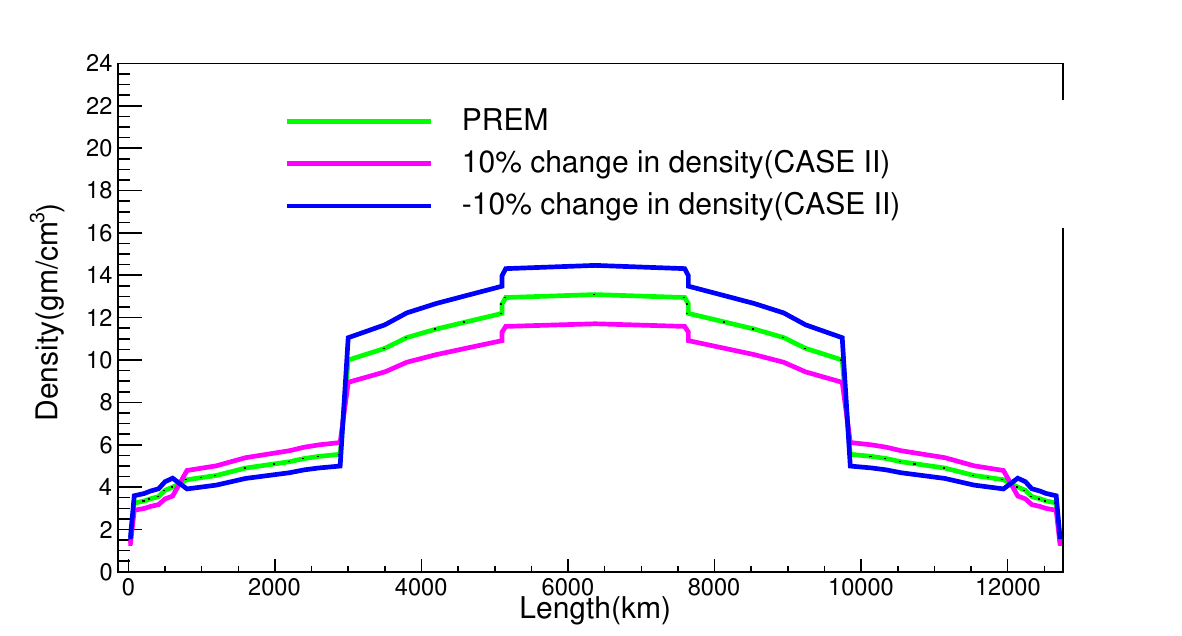}
    \includegraphics[width=0.3\textwidth]{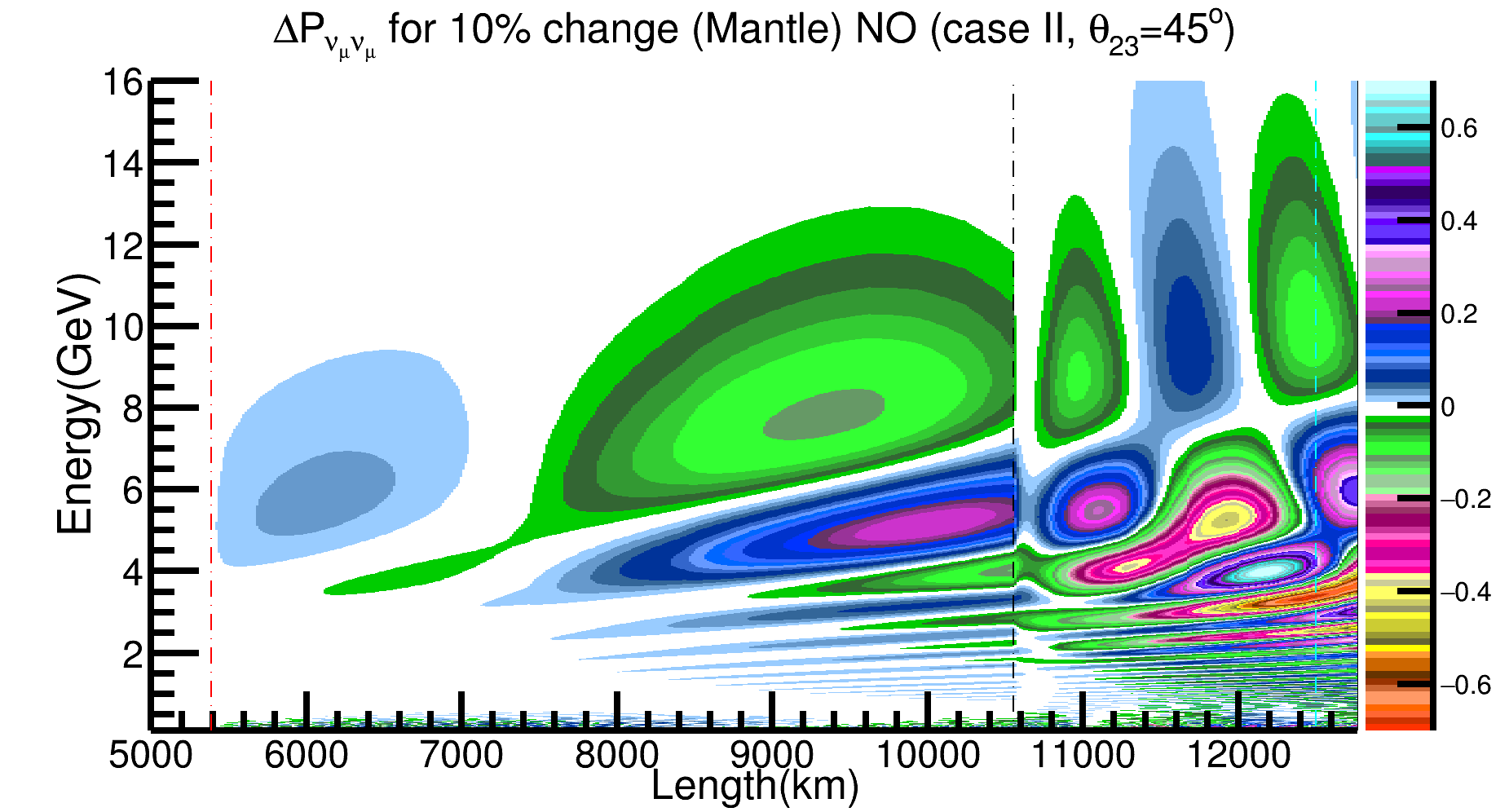} 
    \includegraphics[width=0.3\textwidth]{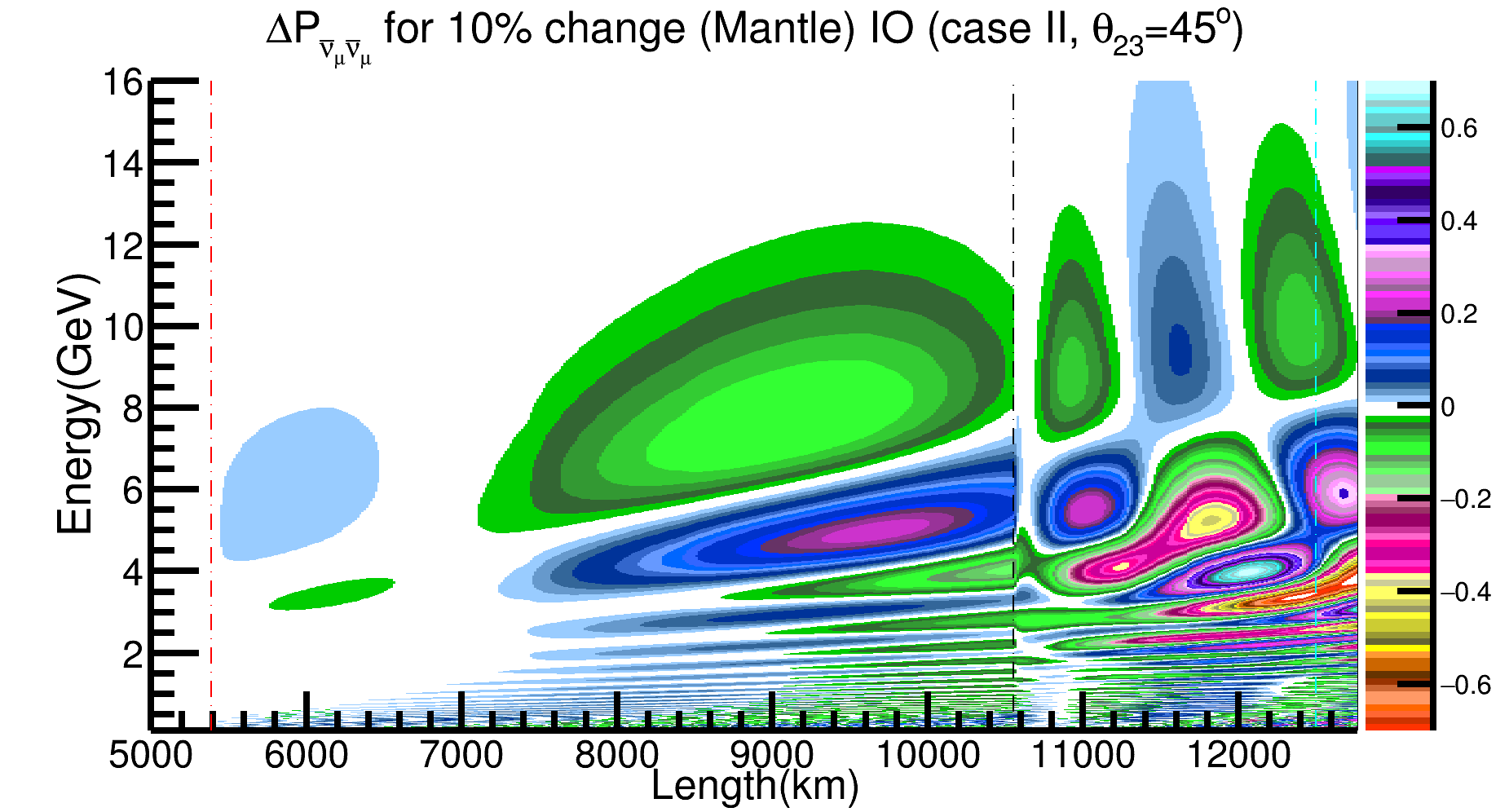}
    \caption{Top panel shows the change in density of mantle according to Case II. Middle panel  shows the effect of density change on the survival probability via $\Delta P_{\nu_{\mu}\nu_{\mu}}=P_{\nu_{\mu}\nu_{\mu}}^{PREM}-P_{\nu_{\mu}\nu_{\mu}}^{newPREM}$ for NO and neutrinos, where $newPREM$ corresponds to density modified case. Lower panel is the same as panel (b) but for IO and antineutrinos.}
    \label{fig:maprob}
\end{figure}

\begin{figure}
    \centering
    \includegraphics[width=0.3\textwidth]{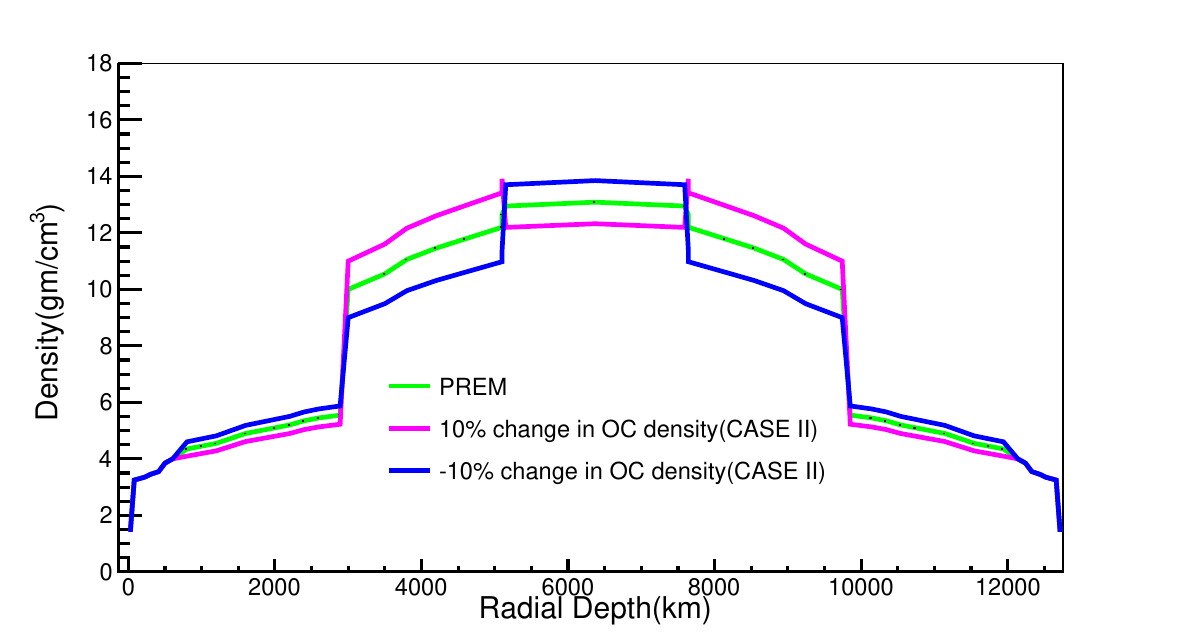}
    \includegraphics[width=0.3\textwidth]{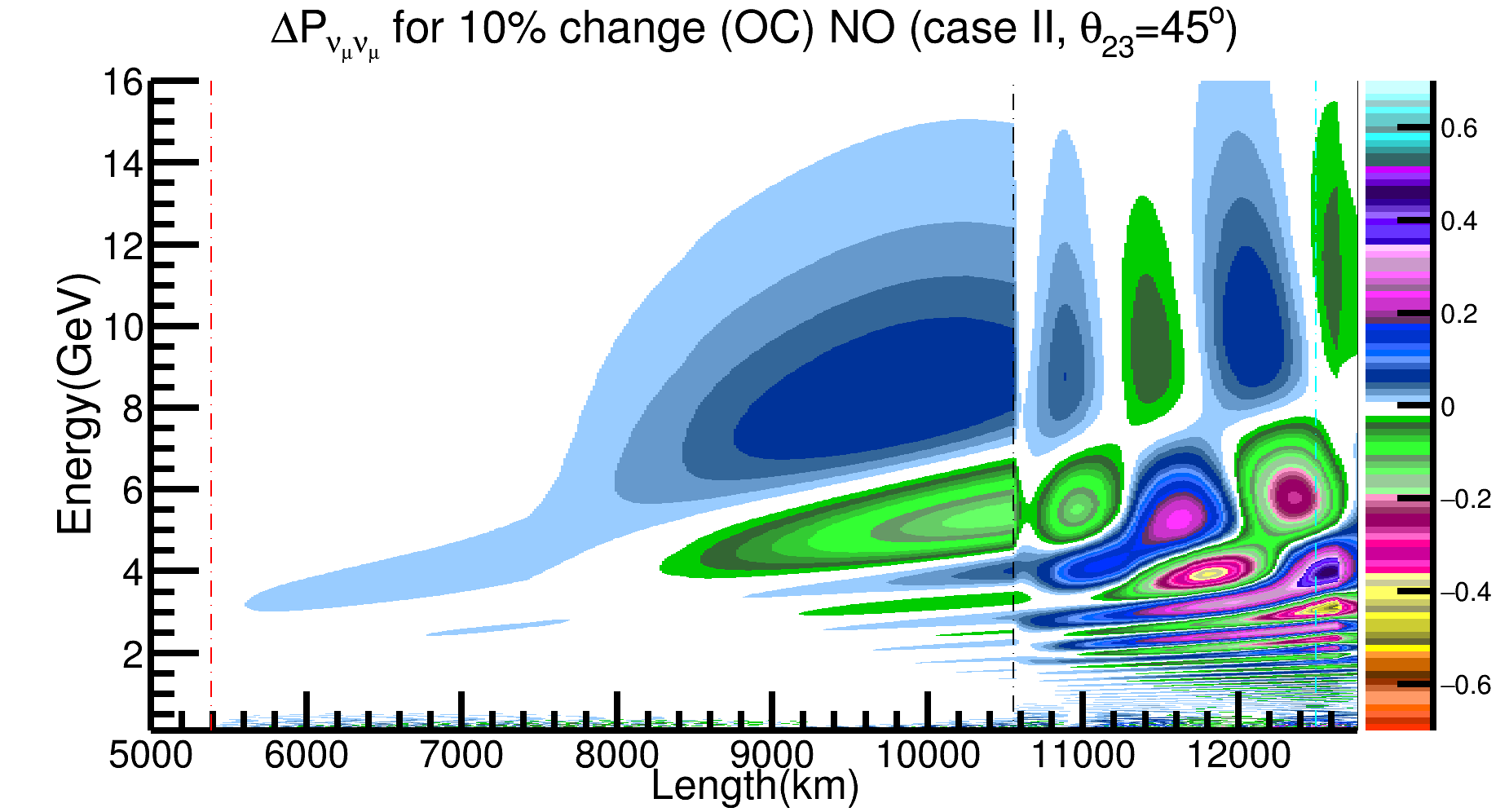}
    \includegraphics[width=0.3\textwidth]{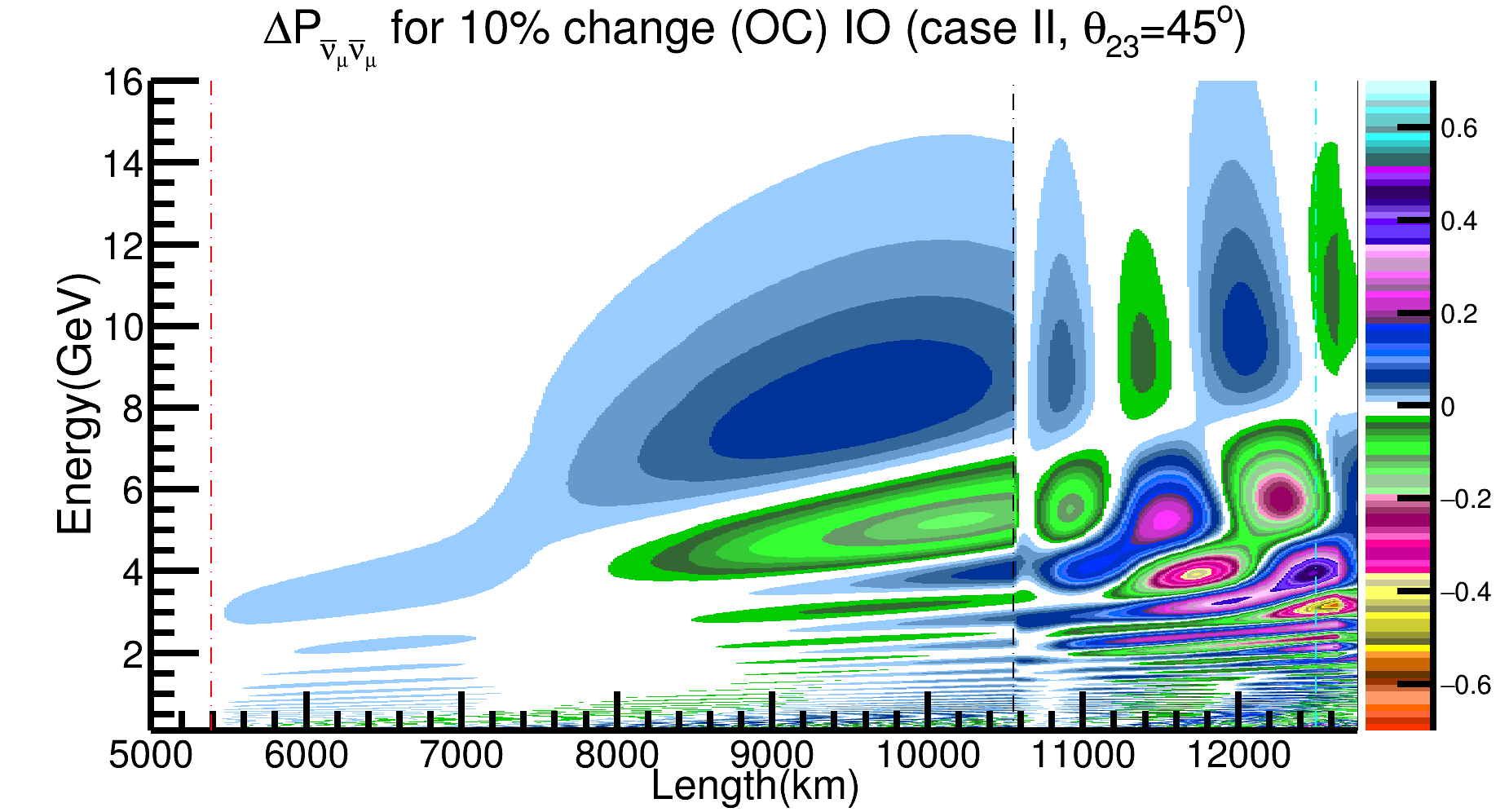}
    \caption{The panels shown here are the same as in Fig. \ref{fig:maprob} but for density change in OC for Case II.}
    \label{fig:ocaprob}
\end{figure}

When we change the layer density by $\pm x\%$ with respect to the PREM profile, we effectively change the mass of the Earth. The mass of the Earth is known to much better precision as compared to its density profile. In our analysis here for this case, we take the Earth mass to be fixed. Therefore, when we increase (decrease) the density in any given layer, we must decrease (increase) the density in some other layers such that the mass of the Earth remains constant. In order to quantify this we do the following. We consider that the density in the three layers - inner core, outer core and mantle could change, while the density of all the other layers of the Earth is kept fixed.  So the Earth mass is given by 
\begin{equation}
    M_{Earth} = \sum_i V_{i}\rho_{i} + \sum_j V_{j}\rho_{j} + \sum_k V_{k}\rho_{k} + M_{\rm fixed},
\end{equation}
where $i$, $j$ and $k$ are labels for the three regions, $\rho_i$, $\rho_j$ and $\rho_k$ are the densities of the individual layers in each of these regions and $V_{i}$, $V_{j}$ and $V_{k}$ are the corresponding volumes of the individual layers. The last term $M_{\rm fixed}$ is the mass of the Earth in the layers whose density is taken as fixed. If we change the density by $x\%$ in any one region, say the region marked by $i$, 
it increases the Earth mass. In order to make the Earth mass constant we need to decrease the density of the other two regions. For simplicity we assume that the density change in the other two regions  are the same, given by $y\%$
according to the following rule,
\begin{equation}
M_{Earth} = (1+x) \sum_i V_{i}\rho_{i} +  (1+y)\bigg ( \sum_j V_j \rho_{j} + \sum_k V_{k}\rho_{k} \bigg )+ M_{\rm fixed}.
\end{equation}
Which gives $y$ in terms of $x$ as follows,
\begin{equation}
    y=-\frac{x \sum_i  V_{i}\rho_{i}}{\sum_j V_{j}\rho_{j} + \sum_k V_{k}\rho_{k}}.
    \label{eq:y}
\end{equation}
In Figs. \ref{fig:maprob} and \ref{fig:ocaprob} we show plots similar to Figs. \ref{fig:mnprob} and \ref{fig:ocnprob} but for the Case II where we keep the Earth mass fixed. In the left panel of both figures we can see that the densities in all layers of the earth get altered. In the left panel Fig. \ref{fig:maprob}, when the mantle density decreases (increases) the inner and outer core densities increase (decrease). Similarly, in the left panel of Fig. \ref{fig:ocaprob}, when the outer core density decreases (increases) the mantle and inner core densities increase (decrease). One can also note from Figs. \ref{fig:maprob} that a small density change in mantle induces large density changes in the core. One the other hand Figs. \ref{fig:ocaprob} shows that a density change in the outer core induces a larger density change in the inner core and a smaller density change in the mantle. The middle and right panels of Figs. \ref{fig:maprob} and \ref{fig:ocaprob} should be compared with the middle and rights panels of Figs. \ref{fig:mnprob} and \ref{fig:ocnprob}. We see that compared to Case I, for Case II the effect of the density change on the oscillation probabilities has increased. For instance comparing middle panels of Figs. \ref{fig:mnprob} and \ref{fig:maprob} shows that there are two ways in which the effect of the density change on the probability has increased - first, the probability change is now over a much wider range of zenith angle and second, even in the zenith angle range $-1 <\cos\theta < -0.4 $ the $|\Delta P_{\nu_\mu \nu_\mu}|$ is larger. The same is true for the right panels. Same is true for the outer core case.

\subsection{Case III}

\begin{figure}
    \centering
    \includegraphics[width=0.3\textwidth]{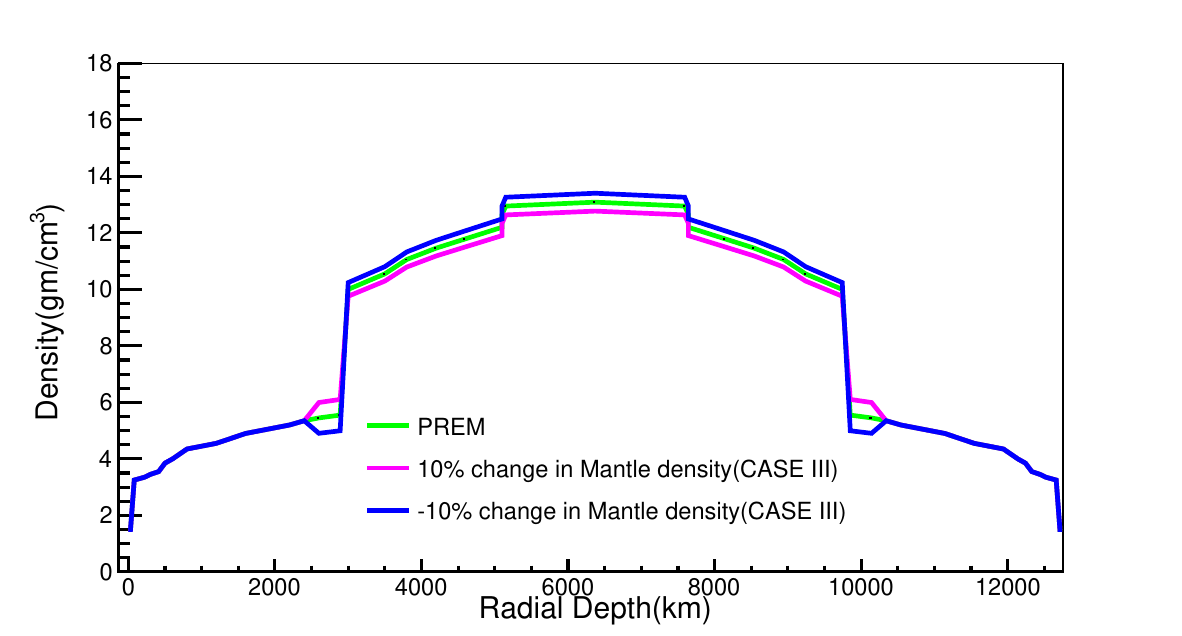}
    \includegraphics[width=0.3\textwidth]{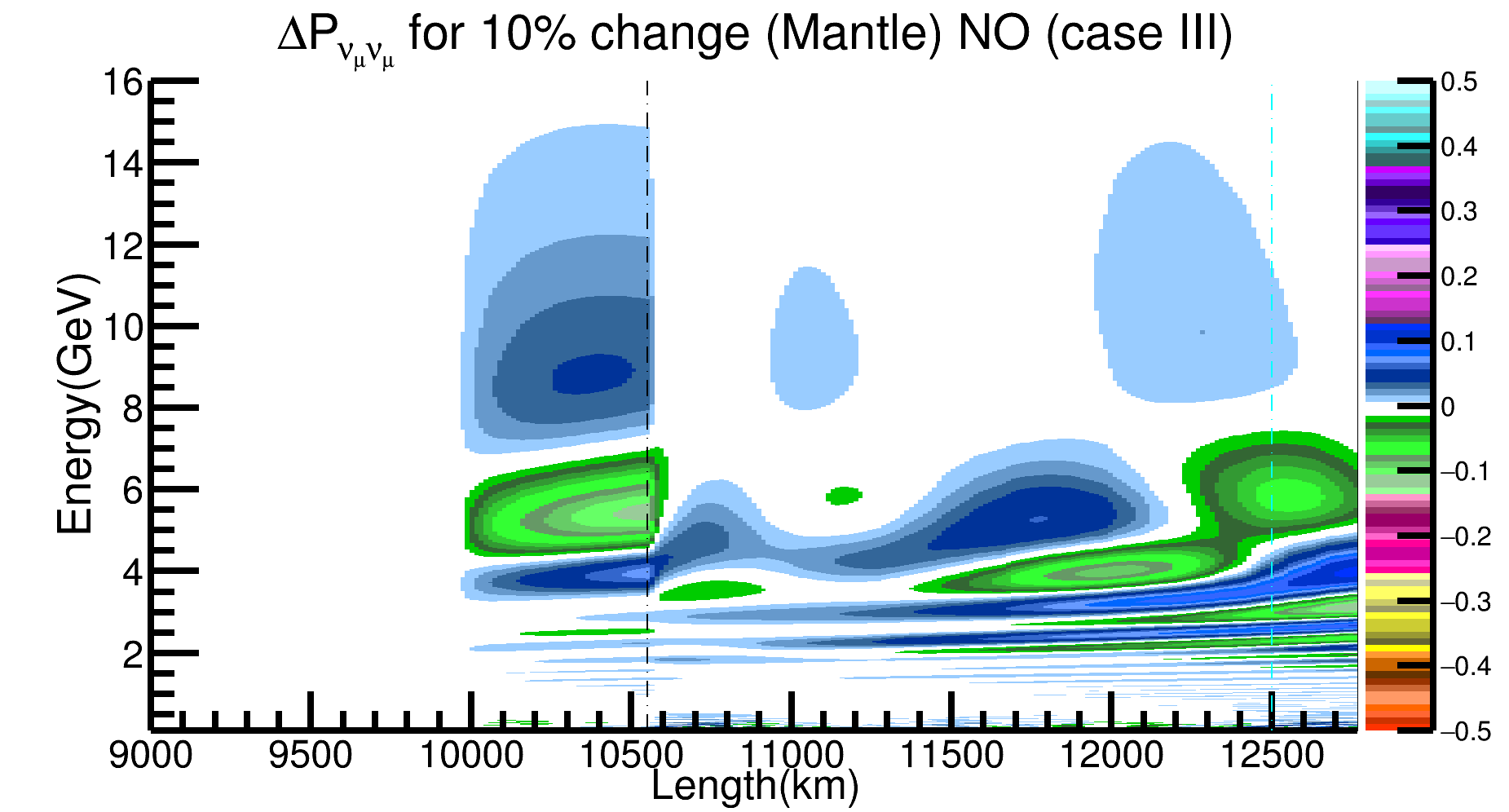}
    \includegraphics[width=0.3\textwidth]{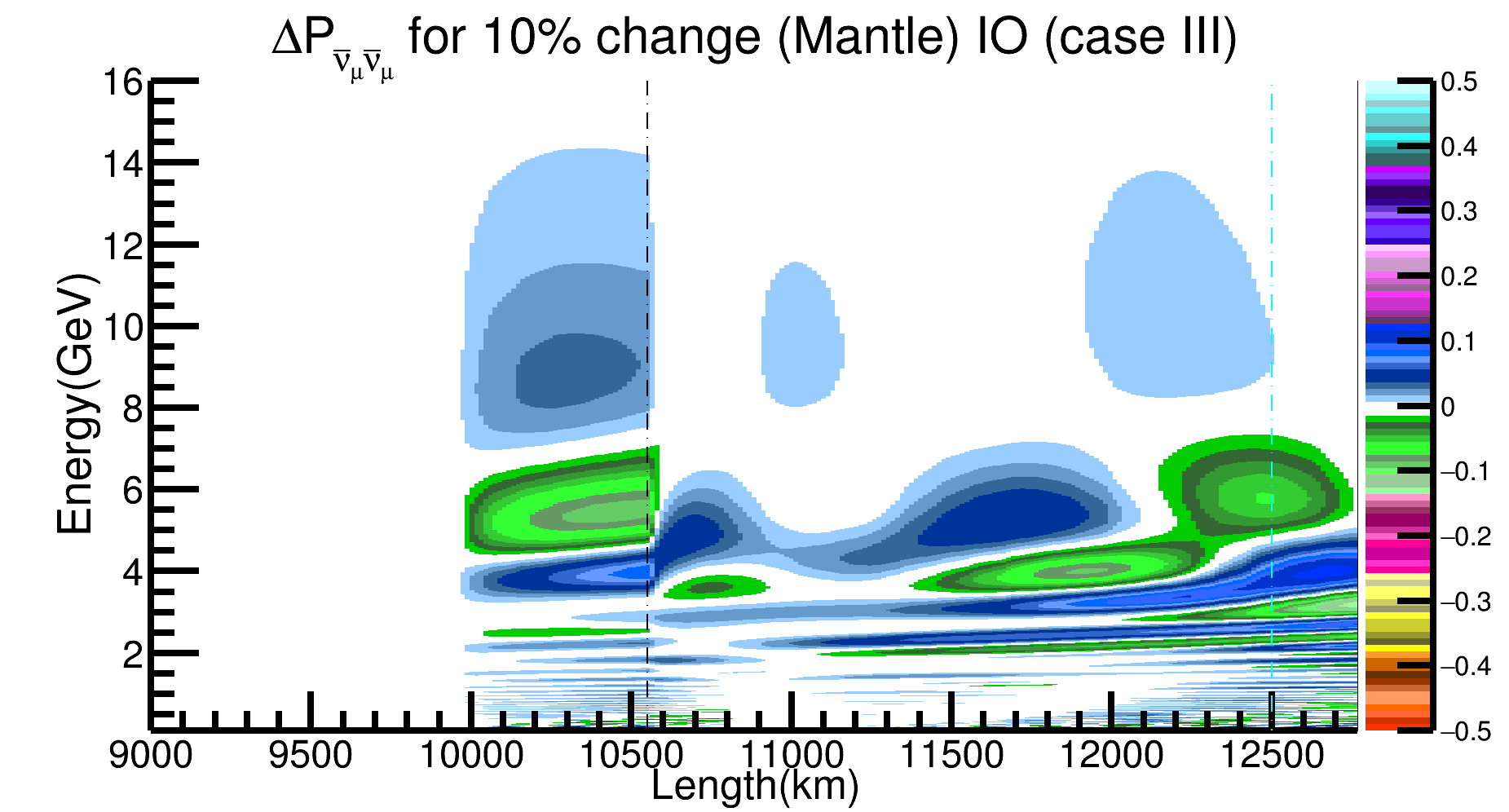}
    \caption{Top panel shows the change in density of mantle according to Case III. Middle panel shows the effect of density change on the survival probability via $\Delta P_{\nu_{\mu}\nu_{\mu}}=P_{\nu_{\mu}\nu_{\mu}}^{PREM}-P_{\nu_{\mu}\nu_{\mu}}^{newPREM}$ for NO and neutrinos, where $newPREM$ corresponds to density modified case. Lower panel is the same as panel middle panel but for IO and antineutrinos.}
    \label{fig:miprob}
\end{figure}

\begin{figure}
    \centering
    \includegraphics[width=0.3\textwidth]{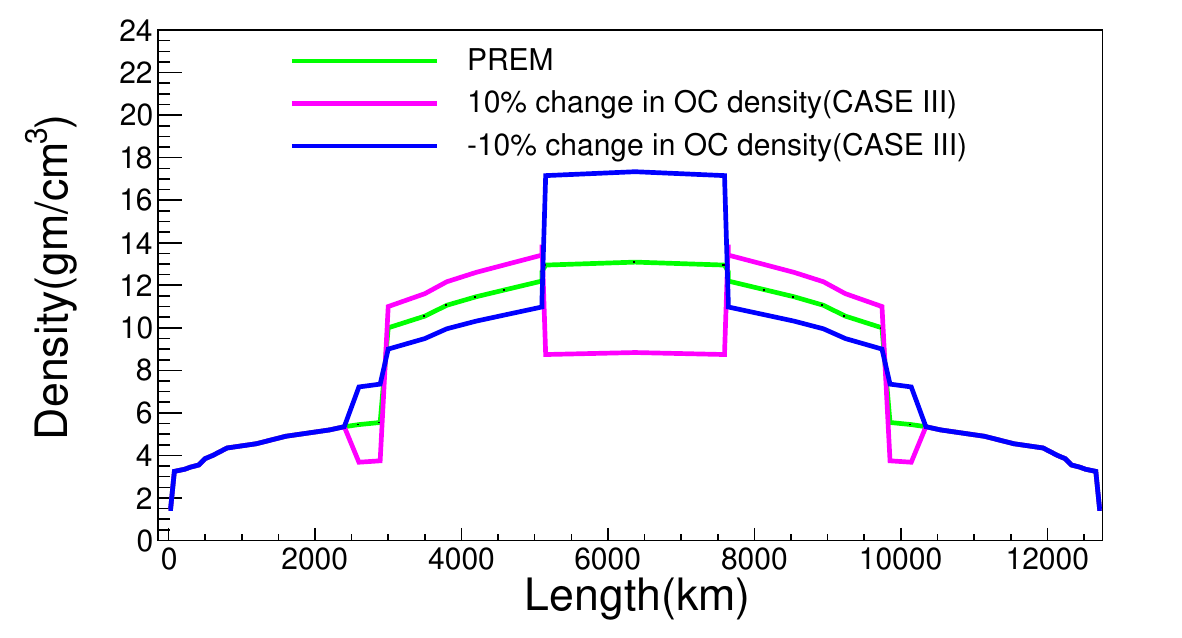}
    \includegraphics[width=0.3\textwidth]{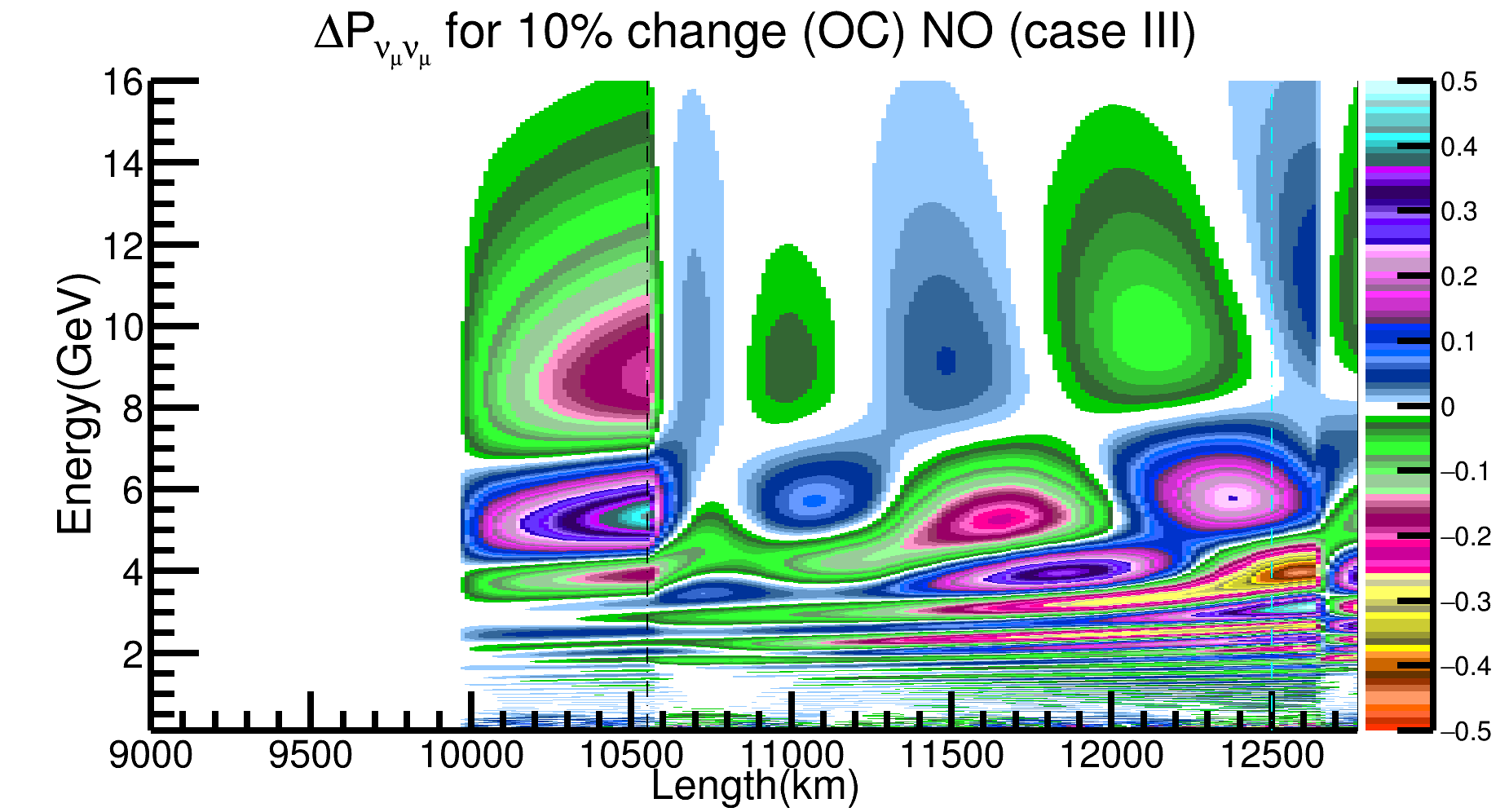}
    \includegraphics[width=0.3\textwidth]{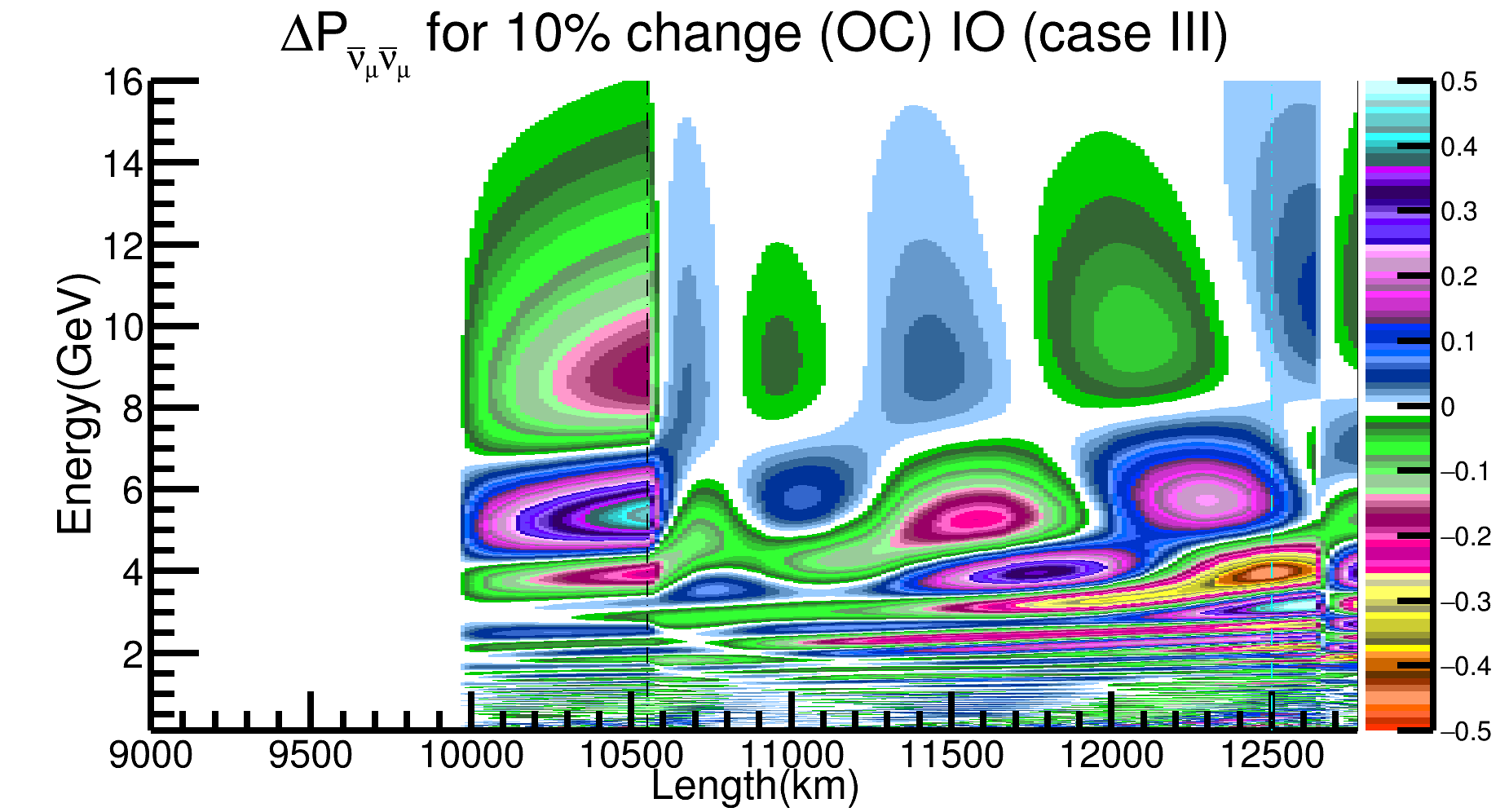}
    \caption{The panels shown here are the same as in Fig. \ref{fig:mnprob} but for density change in OC for Case III.}
    \label{fig:ociprob}
\end{figure}

In Case II, when we change density in any given layer, we make complimentary changes in all layers by the same amount, to compensate for the fixed mass of the Earth. But this is not exactly realistic. This is because not all layers of the Earth have the same uncertainty in their current measurements. In general, the density measurements in the deeper layers is more uncertain than the outer layers of the Earth. In particular, the outer regions are in general better studied and better understood via seismology, and hence the density in these regions suffers from fewer uncertainties. In our analysis for Case III, we take $d > 2200$ km as the inner regions where density measurements are taken to be uncertain. This essentially implies that we break the mantle into the outer mantle region, where the density of the layers is taken to be fixed, and an inner mantle region where the density is taken as variable. Therefore, in this case, the three regions of relevance to our discussion are the inner core, the outer core and the inner mantle. Hence, in this case when we change the density in a given region by $x\%$, we compensate for the fixed Earth mass by changing the density elsewhere by $y\%$, but now only for the layers in the inner core, outer core and inner mantle are changed. The relation between $x$ and $y$ is still given by a relation similar to Eq.~\ref{eq:y} but where the three relevant layers for compensation are only the inner core, outer core and inner mantle. 

We show the density change for inner mantle and outer core in the left panels of Figs.~\ref{fig:miprob} and \ref{fig:ociprob}, respectively. A comparison of Case III figures with the ones presented earlier for Cases I and II will come hand when understanding the expected sensitivity plots for ICAL@INO.

\section{Analysis Method \label{sec:analysis}}

\subsection{Experimental setup and simulation details}
\label{spec}

We give below a brief overview of our simulation framework.

The proposed India-based Neutrino Observatory (INO) to be built in India, will house a 50 kton magnetized Iron CALorimeter (ICAL) \cite{ICAL:2015st}. In this work we refer to this detector as ICAL@INO. The design proposal for ICAL@INO is to have a layered structure, with 5.6 cm thick iron slabs interlaced with RPCs (Resistive Plate Chambers) \cite{ICAL:2015st} as the active detector elements. The detector will be placed inside a 1.5 T magnetic field, making ICAL@INO magnetised. This gives ICAL@INO its charge identification sensitivity, allowing it to observe particles and anti-particles separately. Therefore, this detector will independently and efficiently observe $\nu_{\mu}$ and $\bar{\nu}_{\mu}$ produced in the Earth's atmosphere. The atmospheric $\nu_{\mu}$ ($\bar{\nu}_{\mu}$) will interact with the iron producing $\mu^-$ ($\mu^+$) and hadron(s). The $\mu^-$ ($\mu^+$) produce long track in the detector, while the hadron(s) produce a hadronic shower. Both the muon track and the hadron shower can be observed at ICAL@INO. The magnetic field at ICAL@INO bends the $\mu^-$ and $\mu^+$ in opposite directions, allowing the experiment to record the $\mu^-$ and $\mu^+$ track events separately. The length, curvature and direction of the track can be used to reconstruct the energy and zenith angle of the muon. The hadronic shower can be used to measure the energy of the hadron.  

We use the Honda 3D atmospheric neutrino fluxes computed for the Theni site in India \cite{honda}. Atmospheric neutrino events in ICAL are generated using the GENIE MC \cite{genie} tailored for the ICAL detector \cite{genie-ino}. Events are generated for 1000 years of ICAL running to reduce MC errors and then normalised to 25 years for our analysis. Events from GENIE are generated for unoscillated neutrino fluxes. Relevant neutrino oscillation probabilities are then included via the re-weighting algorithm \cite{Ghosh2013,barger-code}. On these raw Genie events we next implement the event reconstruction efficiencies, charge identification efficiency, energy and angle resolutions on muon events \cite{muon-reso} and energy resolution on hadron events \cite{hadron-resol}. The detector efficiencies and resolutions that we use have been obtained by the INO collaboration using the ICAL detector simulator based on the Geant4 simulation code \cite{geant4}. This gives us events in terms of their reconstructed energy and reconstructed zenith angle. The muon data is then binned in reconstructed muon energy and reconstructed muon zenith angle bins, while the hadron data is binned in reconstructed hadron energy bins only. Therefore, we have a three-pronged binned data and the binning scheme used in this work is shown in Table \ref{table:bin}. 

\begin{table}[h!]
\begin{center}
\scalebox{0.75}{
\begin{tabular}{ |c|c|c|c| } 
 \hline
 Observable & Range & Bin width & No. of bins\\ 
\hline
 $E^{obs}_{\mu}$(GeV)(15 bins) & [0.5,4] & 0.5 & 7\\
 						 & [4,7]    & 1   & 3\\ 
						& [7,11]   & 4   & 1 \\
						& [11,12.5] & 1.5 & 1\\
						&[12.5,15] & 2.5&1\\
						&[15,25] & 5&2\\
$cos(\theta^{obs}_{\mu}$ (21 bins)       & [-1.0,-0.98]&0.02&1\\
                        &[-0.98,-0.43]&0.05&11\\
                        &[-0.43,-0.4]&0.03&1\\
						&[-0.4,0.2]&0.10&2\\
						&[-0.2,1.0]&0.2&6\\
$E^{obs}_{had}$ (GeV)  (4 bins)                  &[0,2]&1&2\\
						&[2,4]&2&1\\
						&[4,15]&11&1\\

 \hline
\end{tabular}}
\caption{The binning scheme in the three observable $E_{\mu}^{obs}$, $\cos\theta_{\mu}^{obs}$ and $E_{had}^{obs}$ used in the analysis.}
\label{table:bin}
\end{center}
\end{table}



\subsection{Oscillation parameters}

The assumed true values for oscillation parameters used for simulating the prospective ICAL data are given in Table \ref{table:2}. These values are compatible with the current best-fit values obtained from global analysis of neutrino oscillation data \cite{Esteban_2020}. Since the value of $\theta_{23}$ and even its true octant is not yet known with any significance, we show results for three different possible true values $\theta_{23}$, $\theta_{23}=42^\circ$, $45^\circ$ and $49^\circ$. We also show results for both mass ordering (NO and IO). We used $\Delta m^{2}_{eff}$ in our analysis, defined as 
\begin{equation}
    \Delta m^{2}_{eff}= \Delta m^{2}_{31}-\bigg(\cos^{2}\theta_{12}-\cos\delta_{CP}\sin\theta_{13}\sin2\theta_{12}\tan\theta_{23}\bigg)\Delta m^{2}_{21}.
\end{equation}
The $\chi^2$ defined below is minimised over $\theta_{23}$ in the range $40^{o}$ to $51^{o}$ and $\Delta m^{2}_{eff}$ in the range given in the Table. We have taken $\delta_{CP}=0^{\circ}$ and kept it fixed in the analysis since the ICAL data is very weakly dependent on $\delta_{CP}$. $\Delta m^{2}_{21}$, $\theta_{13}$ and $\theta_{12}$ are also kept fixed in our analysis at the their values given in the Table.

\begin{table}[h!]
\begin{center}
\scalebox{0.7}{
\begin{tabular}{ |c|c|c|c|c|c|c| } 
 \hline
&$\Delta m^{2}_{21}$(eV$^{2})$  & $\Delta m^{2}_{31}$(eV$^{2})$ & $\sin^{2}\theta_{12}$ & $\sin^{2}\theta_{23}$ & $\sin^{2}2\theta_{13}$ & $\delta_{CP}$\\ 
\hline
true value& 7.42$\times 10^{-5}$ & 2.531$\times 10^{-3}$  & 0.33 & 0.5    & 0.0875  & $0^{\circ}$\\ 				
 \hline
minimization &fixed & $[2.418,2.650] \times 10^{-3}$  & fixed & [40$^\circ, 51^\circ$]    & fixed  & fixed		\\		\hline
\end{tabular}
}
\caption{Assumed true values and minimization ranges of the neutrino oscillation parameters used in the analysis. }
\label{table:2}
\end{center}
\end{table}

\subsection{The $\chi^{2}$ Formula}

We generate data for the PREM profile of the Earth. Then we fit the generated data by a theory, where we modify the Earth density profile within the schemes discussed in the previous section. For statistical analysis of the data we define the following test statistics 
 \begin{equation}    \chi^2 = \chi^2_{\mu^-}+ \chi^2_{\mu^+}\,,
\end{equation}

where,\\

\begin{eqnarray}
\resizebox{1\hsize}{!}{$
\chi^{2}_{\mu^{\pm}} = \sum_{i=1}^{N_{E_{\mu}}}\sum_{j=1}^{N_{\theta_{\mu}}}\sum_{k=1}^{N_{E_{H}}}  2\bigg[\bigg(T^{\pm}_{ij(k)}-D^{\pm}_{ij(k)}\bigg) -D^{\pm}_{ij(k)}\ln \bigg(\frac{T^{\pm}_{ij(k)}}{D^{\pm}_{ij(k)}}\bigg)\bigg] + \sum^{5}_{l^{\pm}=1}\xi^{2}_{l^{\pm}} \,,
$}
\label{eq:chisq}
 \end{eqnarray}
 
where the sum is over muon energy bins ($i=1$ to $N_{E_\mu}$) muon zenith angle bins ($j=1$ to $N_{\theta_\mu}$) and hardon energy bins ($k=1$ to $N_{E_H}$), $D^{\pm}_{ij(k)}$ is the simulated $\mu^\pm$ data binned in muon energy, muon zenith angle and hadron energy bins and $T^{\pm}_{ij(k)}$ is the corresponding systematic uncertainty weighted prediction for a given theoretical model in the same bin and is given as
\begin{equation}
T^{\pm}_{ij(k)} = T^{0\pm}_{ij(k)}\bigg(1+\sum^{5}_{l^{\pm}=1}\pi^{l^{\pm}}_{ij(k)}\xi_{l^{\pm}}\bigg)\,,
\end{equation}
where $T^{0\pm}_{ij(k)}$ gives the number of $\mu^{\pm}$ events without systematic errors in theory. We consider five kinds of systematic errors in muons and antimuon data seperately, giving a total of 10 systematic uncertainties ($\pi^{l^\pm}$) and pulls ($\xi_{l^\pm}$). The systematic uncertainties considered are $20\%$ flux normalization error, $10\%$ cross-section error, $5\%$ tilt error, $5\%$ zenith angle error and $5\%$ overall systematic, in both $\mu^+$ and $\mu^-$ channels

\section{Results}
\label{sec:results}

In this section we present our numerical results and quantify how well ICAL is able to resolve the density profile of the Earth. We present the results for the three Cases mentioned before and separately for the mantle and outer core. In each case the data is generated for the standard PREM density profile. This data is then fitted by changing the density for a given Case as a percentile and the corresponding $\chi^2$ is plotted as a function of this given percentile change. 

\subsection{Case I}
We start with considering Case I, which is the simplest case where we consider percentage density variation in a given layer of the Earth, say mantle or outer core, without putting any other constraint on the density. 

\subsubsection{Mantle}

\begin{figure*}
    
    \begin{minipage}[t]{0.3\textwidth}
   \includegraphics[width=\linewidth]{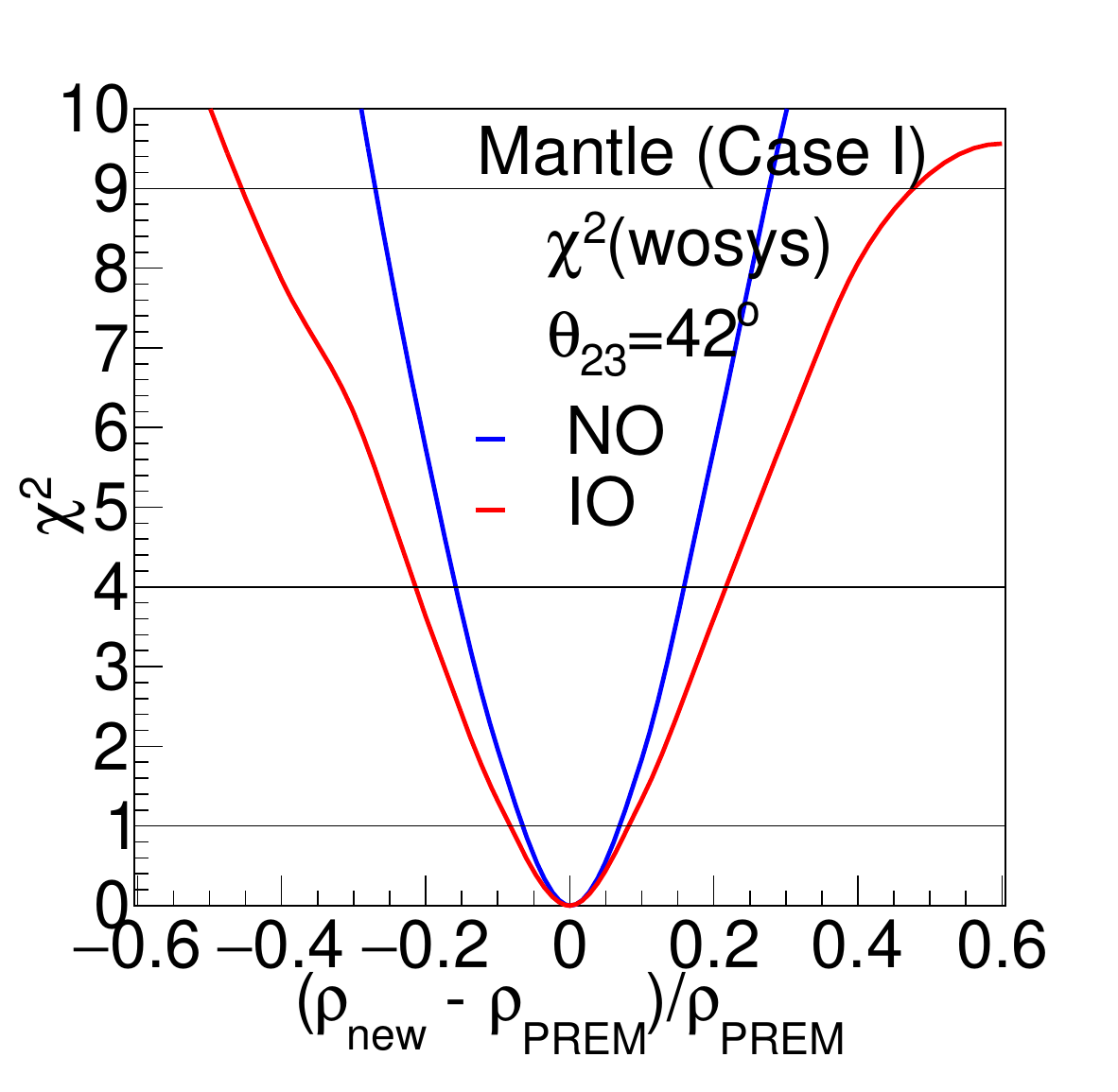} 
   \end{minipage}
   \begin{minipage}[t]{0.3\textwidth}
       \includegraphics[width=\linewidth]{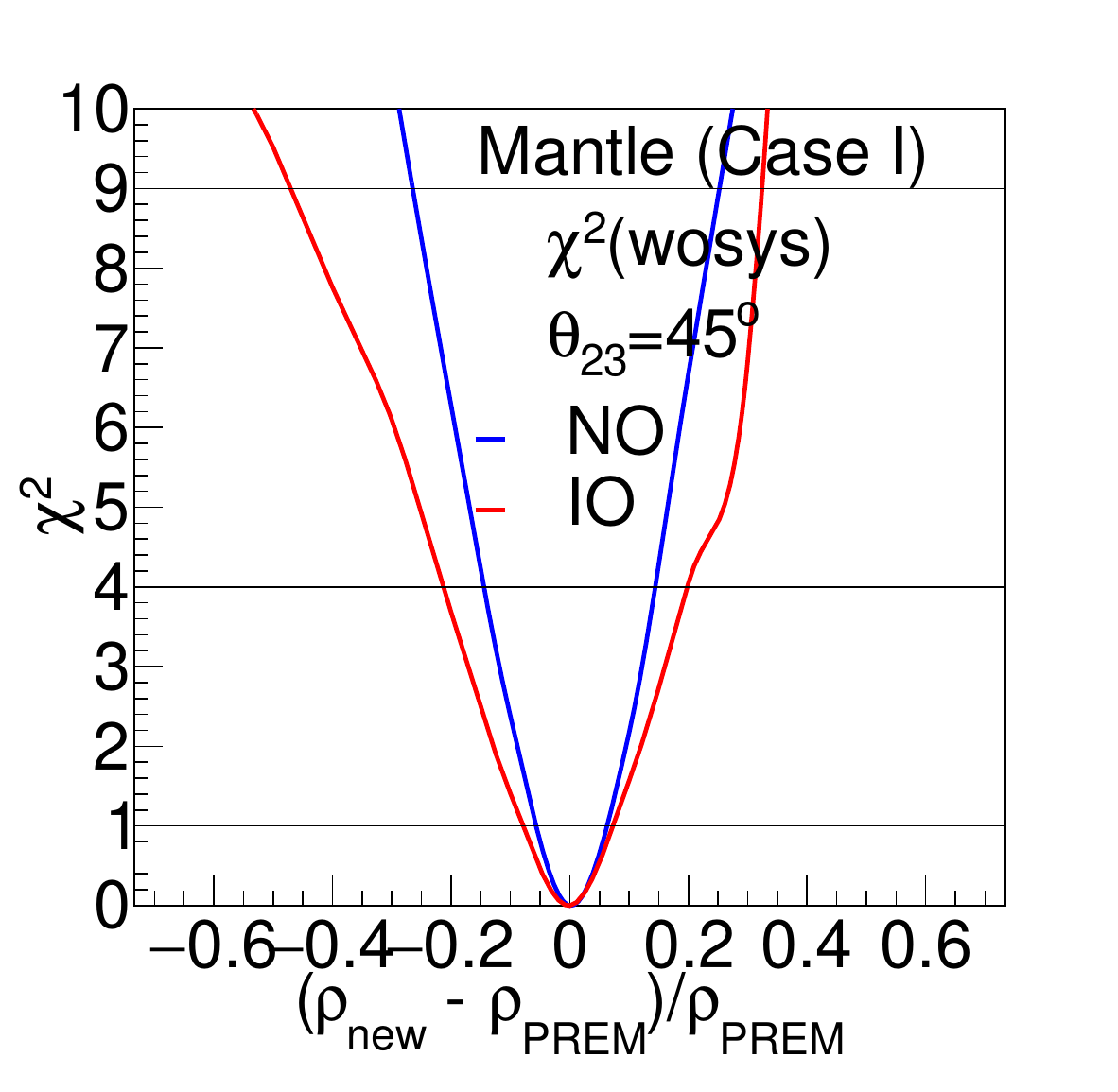}
   \end{minipage}
    \begin{minipage}[t]{0.3\textwidth}
    \includegraphics[width=\linewidth]{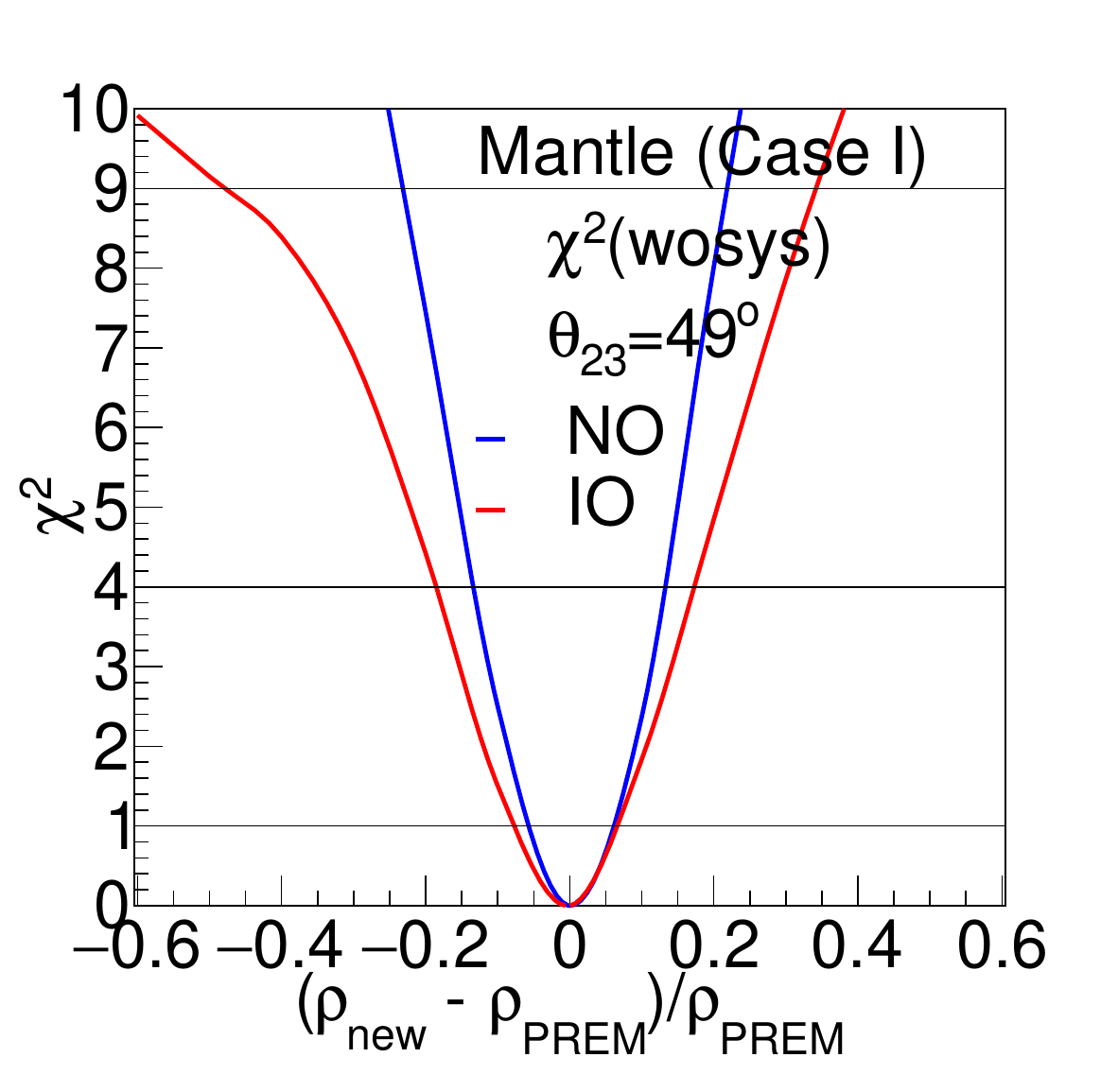}
     \end{minipage}

     \begin{minipage}[t]{0.3\textwidth}
    \includegraphics[width=\linewidth]{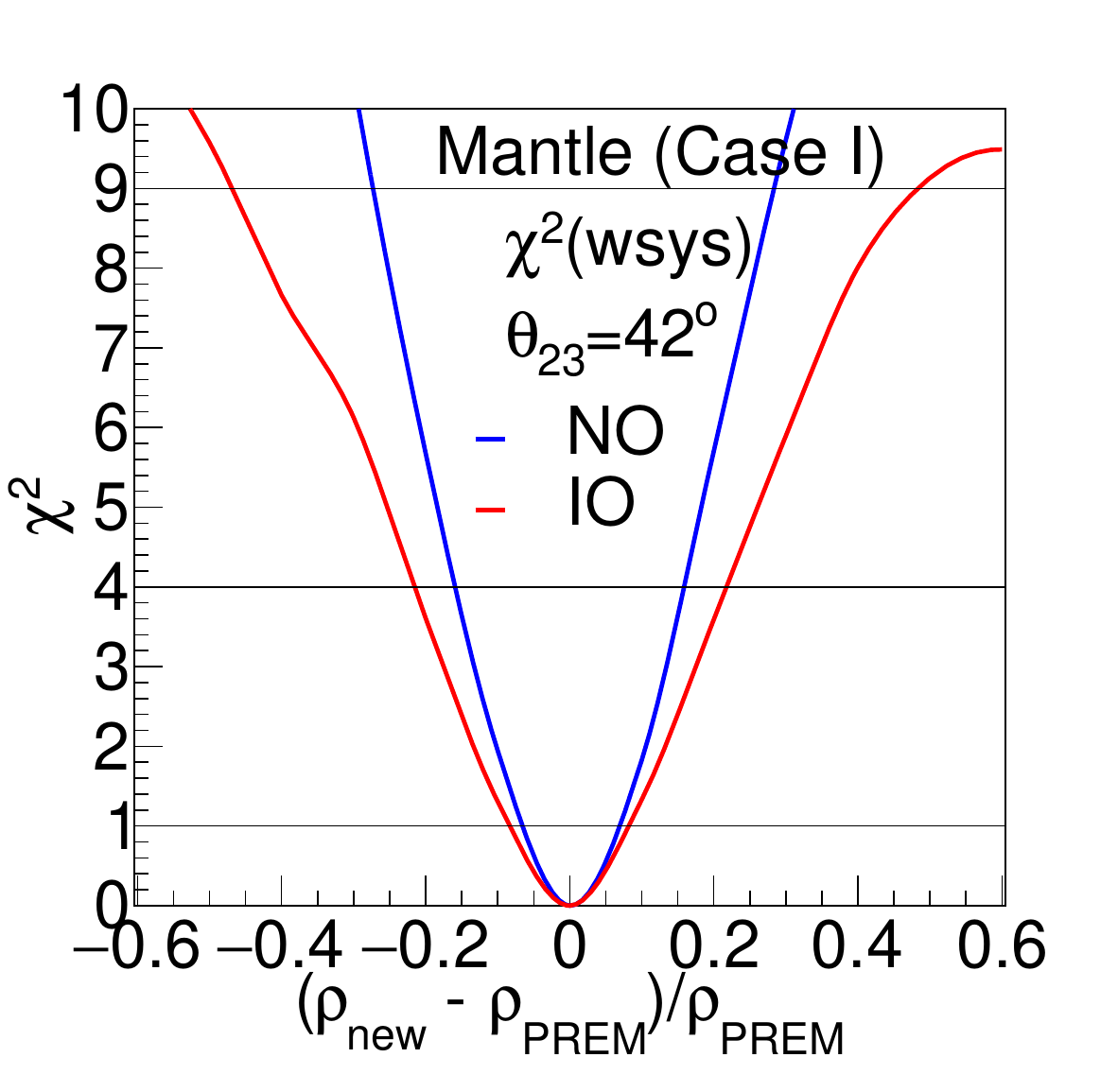} 
    \end{minipage}
    \begin{minipage}[t]{0.3\textwidth}
    \includegraphics[width=\linewidth]{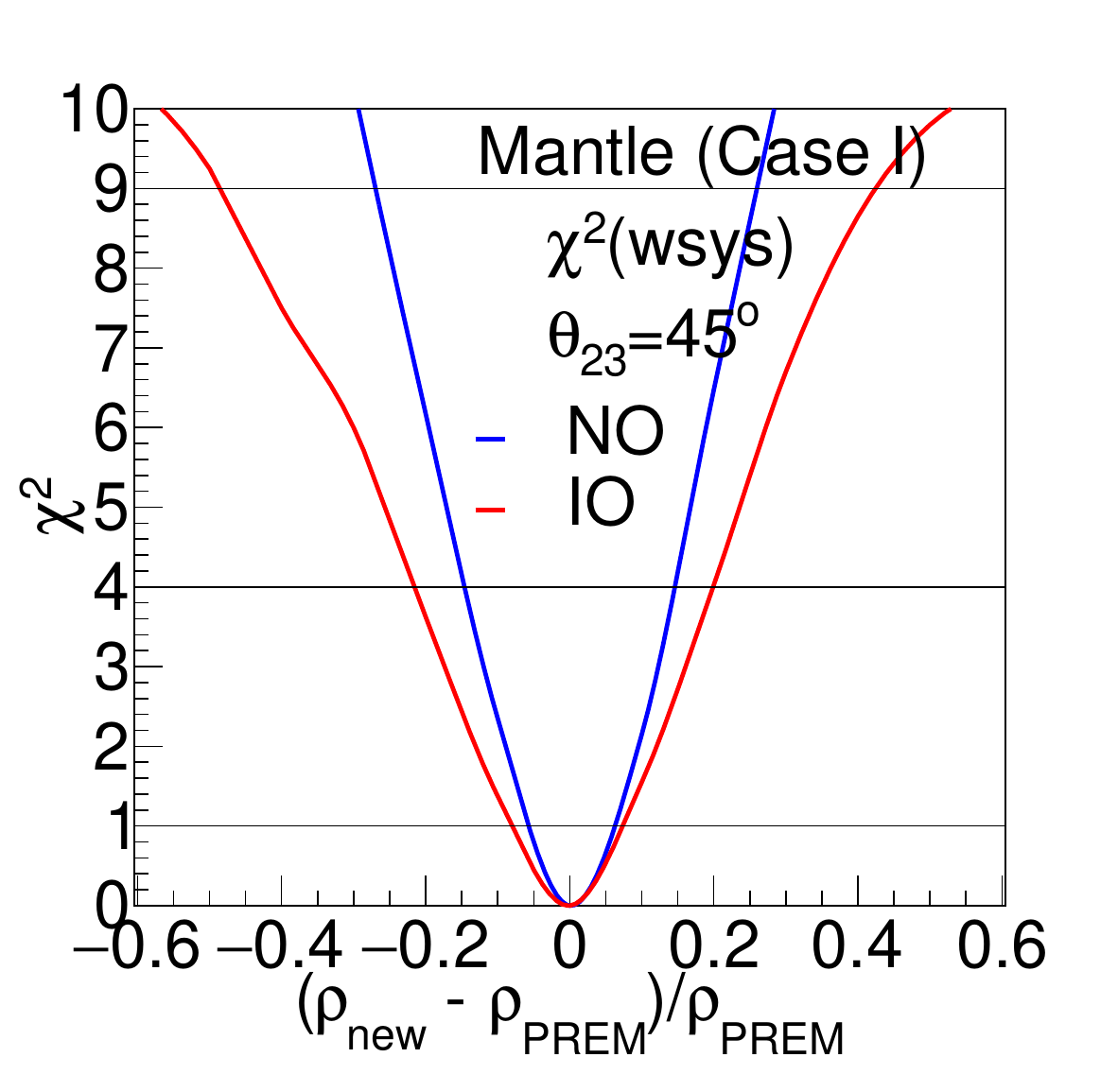}
    \end{minipage}
    \begin{minipage}[t]{0.3\textwidth}
    \includegraphics[width=\linewidth]{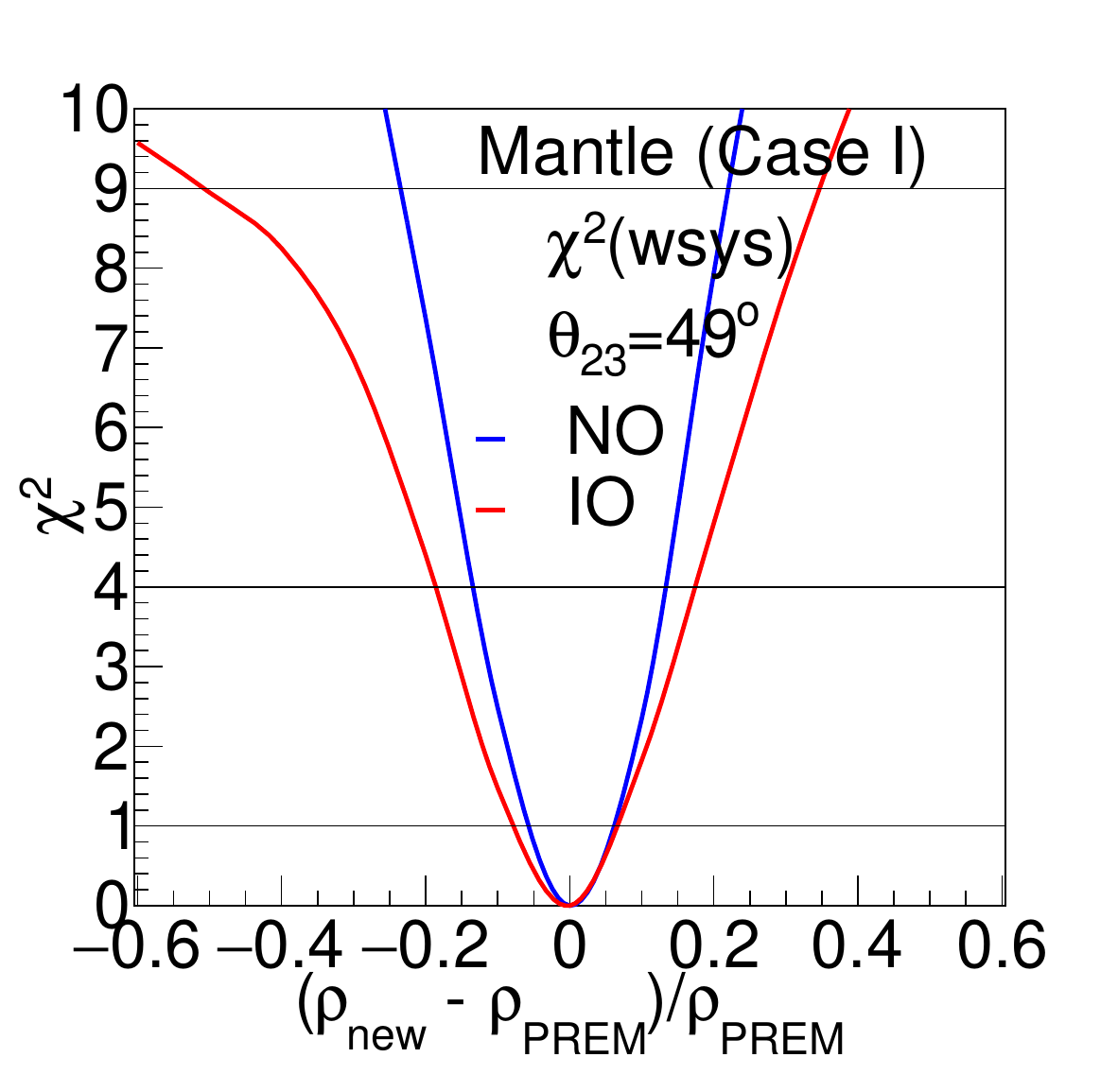}
    \end{minipage}
    
    \caption{The $\chi^{2}$ as a function of percentage change in density in the mantle for Case I. Red lines are for NO and blue lines are for IO. Panel (a) is for  $\theta_{23}=42^{\circ}$, (b) for $\theta_{23}=45^{\circ}$ and (c) for $\theta_{23}=49^{\circ}$. Upper panels are for no systematic uncertainties in the analysis while the lower panels show the $\chi^2$ including systematic uncertainties.}
    \label{fig:mnchiws}
\end{figure*}

Fig. \ref{fig:mnchiws} illustrates the expected sensitivity of ICAL@INO to the density of the mantle in Case I, where no constraint is placed on the Earth's mass. We show results with no systematic errors (upper plots) and with systematic errors (bottom plots). The figures depict the $\chi^2$ as a function of the percentage density variation in the mantle. The results shown are for NO (blue line) and IO (red line) in each panel. In Fig. \ref{fig:mnchiws}, the left panels are for $\theta_{23} = 42 ^{\circ}$, the middle panels are for $\theta_{23} = 45 ^{\circ}$, and the right panels are for  $\theta_{23} = 49 ^{\circ}$. 

\begin{table}[!h]
\centering
\caption{The range of density variation values for which ICAL@INO is sensitivity to the Mantle density at $1\sigma$, $2\sigma$ and 3$\sigma$, for Case I. We show these ranges for 3 choices of $\theta_{23}$ and for both NO and IO. } 
\begin{tabular}{|c|c|c|c|c|c|c|}
\hline
  &  \multicolumn{2}{|c|}{$\theta_{23}=42^{\circ}$} & \multicolumn{2}{|c|}{$\theta_{23}=45^{\circ}$} & \multicolumn{2}{|c|}{$\theta_{23}=49^{\circ}$}\\
\hline
{C.L.}   & NO   & IO    & NO   & IO & NO   & IO\\
 \hline
 1$\sigma$  &  -6.6/7.1 & -8.5/8.4   & -5.4/6.3  & -7.6/8 & -5.7/6.1   & -7.5/6.8\\
 2$\sigma$  &  -15.5/16.4 & -21/22   & -13/15  & -20/20 & -13.3/13.4   & -18.5/17.5\\
 3$\sigma$  & -26/29.6  & -46/49   & -24/28  & -46/44 & -23.5/22   & -50/34\\
\hline
\end{tabular}
\label{tab:case1mantle}
\end{table}


In Fig. \ref{fig:mnchiws}, we see that the $\chi^2$ curve is symmetric for small percentage changes in mantle density. However, as the density change increases, the curves become asymmetric with respect to zero. This asymmetry is seen to be largest for $\theta_{23}=49 ^{\circ}$. We also notice that the sensitivity for NO is generally higher than for IO because the neutrino data is statistically stronger than the antineutrino data. Finally, a comparison of upper and lower panels reveals that the sensitivity gets slightly worse with systematic uncertainties. In Table \ref{tab:case1mantle} we show the percentage uncertainty expected in the density in Mantle for Case 1 within $1\sigma$, $2\sigma$ and $3\sigma$ C.L. For example, the first column shows that if $\theta_{23}=42^\circ$, then ICAL@INO can determine the density of mantle with $-15.5\%$ and $+16.4\%$ uncertainty at 2$\sigma$ C.L.

\subsubsection{Outer Core (OC)}

\begin{figure*}
    
  \includegraphics[width=0.3\textwidth]{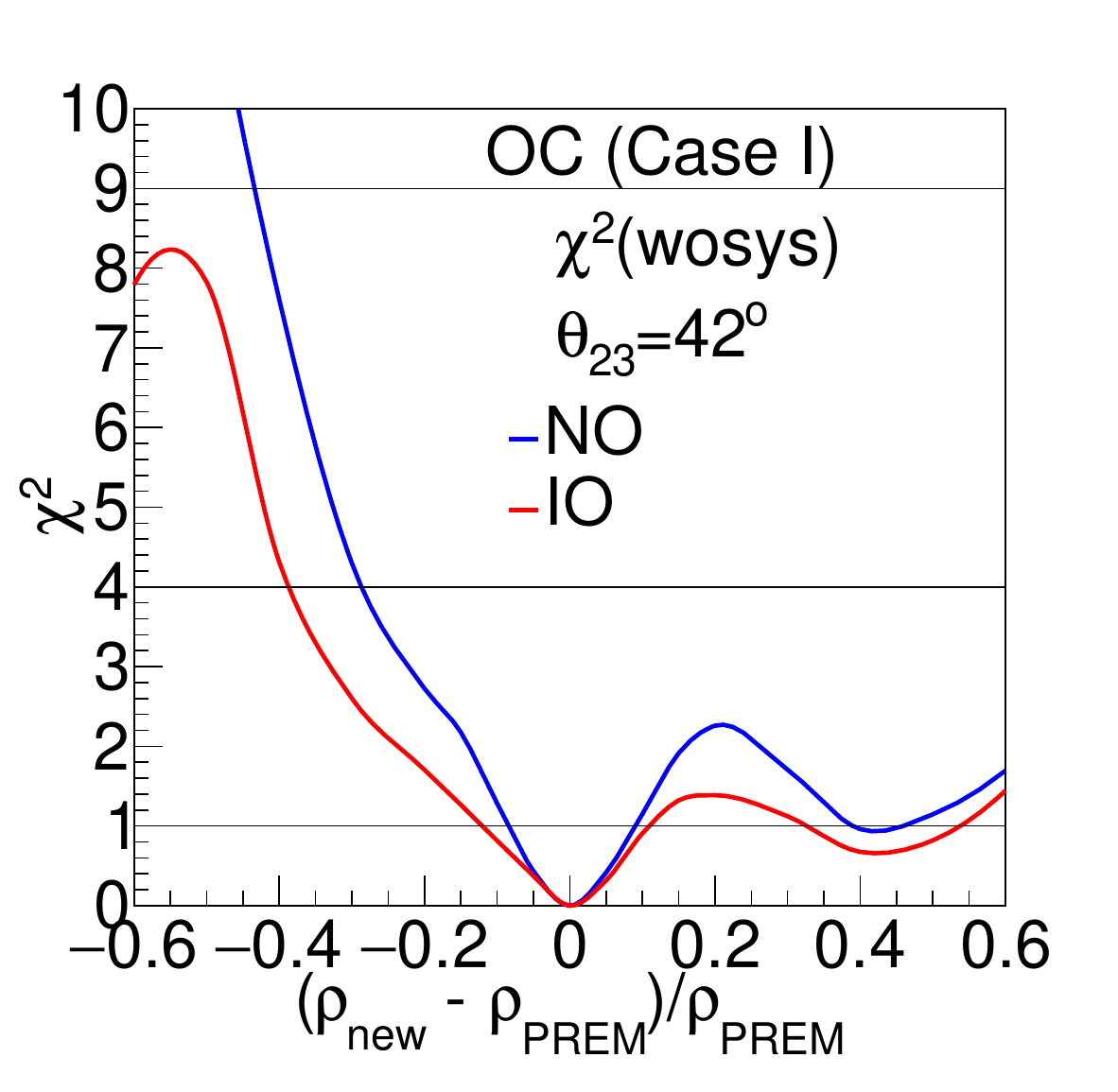}
   \includegraphics[width=0.3\textwidth]{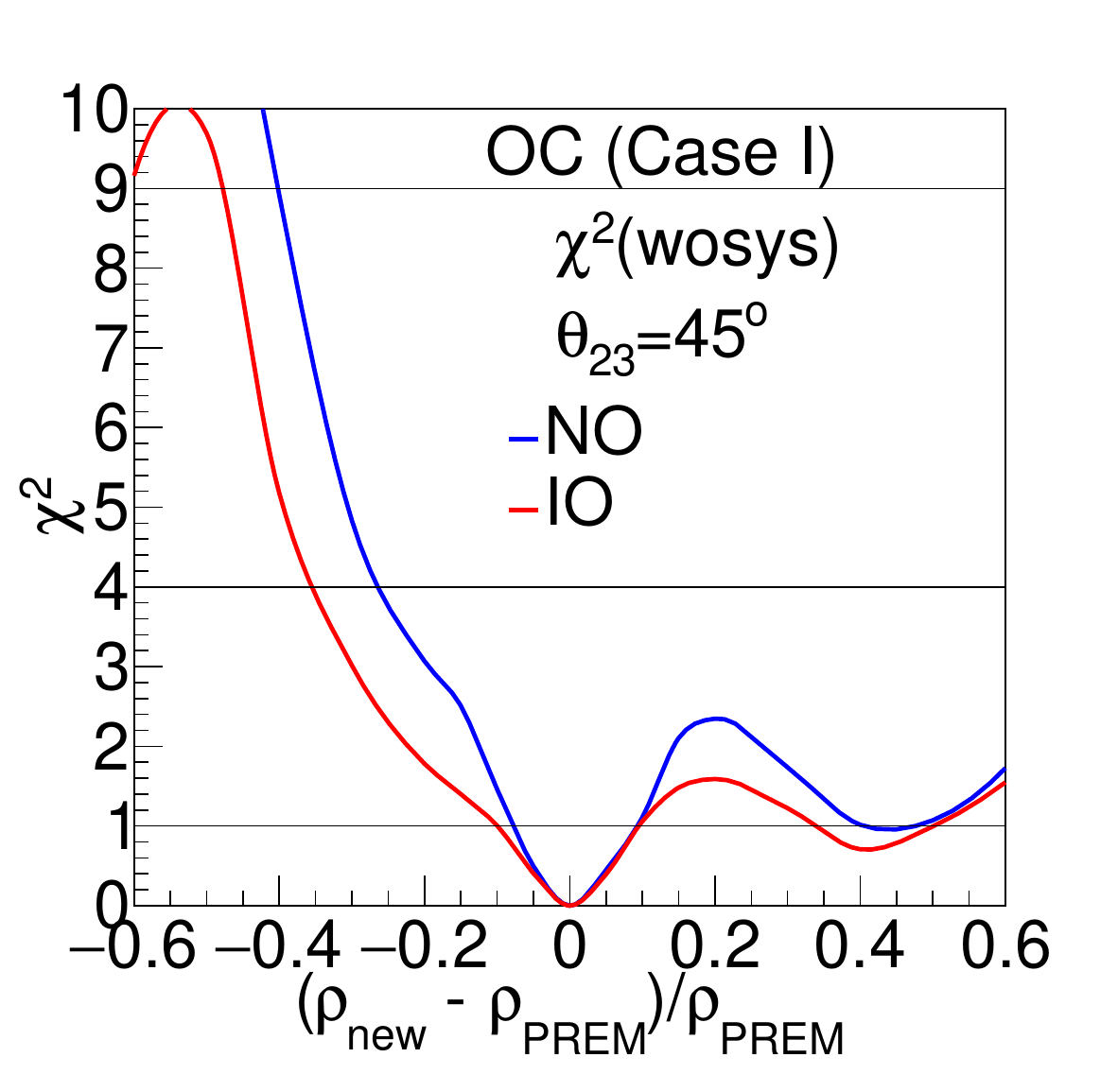}
    \includegraphics[width=0.3\textwidth]{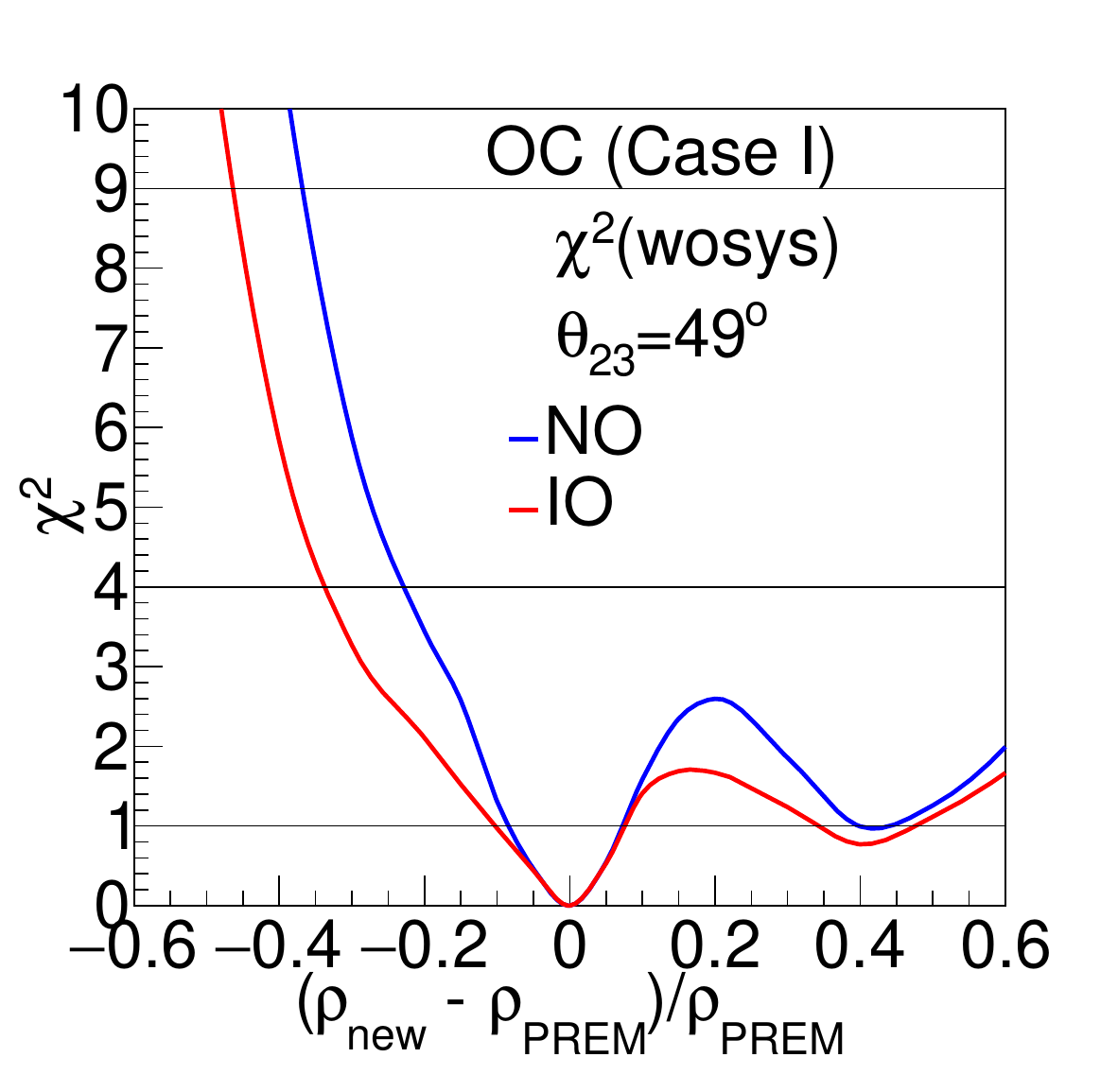}
    
    \includegraphics[width=0.3\textwidth]{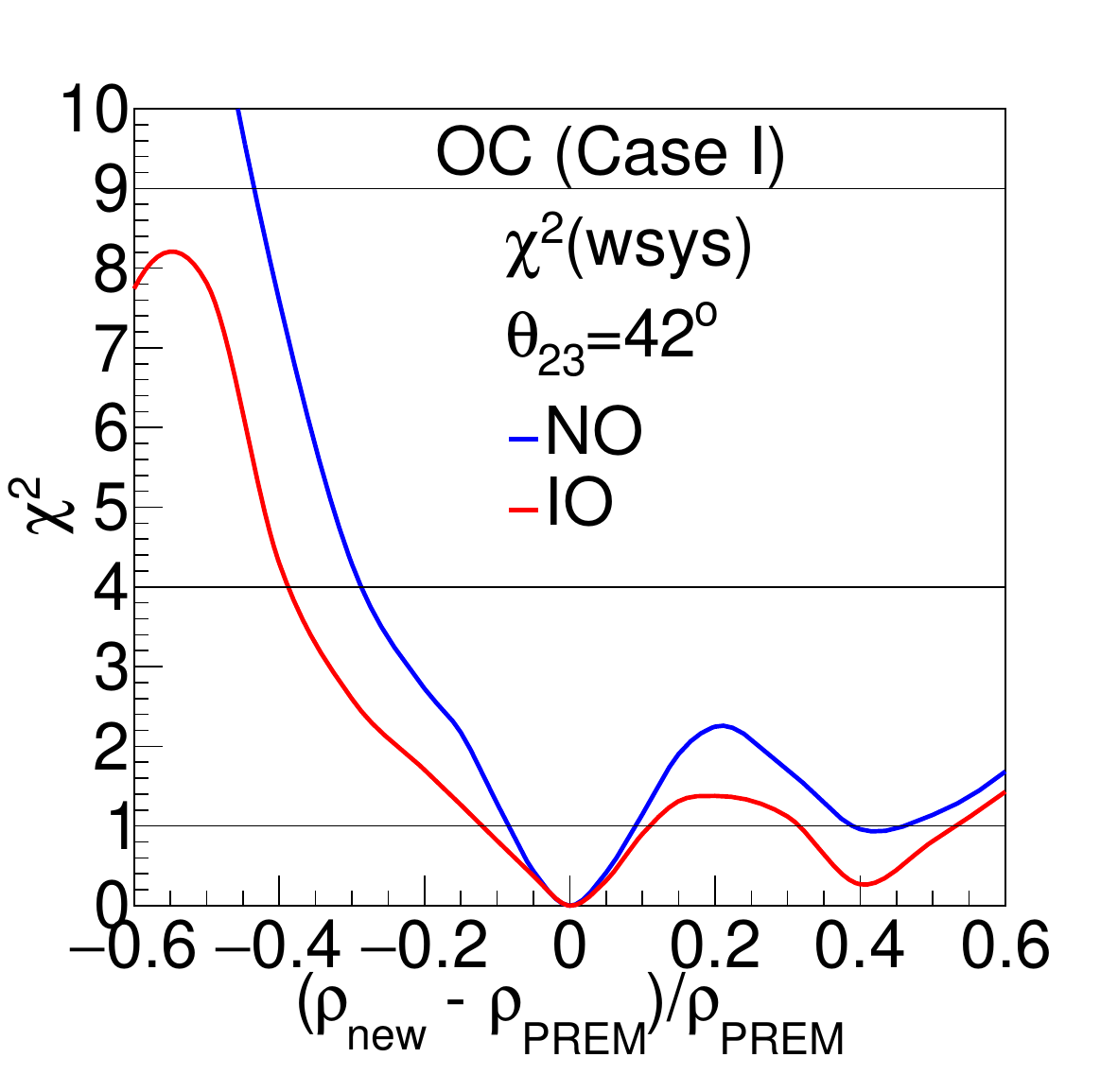}
   \includegraphics[width=0.3\textwidth]{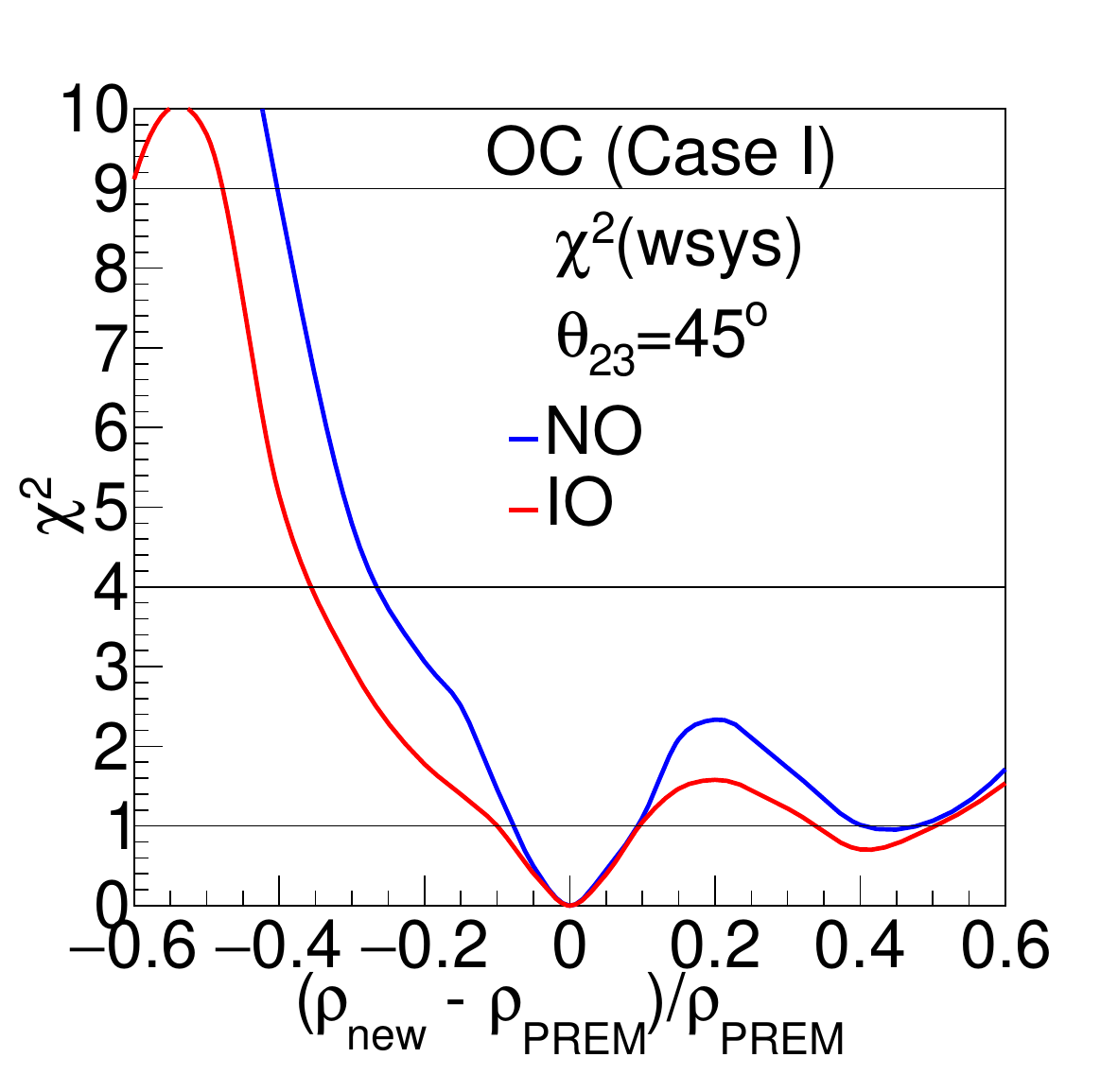}
   \includegraphics[width=0.3\textwidth]{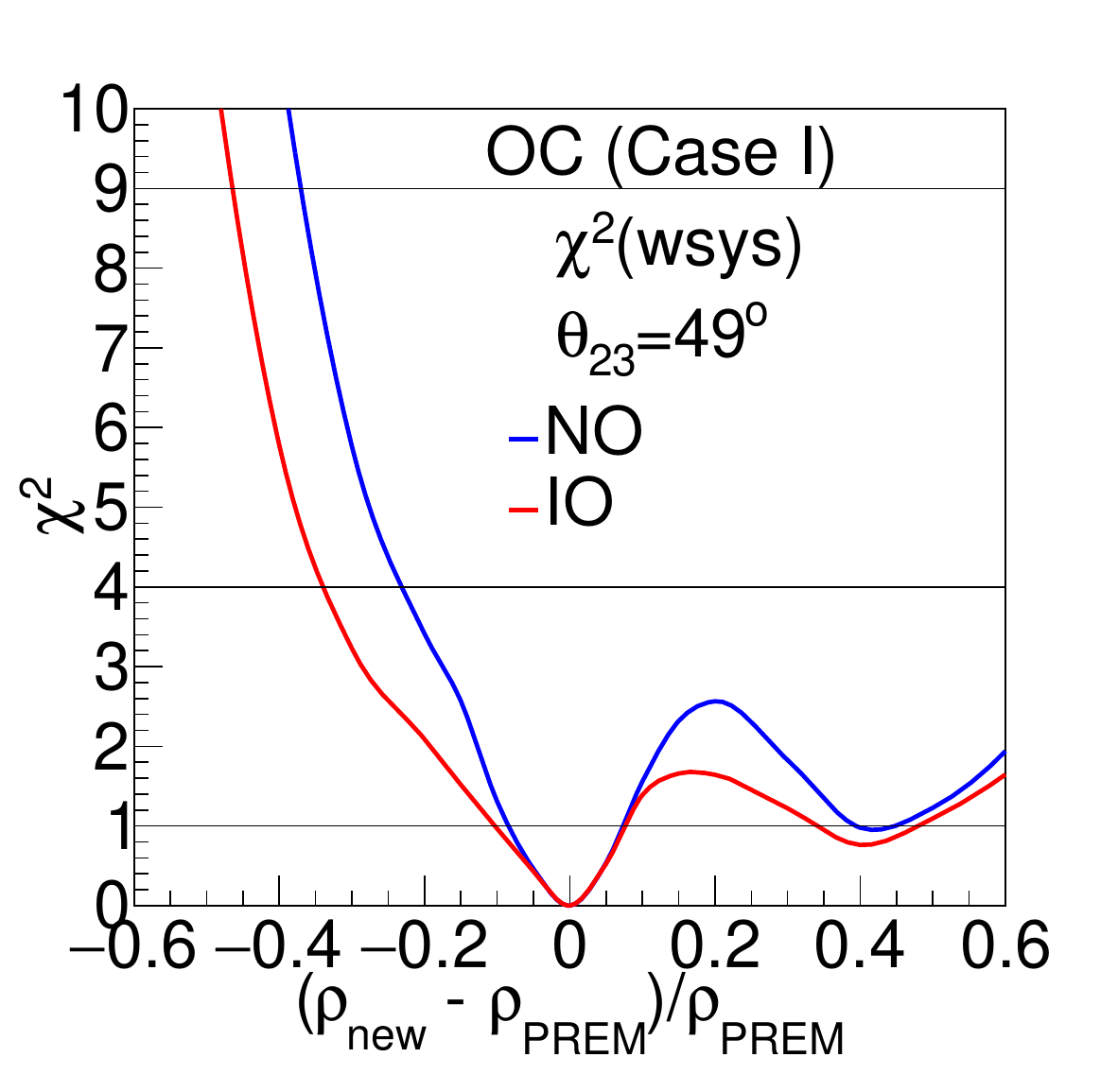}
    \caption{Same as Fig. \ref{fig:mnchiws} but for percentage density change in OC.}
    \label{fig:ocnchiws}
\end{figure*}

Fig. \ref{fig:ocnchiws} depicts the sensitivity of ICAL@INO to the density of the OC for Case I. Similar to Fig. \ref{fig:mnchiws}, we show our results for both without systematic errors (upper plots) and with systematic errors (bottom plots), for NO and IO as well as for 3 choices of $\theta_{23}$. The figures show $\chi^{2}$ as a function of percentage density variation in OC density without any constraint on the Earth's mass.

 \begin{table}[!h]
\centering
\caption{\label{tab:case1oc} The range of density variation values for which ICAL@INO is sensitive to the OC density at $1\sigma$, $2\sigma$ and 3$\sigma$, for Case I. We show these ranges for 3 choices of $\theta_{23}$ and for both NO and IO. } 
\begin{tabular}{|c|c|c|c|c|c|c|}
\hline
  &  \multicolumn{2}{|c|}{$\theta_{23}=42^{\circ}$} & \multicolumn{2}{|c|}{$\theta_{23}=45^{\circ}$} & \multicolumn{2}{|c|}{$\theta_{23}=49^{\circ}$}\\
\hline
{C.L.}   & NO   & IO    & NO   & IO & NO   & IO\\
 \hline
 1$\sigma$  &  -8.1/8.7 & -11.6/10.4   & -7.5/9.1  & -9.6/9.4 & -8.3/7.4   & -9.9/7.4\\
 2$\sigma$  &  -29/ & -39/   & -26/  & -34/ & -23/   & -33/\\
 3$\sigma$  & -44/  & -/   & -39/  & -47/ & -37/   & -46/\\
\hline
\end{tabular}
\label{tab:case1mantle}
\end{table}


We notice features that rather different as compared to the case with the Mantle. In particular, we see that for the case of OC, the value of $\chi^2$ does not monotonically increase with the percentage change in density. Instead, the value reaches a maximum and then falls, oscillating with the percentage change in density. We see that both the magnitude of $\chi^2$ and the position of its maxima depends on the value of $\theta_{23}$. The above is true for both NO and IO cases. We also notice that, unlike in the case of the mantle, the $\chi^2$ is now very asymmetric for positive and negative variation in density. Indeed, the $\chi^{2}$ is significantly lower for positive variations as compared to negative variations in OC density, and never even reaches $\chi^{2}=4$ for any of the $\theta_{23}$ cases and for both NO and IO. The $\chi^{2}$ is seen to be lower for IO as compared to NO for all plots. Table \ref{tab:case1oc} gives the range of density variation values for which ICAL@INO is sensitive to the OC density at $1\sigma$, $2\sigma$ and 3$\sigma$, for Case I. We show these ranges for 3 choices of $\theta_{23}$ and for both NO and IO.

\subsection{Case II}

For the Case II which we described in Sec 3.2, we put the constraint that the total mass of the Earth is fixed. This constraint therefore, propagates the effect of change of density in one layer to all other density layers of the Earth.

 \subsubsection{Mantle}

\begin{figure*}
    \centering
    \includegraphics[width=0.3\textwidth]{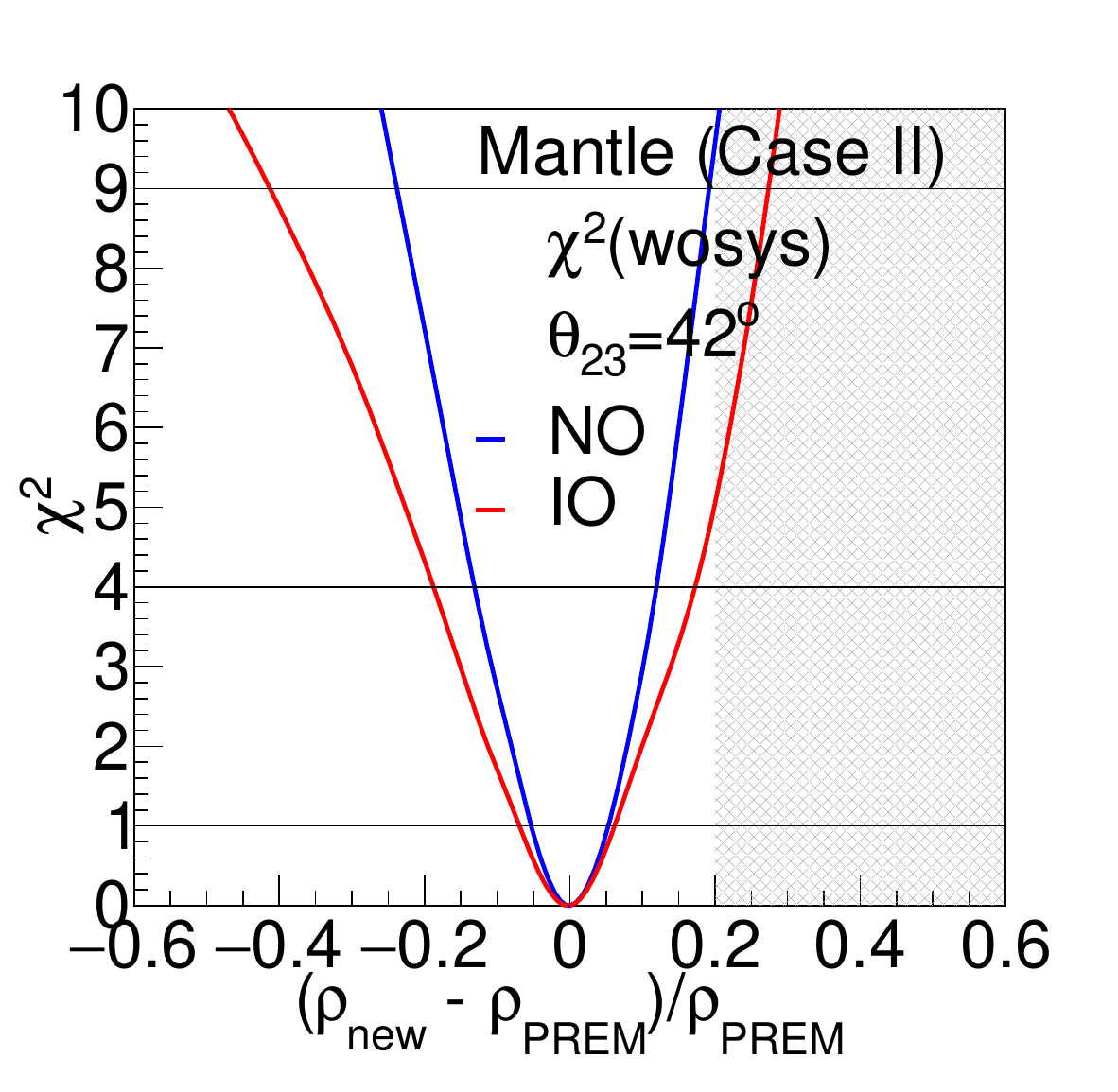} 
   \includegraphics[width=0.3\textwidth]{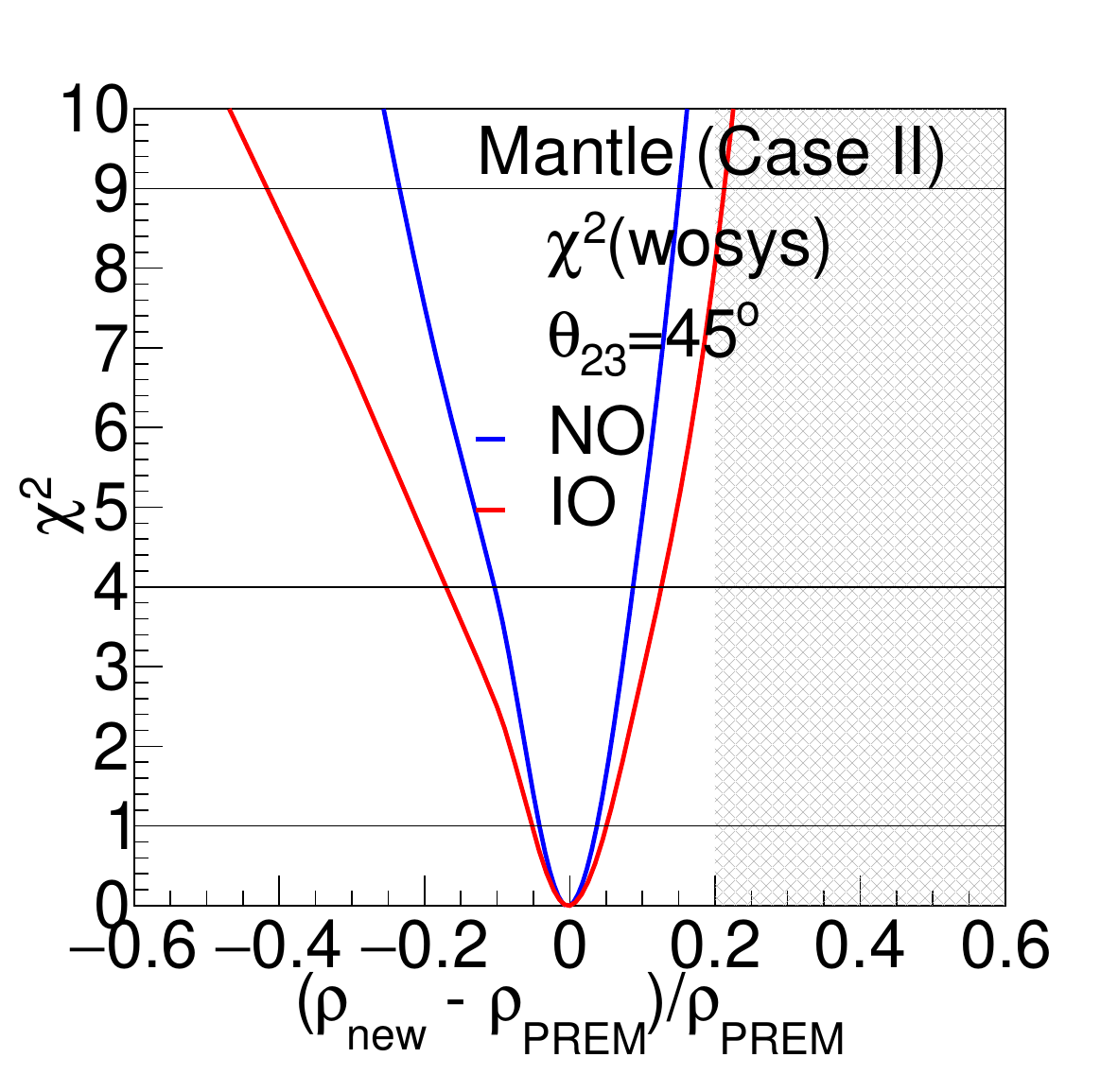}
   \includegraphics[width=0.3\textwidth]{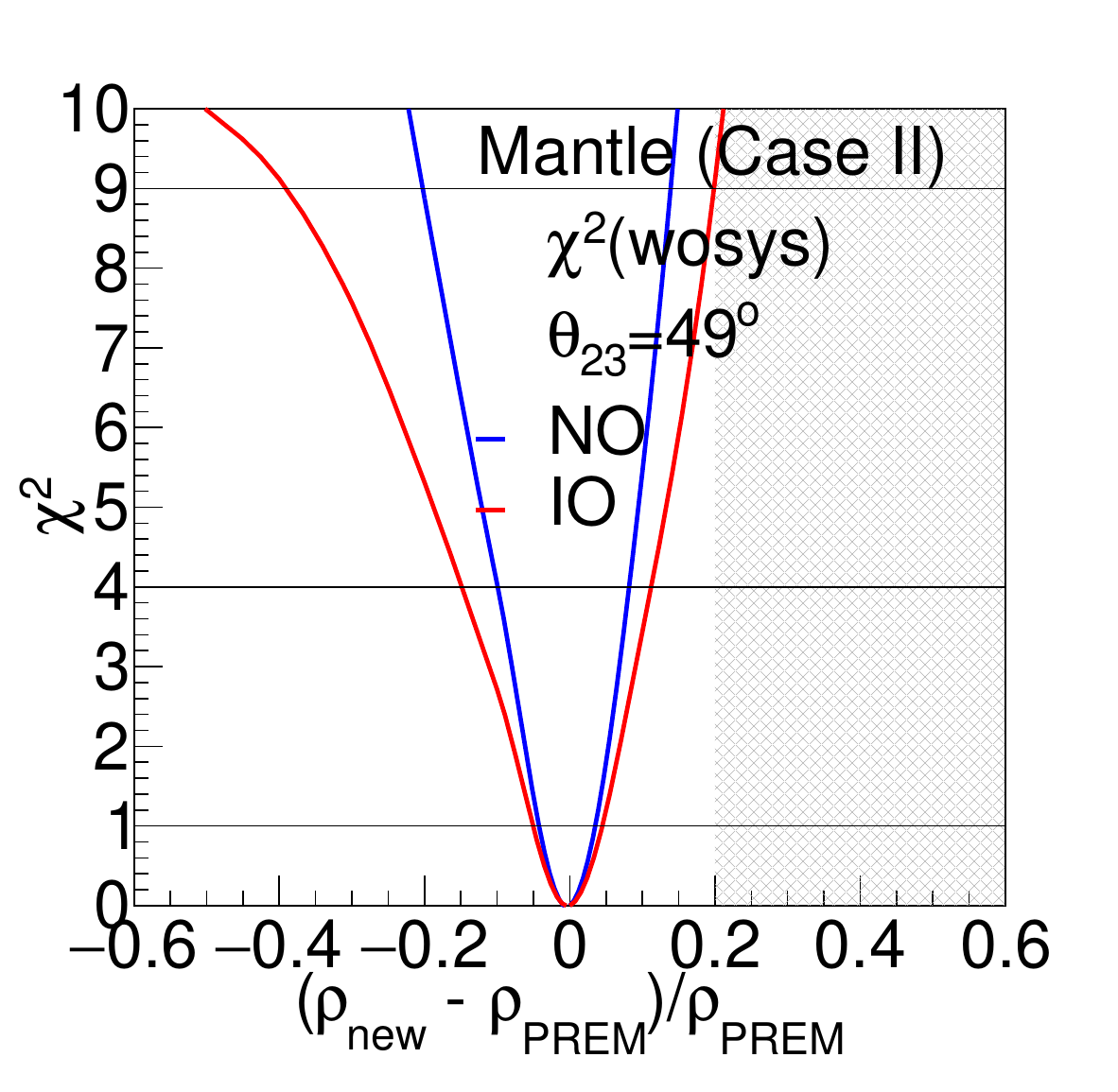}
    
    \includegraphics[width=0.3\textwidth]{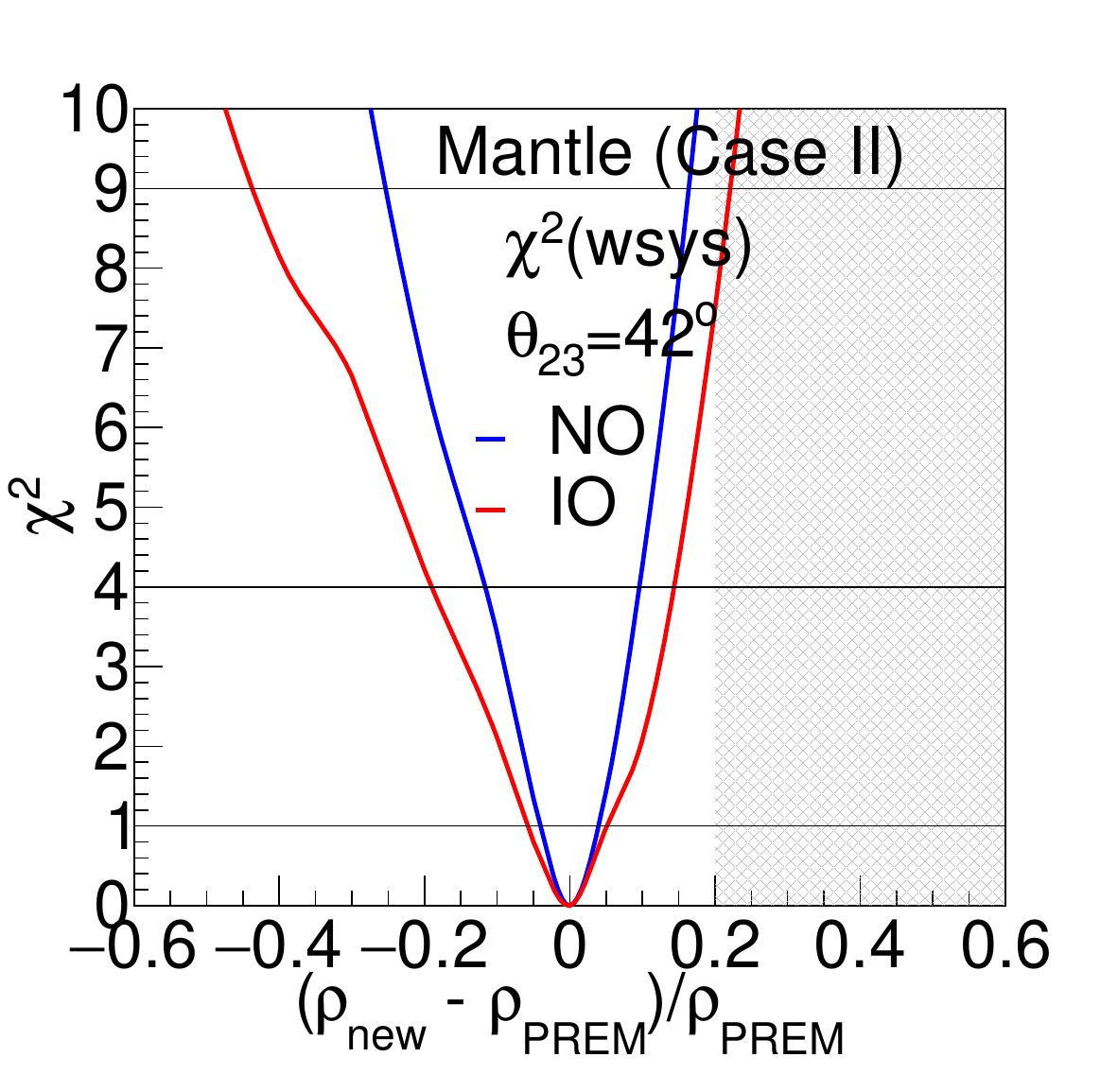} 
   \includegraphics[width=0.3\textwidth]{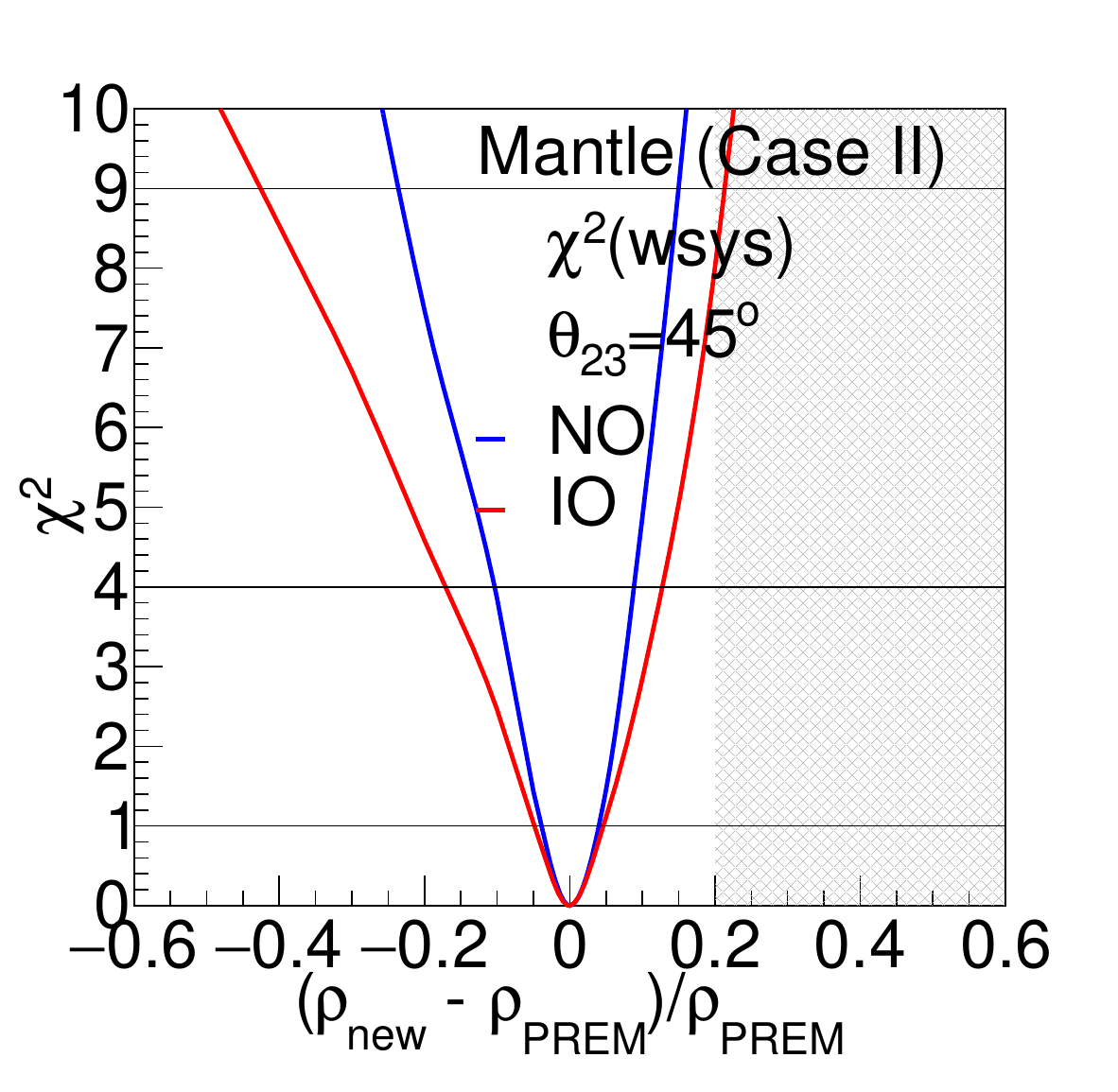}
    \includegraphics[width=0.3\textwidth]{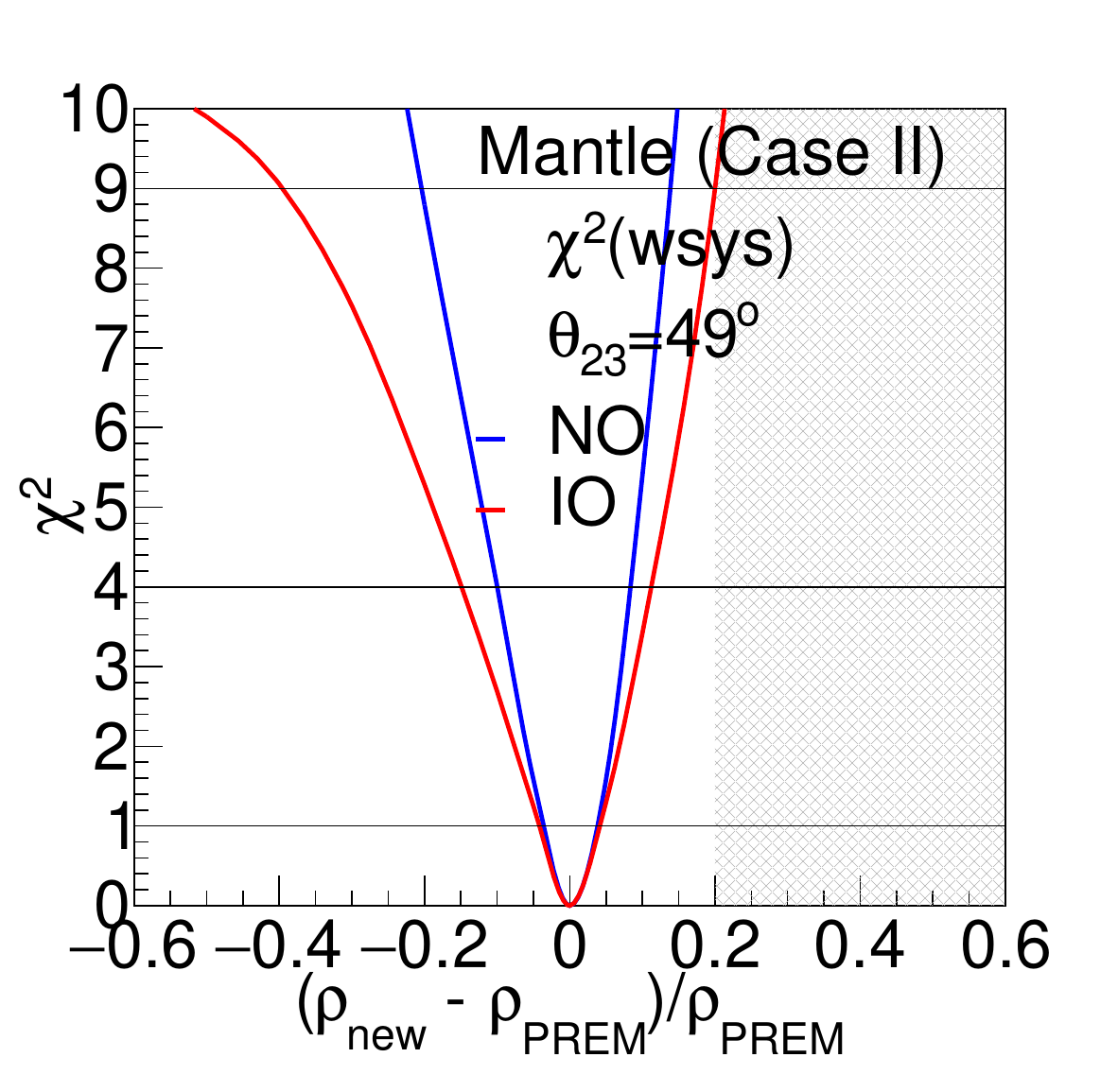}
    \caption{The $\chi^{2}$ as a function of percentage change in density in the mantle for Case II. Red lines are for NO and blue lines are for IO. Panel (a) is for  $\theta_{23}=42^{\circ}$ (b) for $\theta_{23}=45^{\circ}$ and (c) for $\theta_{23}=49^{\circ}$. Upper panels are for no systematic uncertainties in the analysis while the lower panels show the $\chi^2$ including systematic uncertainties.}
    \label{fig:machiws}
\end{figure*}

In Fig. \ref{fig:machiws} we present results on the sensitivity of ICAL@INO to density of the mantle when Earth total mass constraint is implemented and the mantle density variation is compensated by corresponding change in all other layers of the Earth. We consider a simplistic scenario where we make the compensation by taking equal percentage changes in rest of the layers. As before, we show results for without systematic uncertainties and with systematic uncertainties, for NO and IO, as well as for three choices of $\theta_{23}$. 

In our analysis, the constraint on earth mass alters the shape of the $\chi^{2}$ plots, as shown in Fig. \ref{fig:machiws}. For mantle, we lose symmetry in IO plots and a little asymmetry also comes in NO plots. The $\chi^{2}$ values shows marked improvement as compared to Case I. Table \ref{tab:case2man} gives the range of density variation values for which ICAL@INO is sensitive to the mantle density at $1\sigma$, $2\sigma$ and 3$\sigma$, for Case II. We show these ranges for 3 choices of $\theta_{23}$ and for both NO and IO. We can see the improvement in the expected sensitivity of ICAL@INO in comparison to Case I. 


\begin{table}[!h]
\centering
\caption{\label{tab:case2man} The range of density variation values for which ICAL@INO is sensitivity to the Mantle density at $1\sigma$, $2\sigma$ and 3$\sigma$, for Case II. We show these ranges for 3 choices of $\theta_{23}$ and for both NO and IO. }
\scalebox{0.85}{
\begin{tabular}{|c|c|c|c|c|c|c|}
\hline
  &  \multicolumn{2}{|c|}{$\theta_{23}=42^{\circ}$} & \multicolumn{2}{|c|}{$\theta_{23}=45^{\circ}$} & \multicolumn{2}{|c|}{$\theta_{23}=49^{\circ}$}\\
\hline
{Mantle}   & NO   & IO    & NO   & IO & NO   & IO\\
\hline
 1$\sigma$  &  -4.7/3.1 & -6.5/4.4   & -3.9/3.8  & -5/4.4 & -3.7/3.9   & -4.25/4.2 \\
 2$\sigma$  &  -11.9/8.7 & -19.2/13.9   & -10.9/8.8 & -17/12.5 & -9.9/8  & -15/11.4\\
 3$\sigma$  &  -25.5/15.5  &    -43/21.7 & -24.2/14.6  & -43/20.6 & -20.5/13.5   & -39.6/19.7\\
\hline
\end{tabular}
}
\end{table}


\subsubsection{OC}

\begin{figure*}
    \centering
    \includegraphics[width=0.3\textwidth]{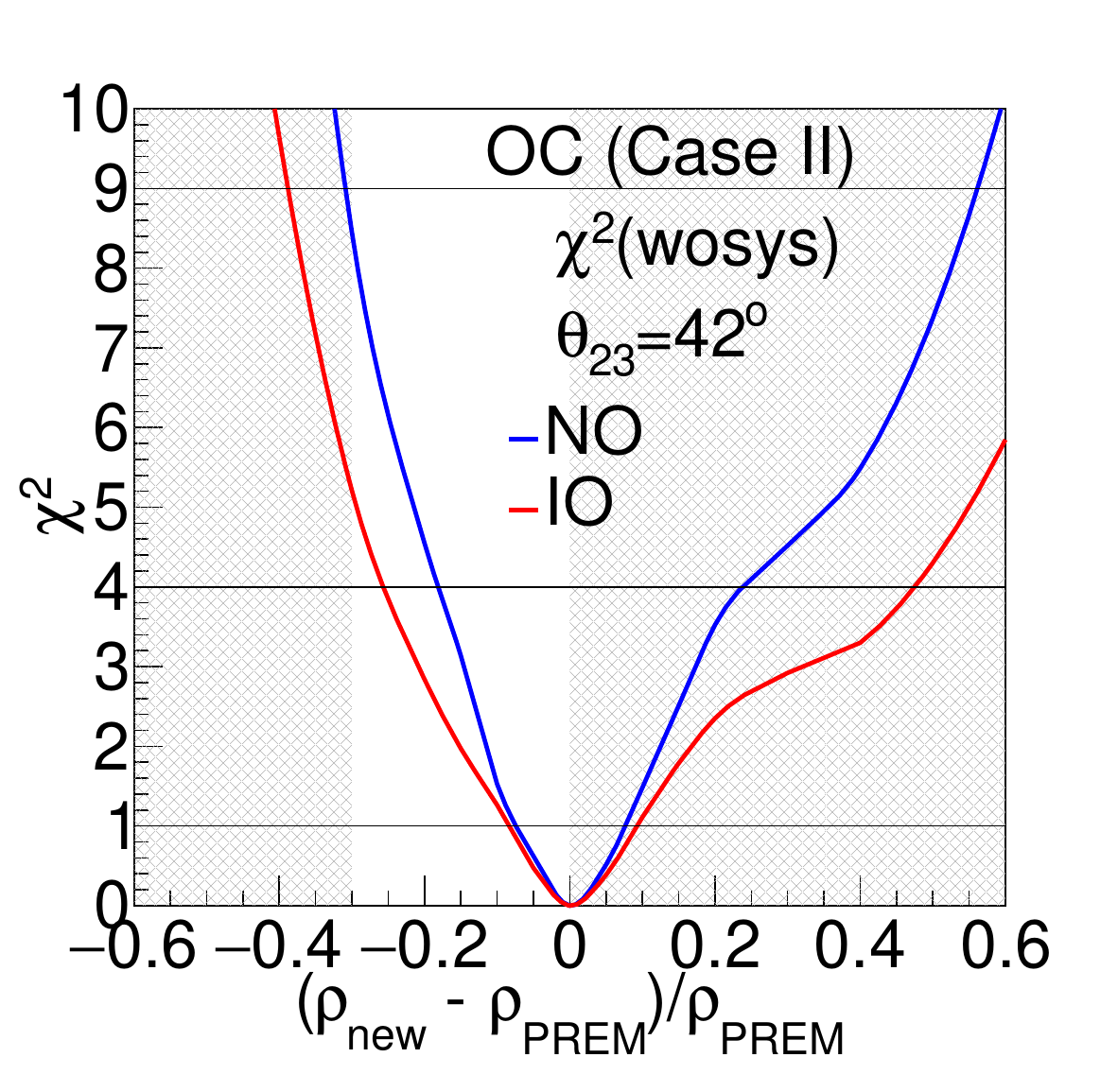}
    \includegraphics[width=0.3\textwidth]{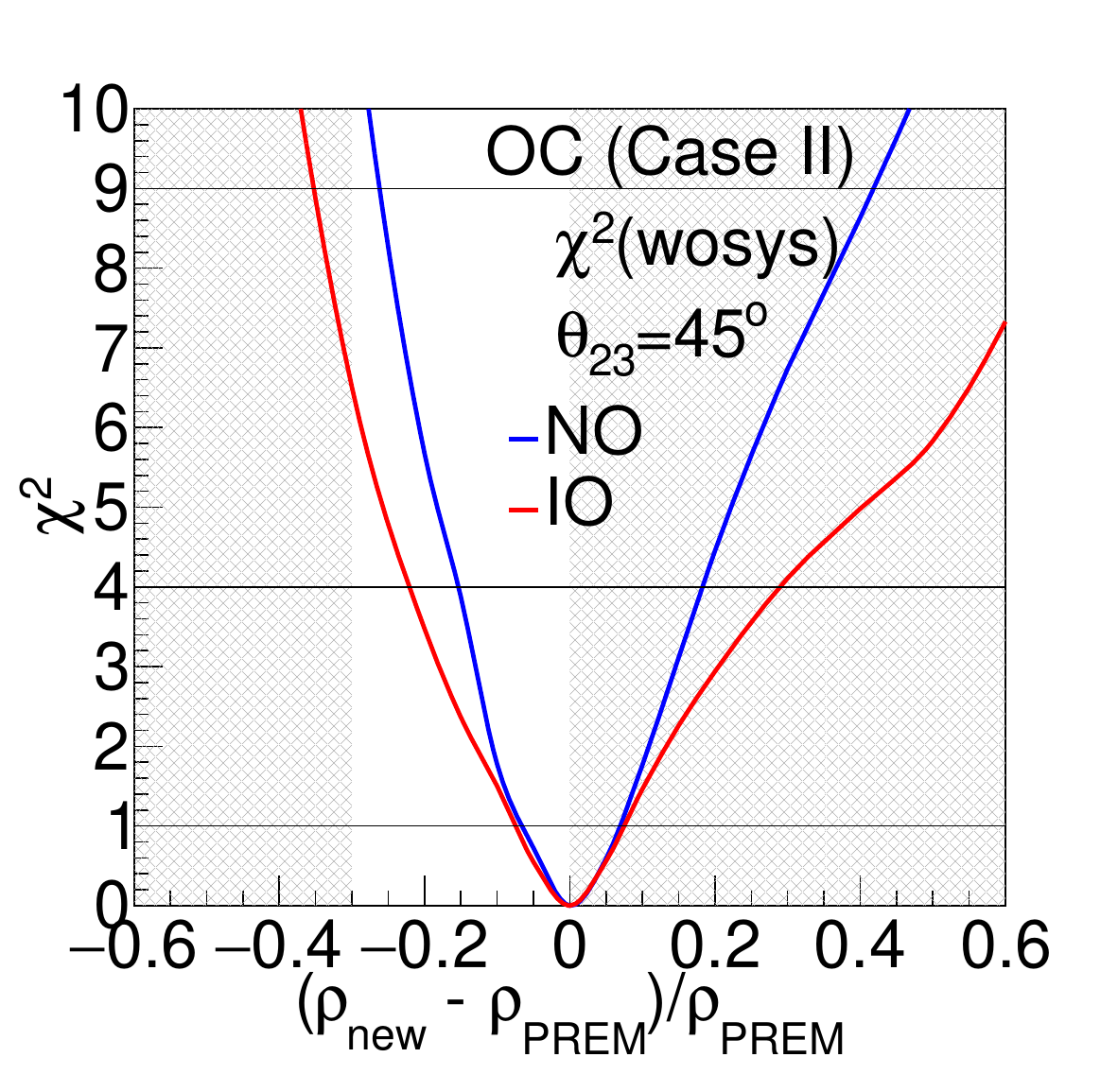}
    \includegraphics[width=0.3\textwidth]{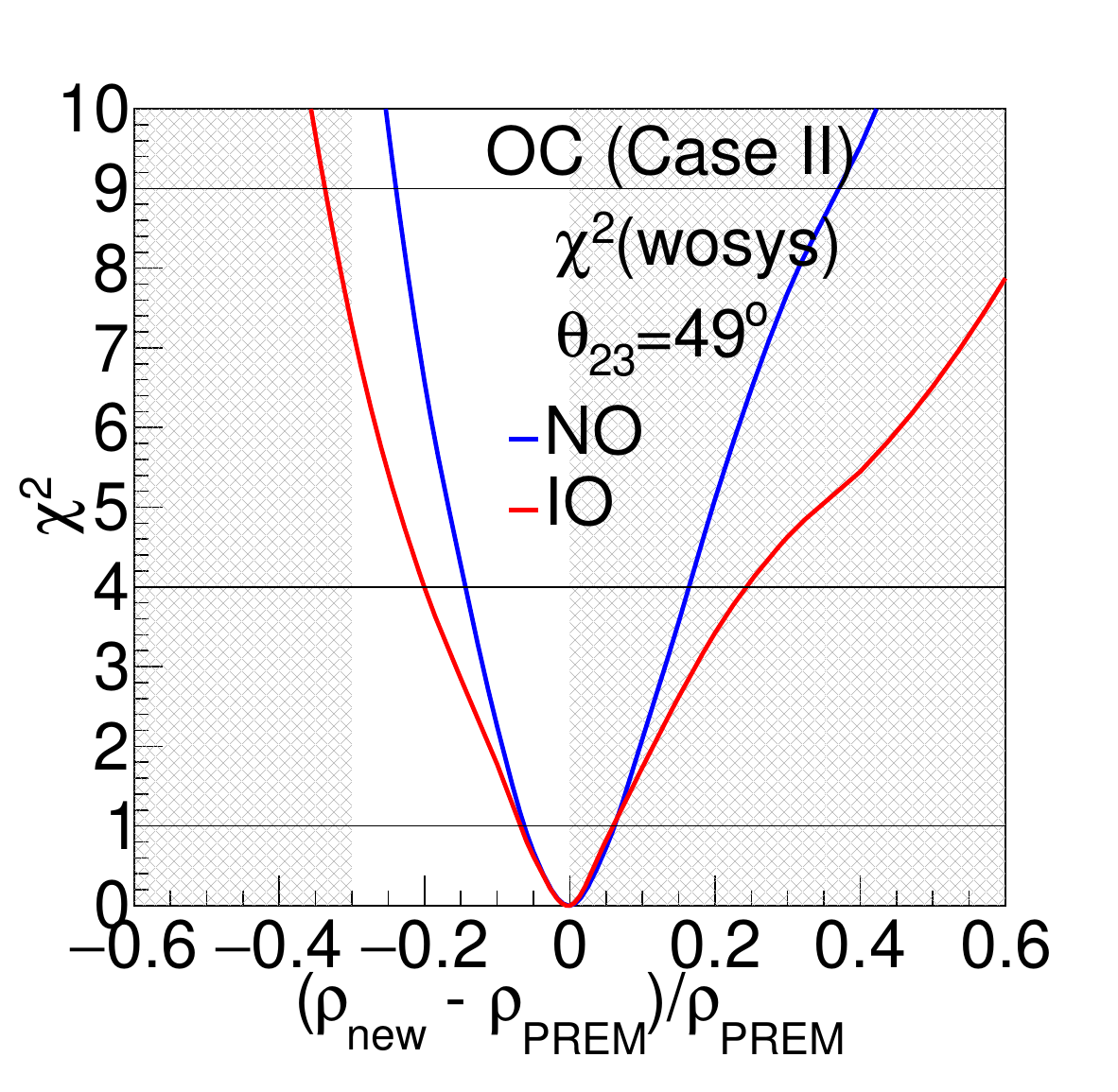}
    
    \includegraphics[width=0.3\textwidth]{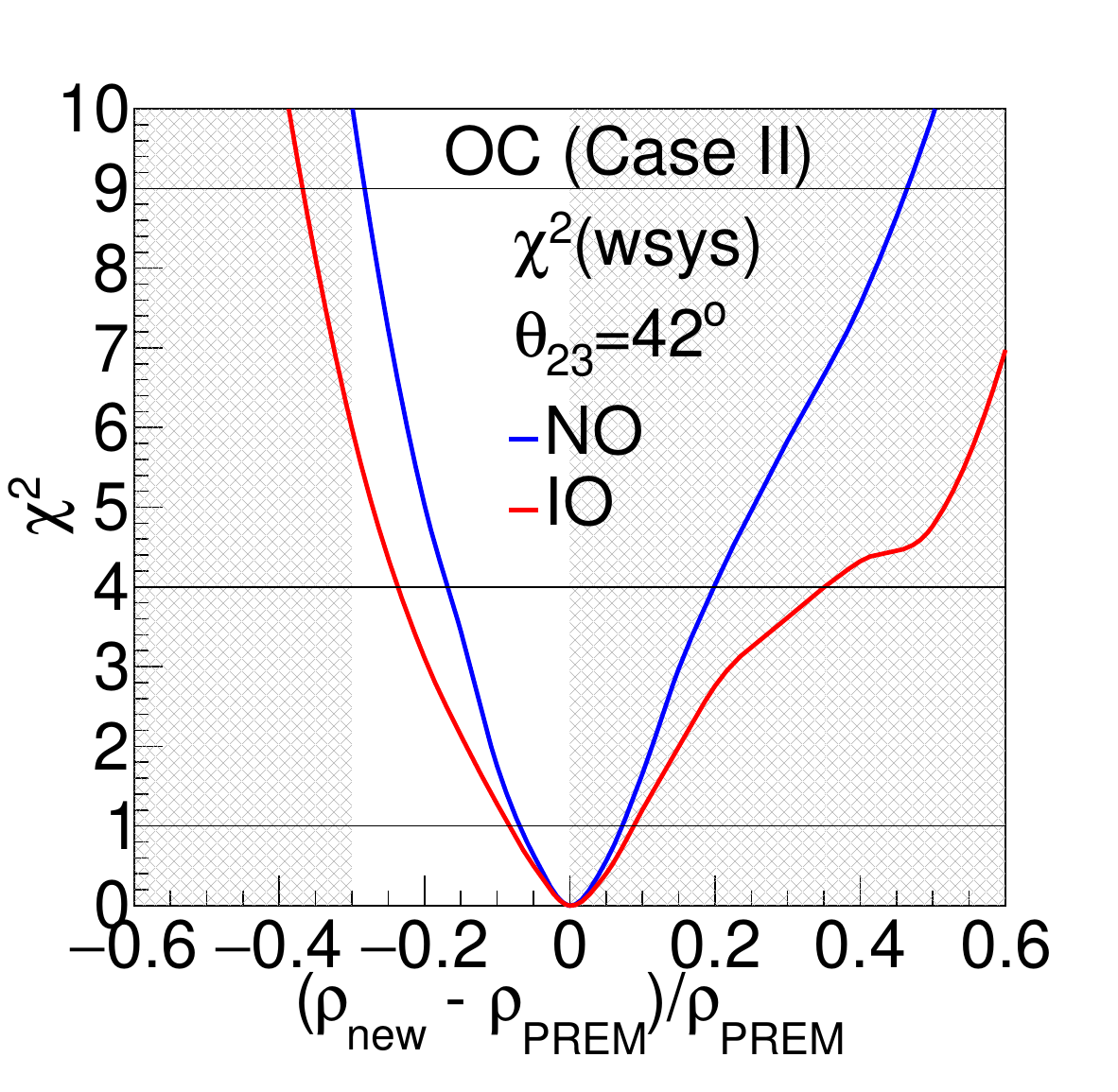} 
    \includegraphics[width=0.3\textwidth]{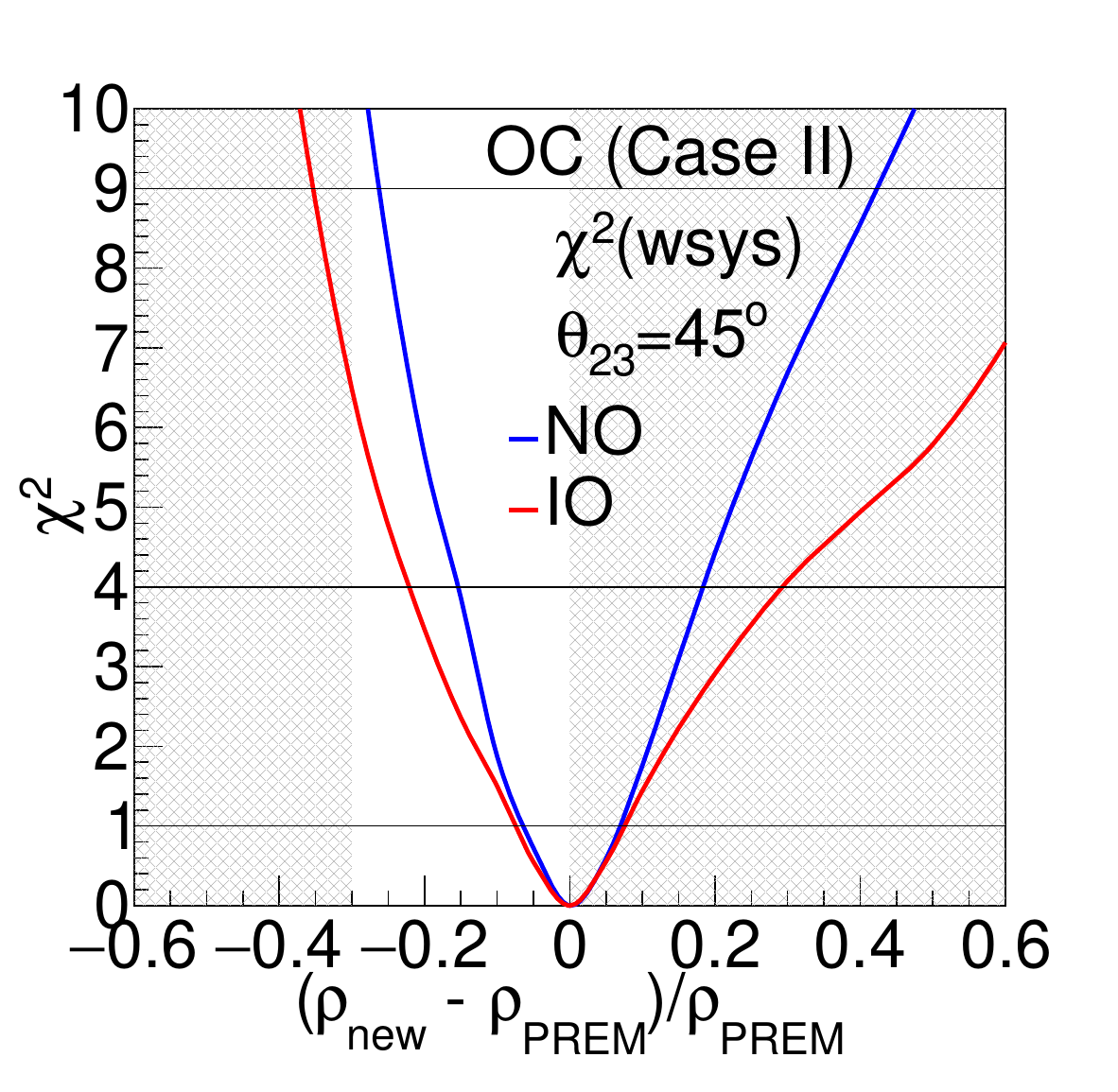} 
    \includegraphics[width=0.3\textwidth]{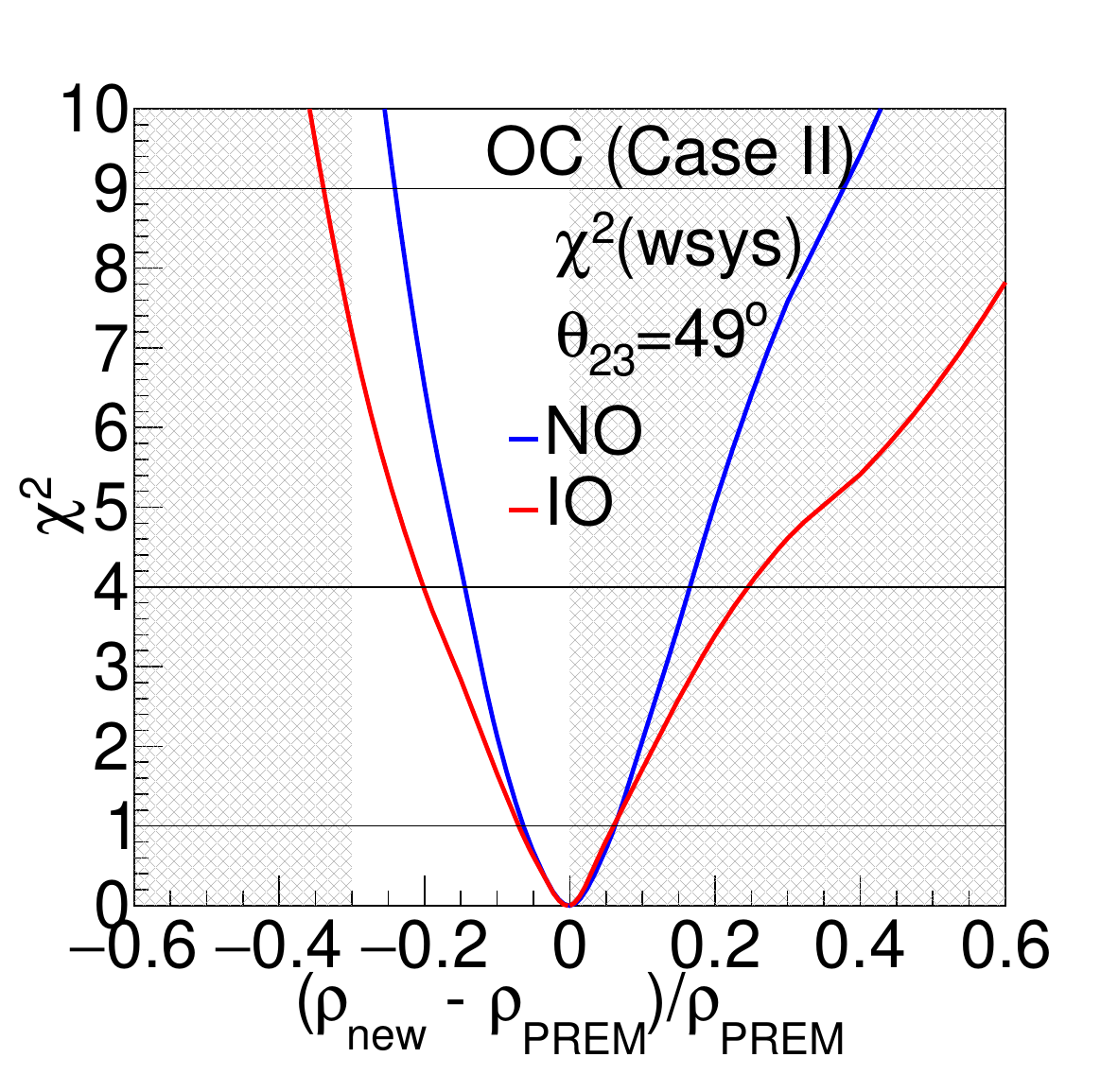}

    \caption{All panels show plots of $\chi{2}$ as a function of percentage change in OC density for Case II. Panel (a) is for $\theta_{23}=42^{\circ}$, (b) for $\theta_{23}=45^{\circ}$, and (c) for $\theta_{23}=49^{\circ}$, both without (upper plots) and with (bottom plots) systematic uncertainties in the $\chi^{2}$ analysis. Blue lines are for NO and red lines are for IO.}
    \label{fig:ocachiws}
\end{figure*}

In Fig.   \ref{fig:ocachiws} we present results on the sensitivity of ICAL@INO to OC density when Earth total mass constraint is implemented and OC density variation is compensated with corresponding change in all other layers of earth with equal percentage. Upper panels of Fig. \ref{fig:ocachiws} are without systematic uncertainties while the lower panels are with systematic uncertainties in the $\chi^{2}$ analysis. Shown as curves for both NO and IO for three choices of $\theta_{23}$.

\begin{table}[!h]
\centering
\caption{\label{tab:case2oc} The range of density variation values for which ICAL@INO is sensitivity to the OC density at $1\sigma$, $2\sigma$ and 3$\sigma$, for Case II. We show these ranges for 3 choices of $\theta_{23}$ and for both NO and IO. } 
\scalebox{0.85}{
\begin{tabular}{|c|c|c|c|c|c|c|}
\hline
  &  \multicolumn{2}{|c|}{$\theta_{23}=42^{\circ}$} & \multicolumn{2}{|c|}{$\theta_{23}=45^{\circ}$} & \multicolumn{2}{|c|}{$\theta_{23}=49^{\circ}$}\\
\hline
{C.L.}   & NO   & IO    & NO   & IO & NO   & IO\\
 \hline
 1$\sigma$  &  -7.2/6.7 & -8.5/8.5   & -6.7/6.4  & -7.4/7.4 & -6.4/5.9   & -6.9/5.9\\
 2$\sigma$  &  -16.8/19.6 & -24/34   & -15.8/17.7  & -22/29 & -14.7/16.5   & -20.1/24.5\\
 3$\sigma$  & -28/46  & -37/   & -26/41  & -35/ & -24.5/37.4   & -34.5/\\
\hline
\end{tabular}
}
\label{tab:case1mantle}
\end{table}

Examination of the plots in Fig. \ref{fig:ocachiws} shows that the $\chi^{2}$ increases significantly for both positive and negative density changes of density in OC. The oscillatory part of the positive side has turned into a steady increasing curve. The reason for this is that increasing OC density decreases mantle density, and we have seen that decreasing mantle density gives a sharp increase in the $\chi^2$, so adding these effects gives us a continued increase for positive OC density change. A decrease in OC density also results in a sharp and steady increase in $\chi^{2}$. Because of the contribution from other layers, particularly the mantle, overall $\chi^{2}$ values are higher than in the previous case. Table \ref{tab:case2oc} gives the range of density variation values for which ICAL@INO is sensitive to the outer core density at $1\sigma$, $2\sigma$ and 3$\sigma$, for Case II. We show these ranges for 3 choices of $\theta_{23}$ and for both NO and IO.

\subsection{Case III}

In this case we take into account the Earth mass constraint by compensating for the density change in any given layer by suitable changes to layers only in the ``inner part" of the Earth. In particular, we change the density of only the inner core, outer core and inner mantle corresponding to $d>2200$ km.

\subsubsection{Mantle}

\begin{figure*}
    \centering
     \includegraphics[width=0.3\textwidth]{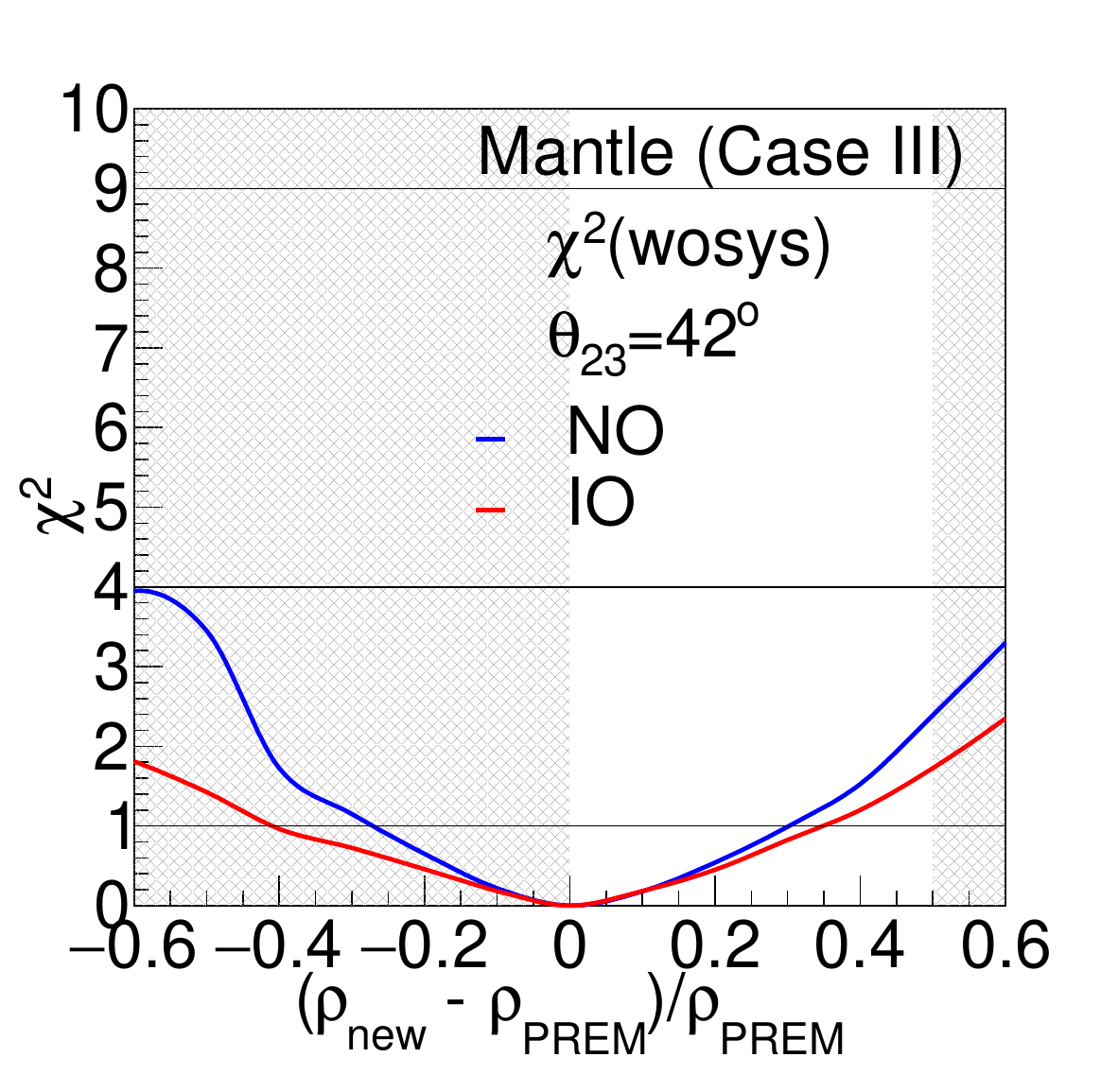}
   \includegraphics[width=0.3\textwidth]{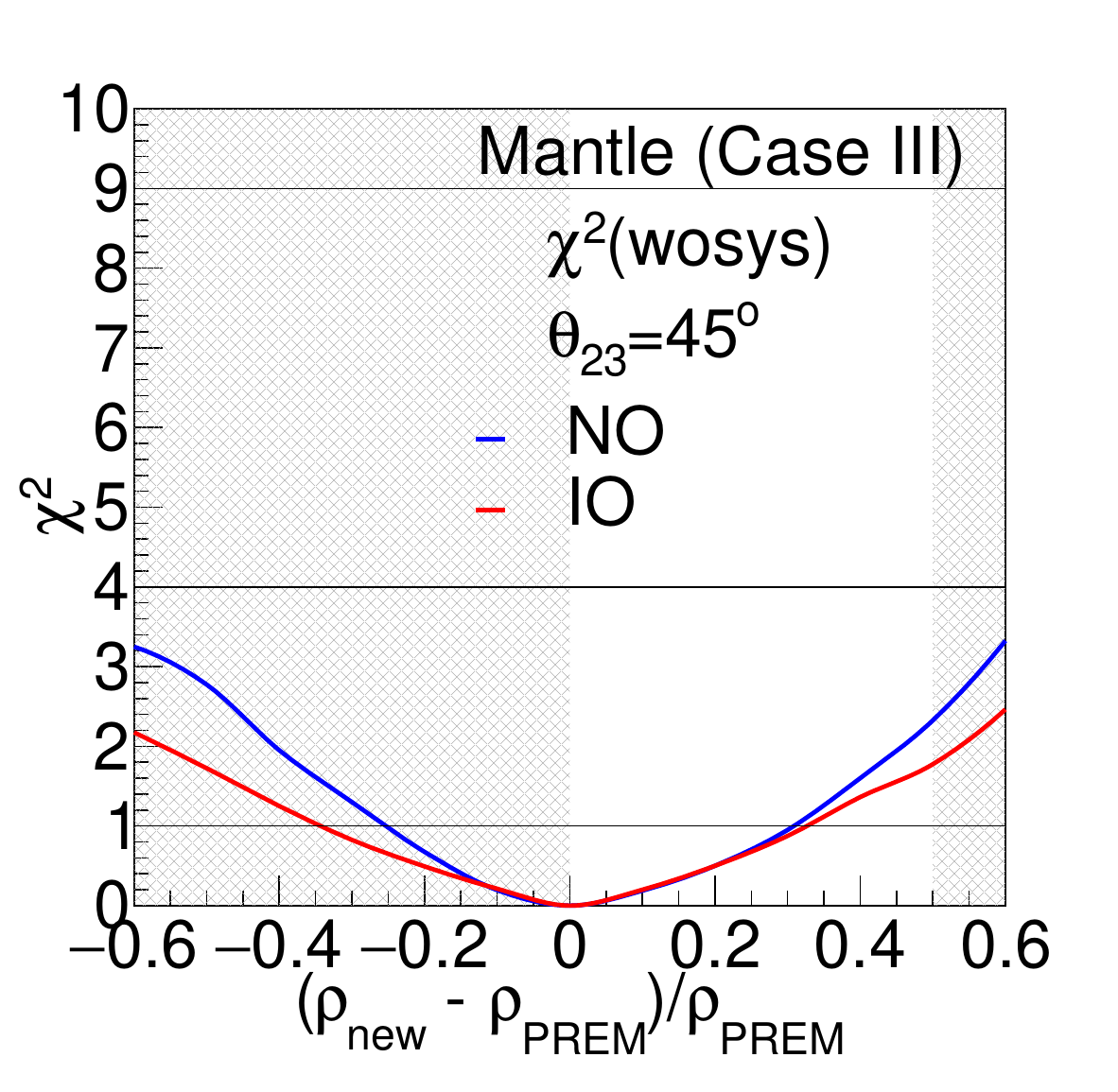}
    \includegraphics[width=0.3\textwidth]{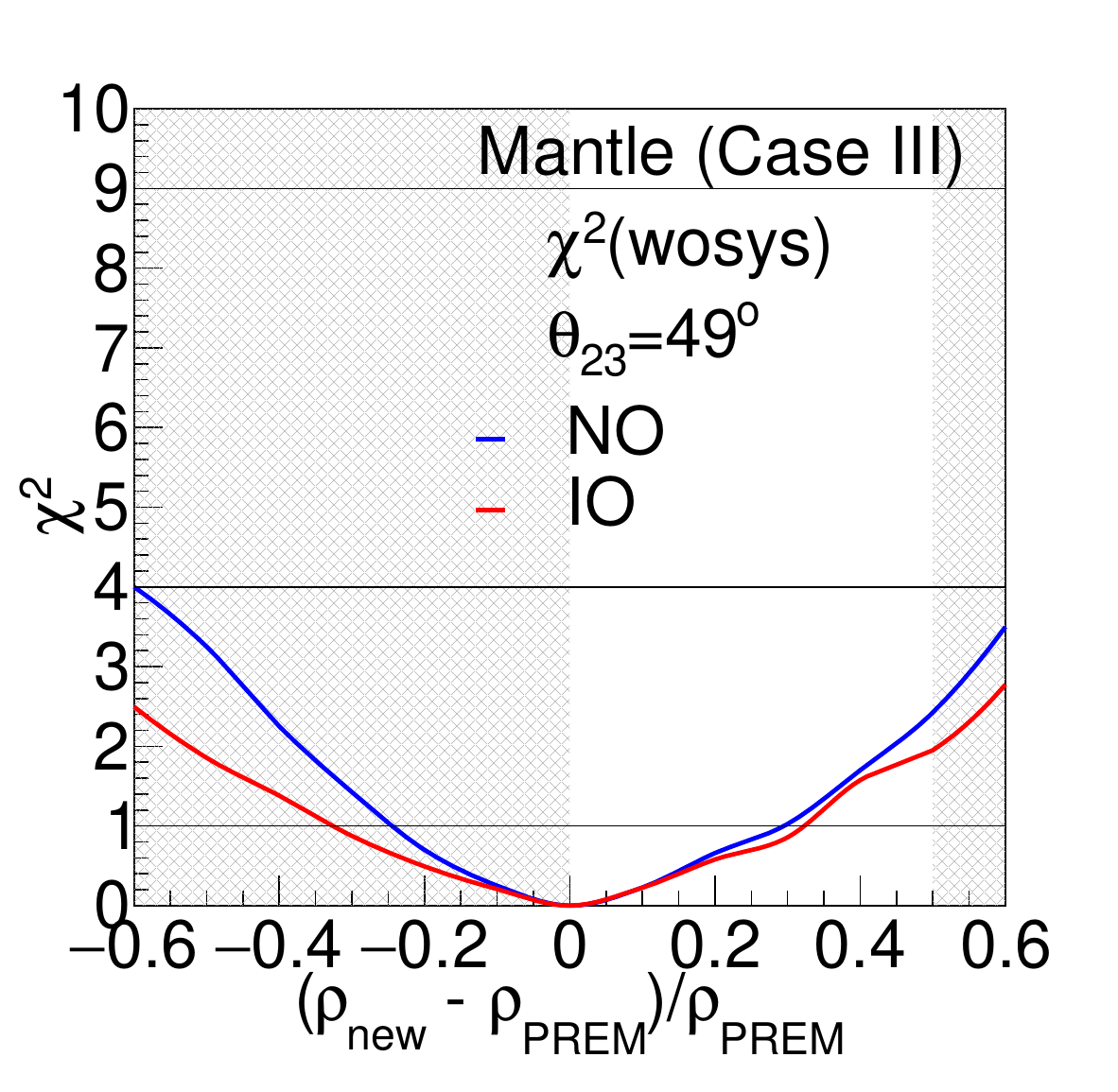}
    
    \includegraphics[width=0.3\textwidth]{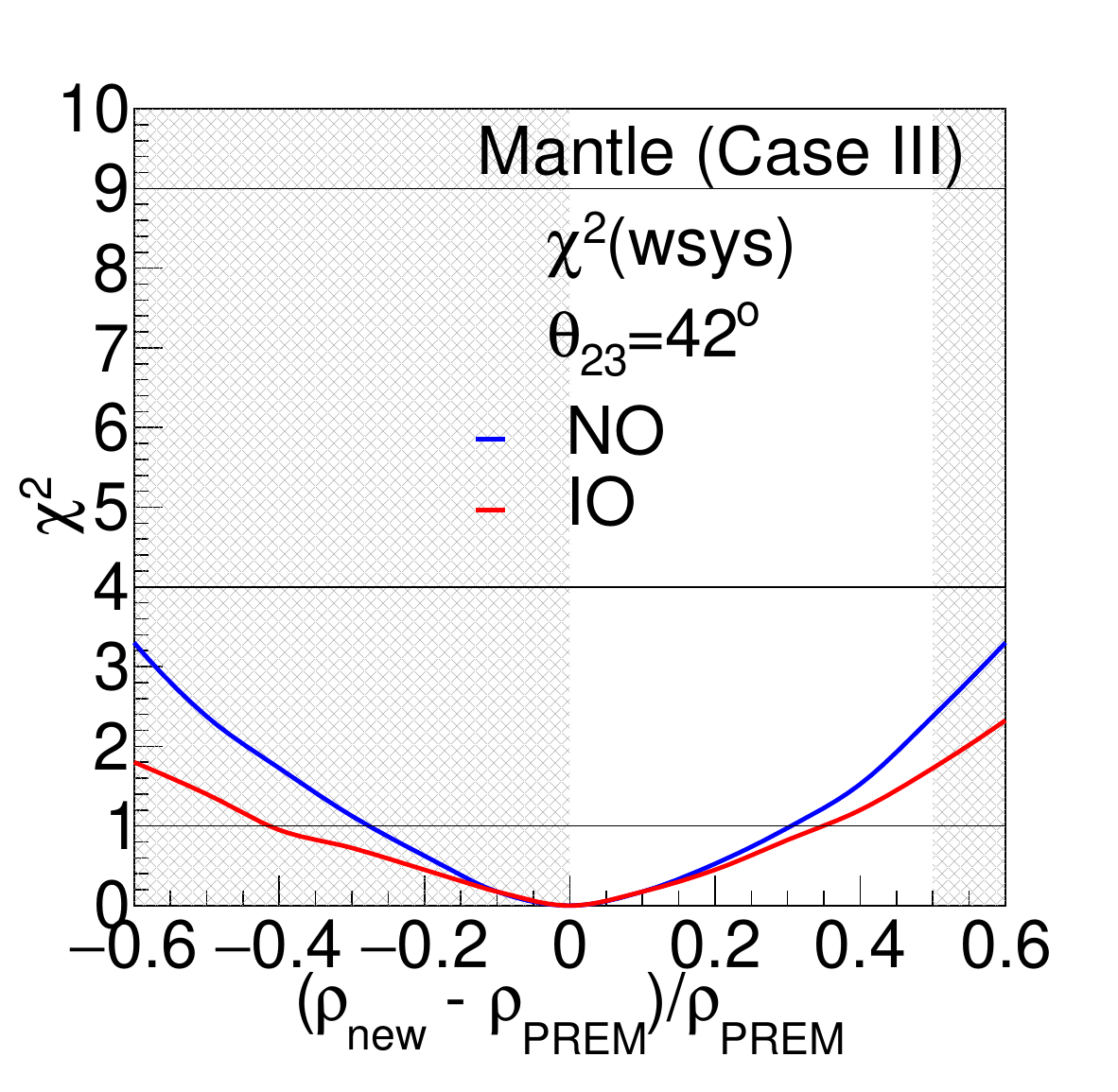}
    \includegraphics[width=0.3\textwidth]{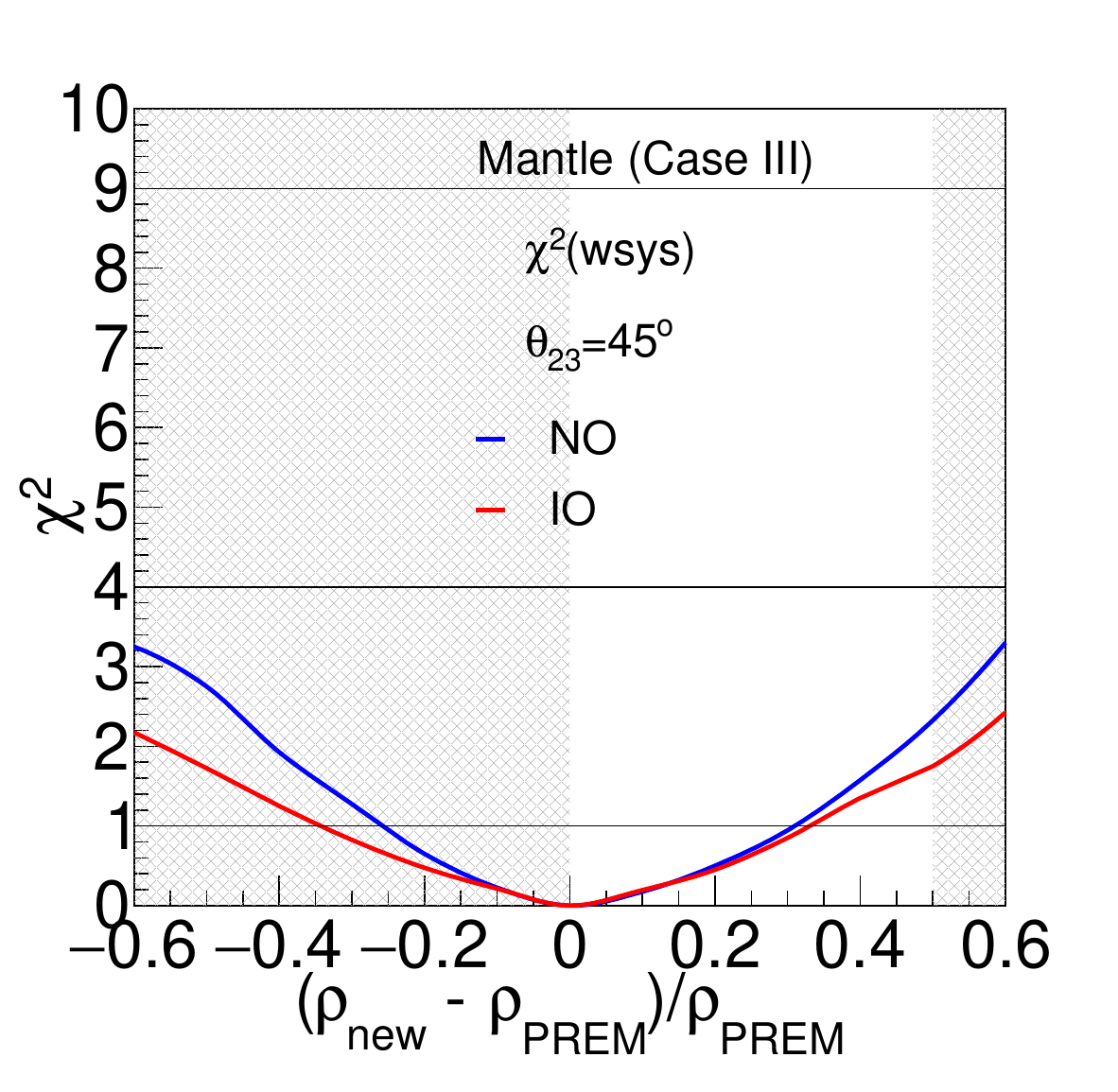}
    \includegraphics[width=0.3\textwidth]{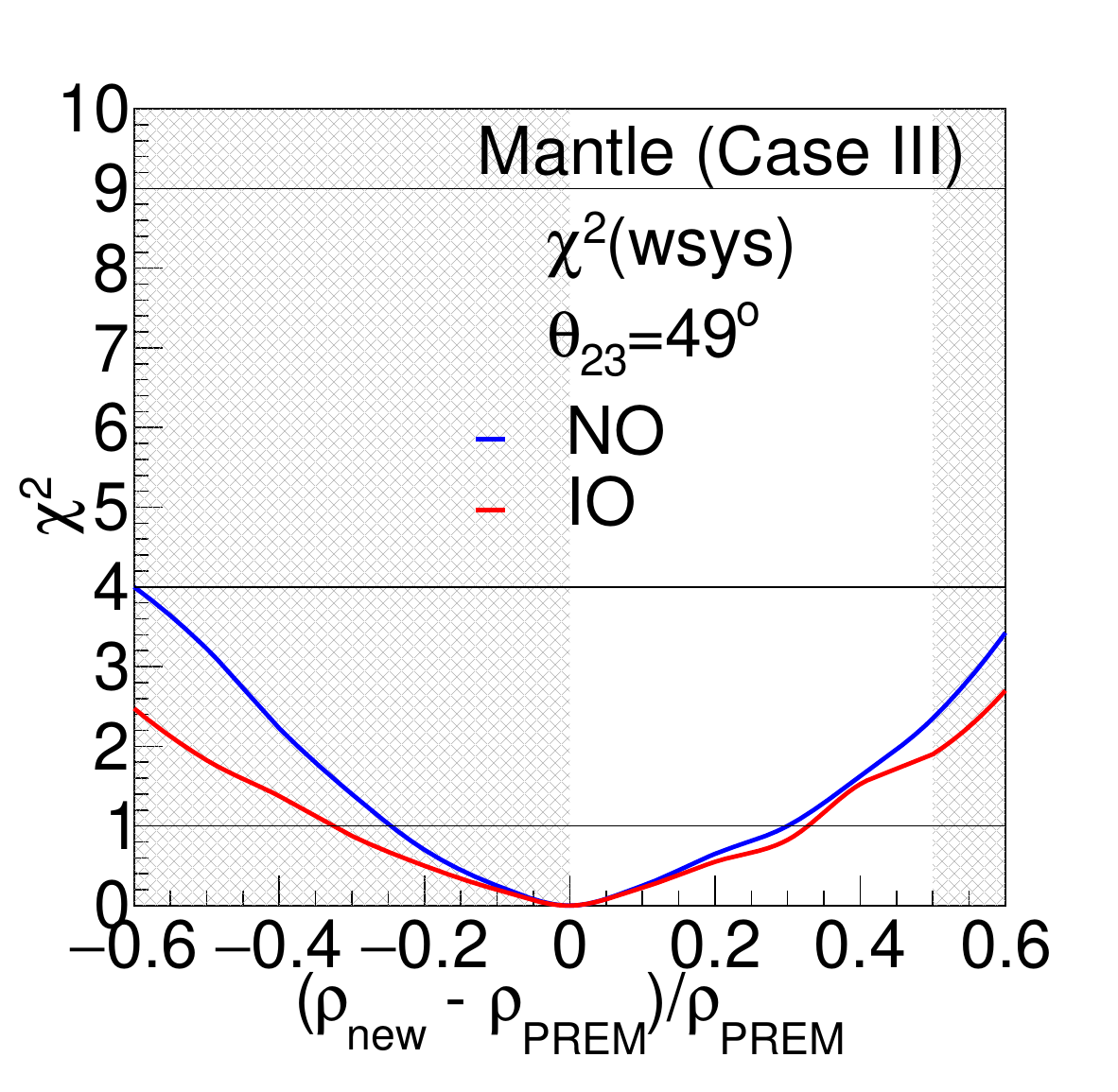}
    \caption{All three panels has plots for $\chi^{2}$ as a function of percentage change in density of Mantle for case III (putting Earth mass constrain in inner layers). Where (a) is for  $\theta_{23}=42^{\circ}$ (b) for $\theta_{23}=45^{\circ}$ and (c) for $\theta_{23}=49^{\circ}$ without (upper plots) and with (bottom plots) systematic uncertainties in $\chi^{2}$ analysis.The Blue line is for NO and Red line for IO as a known MO case.}
    \label{fig:michiws}
\end{figure*}

We start by studying the expected sensitivity of ICAL@INO to the change in density in the mantle region for this case. The results are shown in Fig. \ref{fig:michiws} and Table \ref{table:mtlinchi}. We notice that the $\chi^2$ expected for this case is considerably lower than that for Case II but mildly higher than for Case I. Comparing the results in the lower and upper panels of the figure we see that improvement in systematics is not expected to bring any drastic improvement to the sensitivity.

\begin{table}[!h]
\centering
\caption{The range of density variation values for which ICAL@INO is sensitive to the Mantle density at $1\sigma$, for Case III. We show these ranges for 3 choices of $\theta_{23}$ and for both NO and IO.} 
\begin{tabular}{|c|c|c|c|c|c|c|}
\hline
  &  \multicolumn{2}{|c|}{$\theta_{23}=42^{\circ}$} & \multicolumn{2}{|c|}{$\theta_{23}=45^{\circ}$} & \multicolumn{2}{|c|}{$\theta_{23}=49^{\circ}$}\\
\hline  
{Mantle}   & NO   & IO    & NO   & IO & NO   & IO\\
\hline
 1$\sigma$  &  -27/30 & -41/34& -26/30  & -34/33 & -24.0/30   & -32/33 \\
\hline
\end{tabular}
\label{table:mtlinchi}
\end{table}

\subsubsection{OC}

\begin{figure*}
    \centering
     \includegraphics[width=0.3\textwidth]{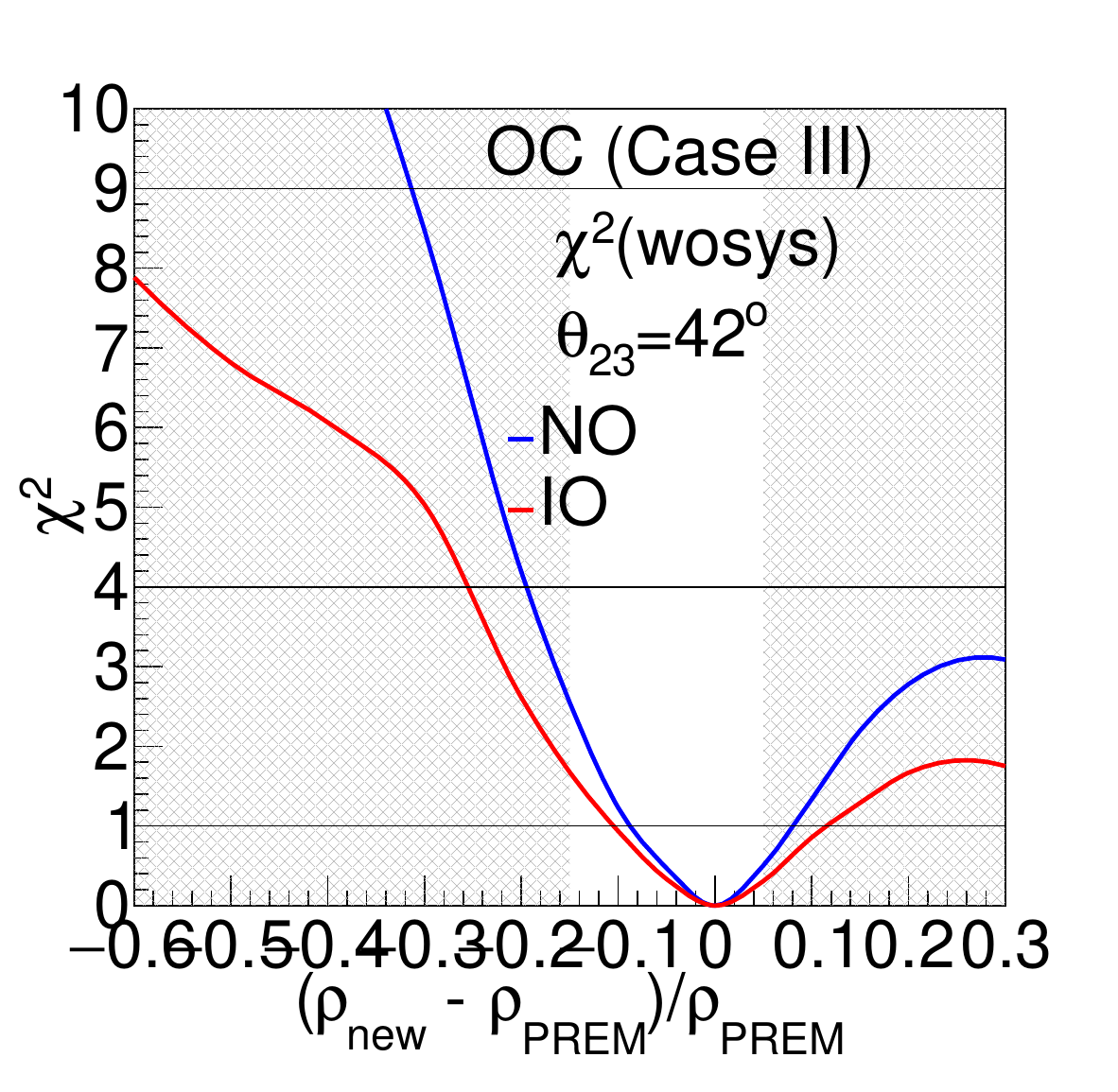}
  \includegraphics[width=0.3\textwidth]{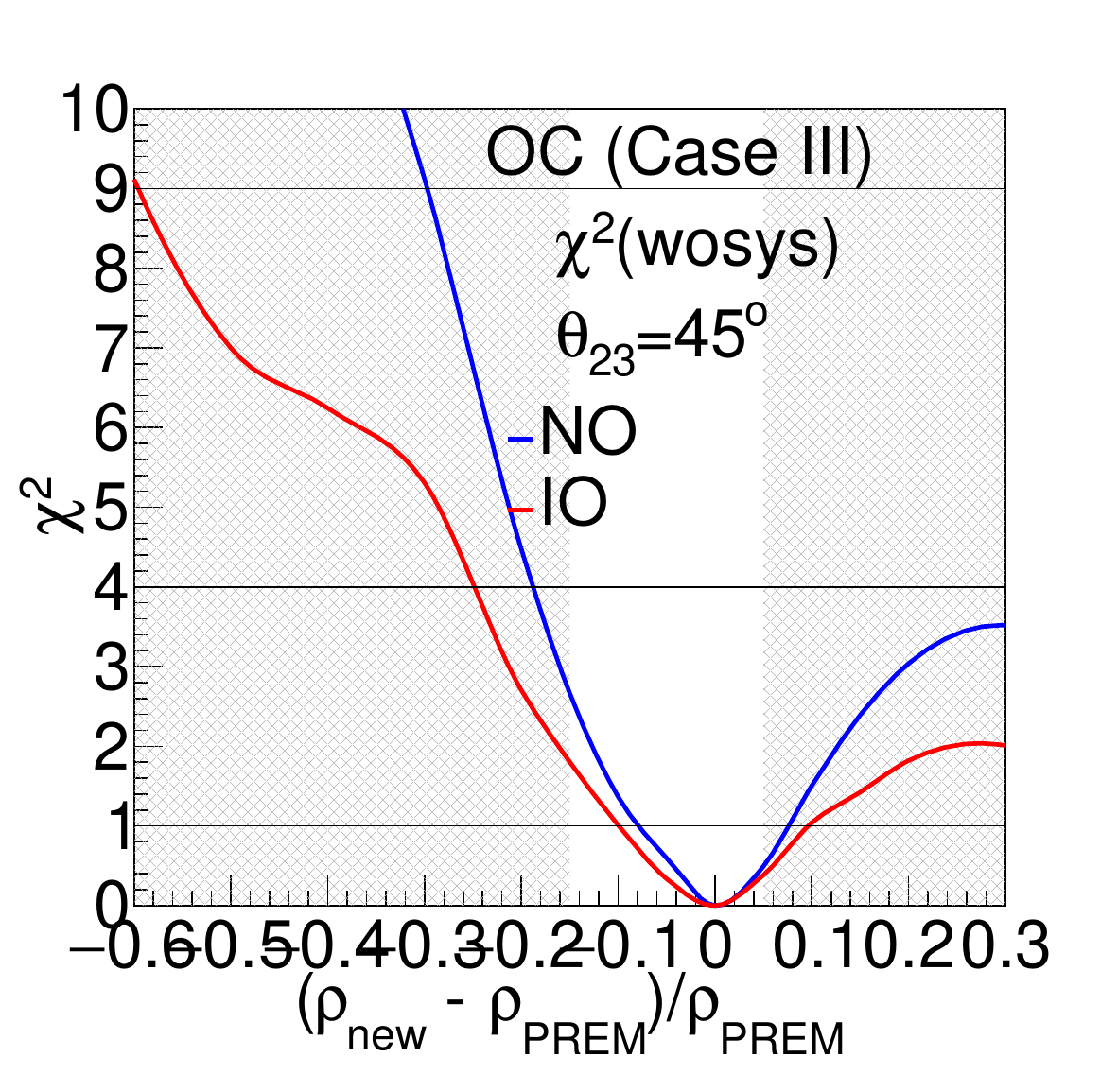}
    \includegraphics[width=0.3\textwidth]{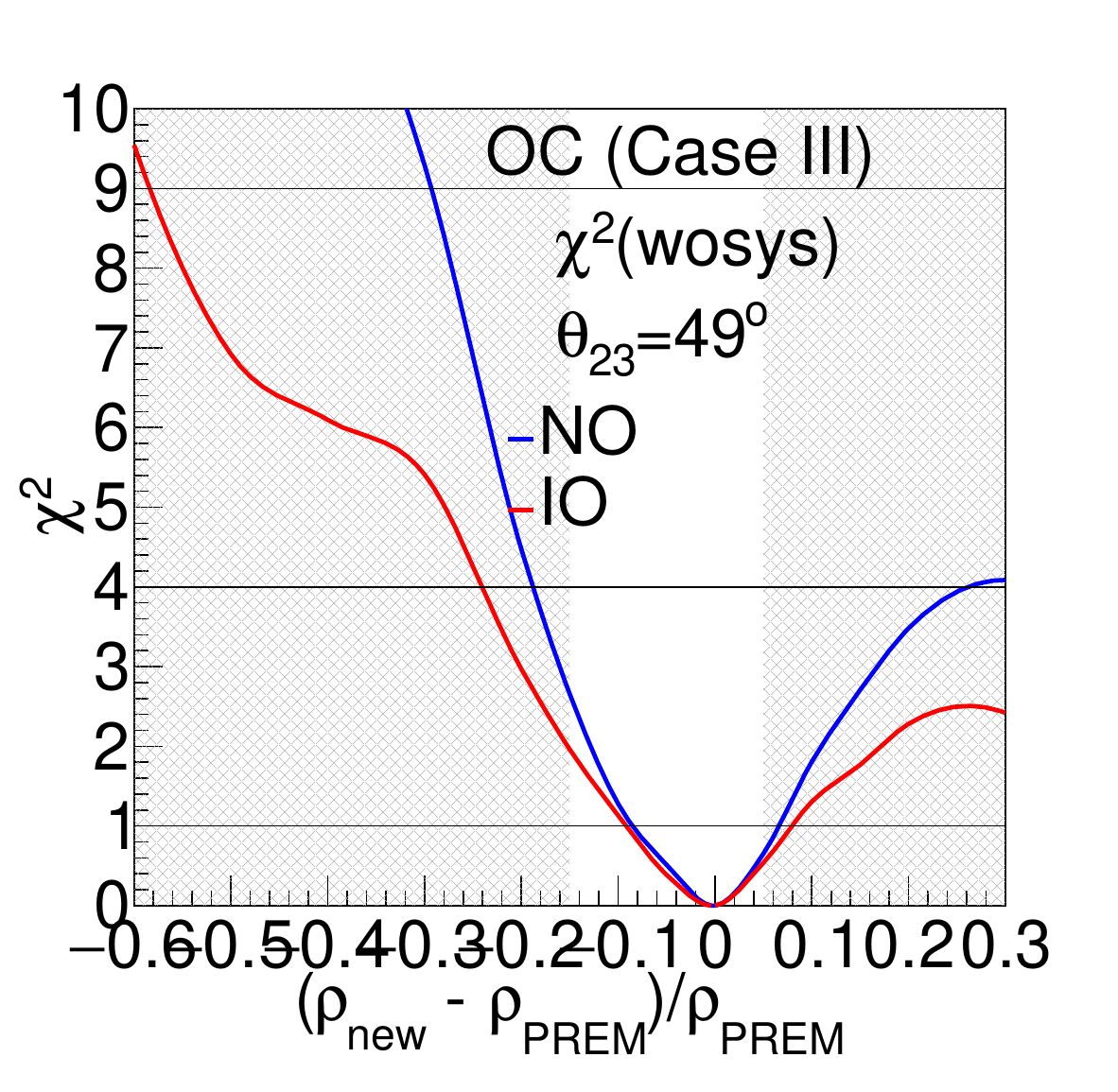}
    
    \includegraphics[width=0.3\textwidth]{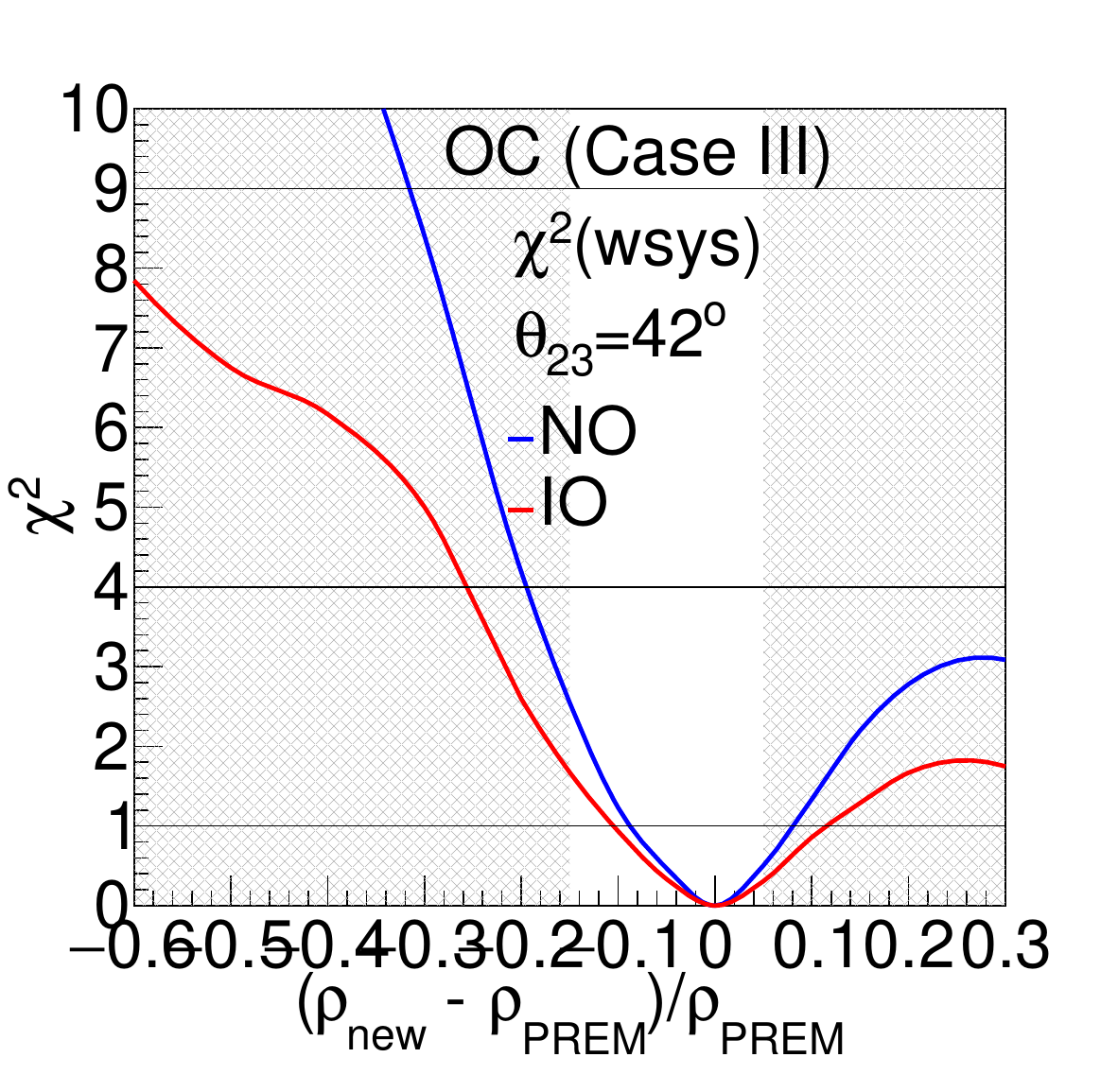}
    \includegraphics[width=0.3\textwidth]{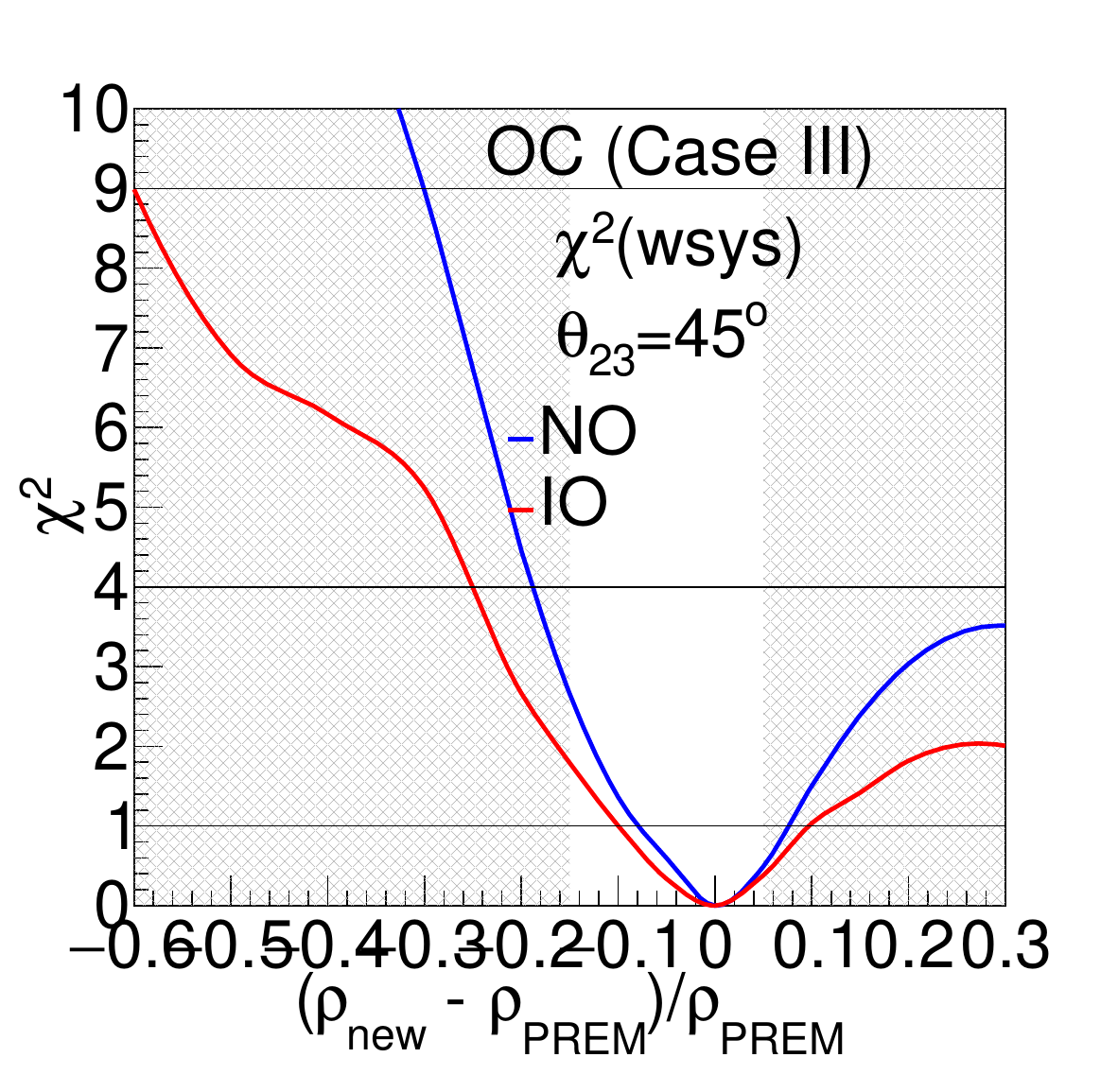}
    \includegraphics[width=0.3\textwidth]{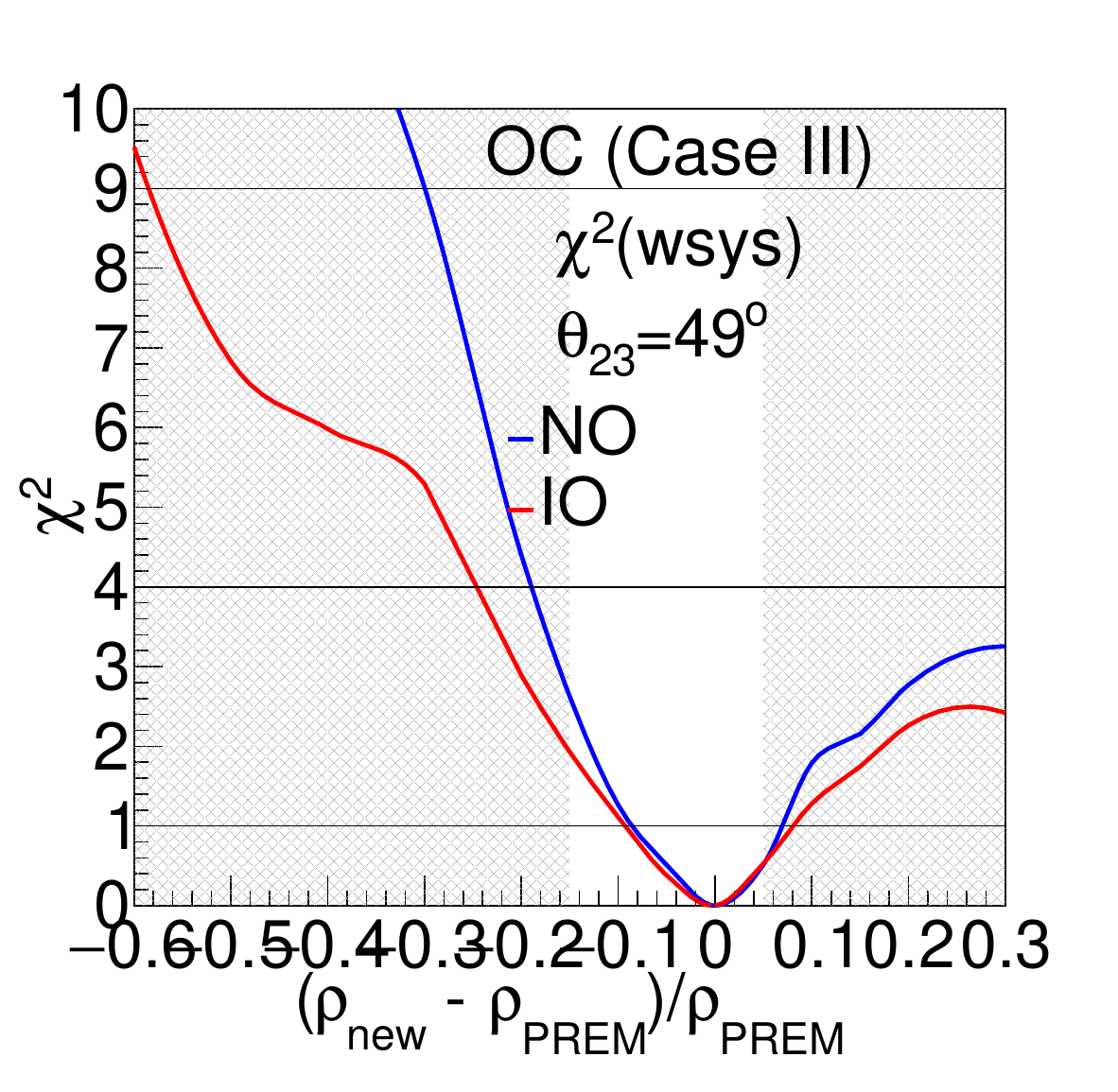}
    \caption{All three panels has plots for $\chi^{2}$ as a function of percentage change in density of OC for case III (putting Earth mass constrain in inner layers). Where (a) is for  $\theta_{23}=42^{\circ}$ (b) for $\theta_{23}=45^{\circ}$ and (c) for $\theta_{23}=49^{\circ}$ without (upper plots) and with (bottom plots) systematic uncertainties in $\chi^{2}$ analysis.The Blue line is for NO and Red line for IO as a known MO case.}
    \label{fig:ocichiws}
\end{figure*}

\begin{table}[!h]
\centering
\caption{The range of density variation values for which ICAL@INO is sensitivity to the Mantle density at $1\sigma$, $2\sigma$ and 3$\sigma$, for Case I. We show these ranges for 3 choices of $\theta_{23}$ and for both NO and IO. } 
\begin{tabular}{|c|c|c|c|c|c|c|}
\hline
  &  \multicolumn{2}{|c|}{$\theta_{23}=42^{\circ}$} & \multicolumn{2}{|c|}{$\theta_{23}=45^{\circ}$} & \multicolumn{2}{|c|}{$\theta_{23}=49^{\circ}$}\\
\hline
{C.L.}   & NO   & IO    & NO   & IO & NO   & IO\\
 \hline
 1$\sigma$  &  -8.8/7.7 & -10.5/10.8   & -7.8/7.1  & -9.9/9 & -8.7/6.8   & -9.4/7.8\\
 2$\sigma$  &  -19.7/ & -25.5/   & -18/  & -24/ & -18.7/   & -24.5/\\
 3$\sigma$  & -31.6/  & -/-  & -30/  & -59/ & -29.6/   & -58/\\
\hline
\end{tabular}
\label{tab:case1mantle}
\end{table}

\begin{table}[!h]
\centering
\caption{The range of density variation values for which ICAL is sensitivity to the OC density at $1\sigma$, $2\sigma$ and 3$\sigma$, for Case III. We show these ranges for 3 choices of $\theta_{23}$. } 
\begin{tabular}{|c|c|c|c|c|c|c|}
\hline
  &  \multicolumn{2}{|c|}{$\theta_{23}=42^{\circ}$} & \multicolumn{2}{|c|}{$\theta_{23}=45^{\circ}$} & \multicolumn{2}{|c|}{$\theta_{23}=49^{\circ}$}\\
\hline
{OC}   & NO   & IO    & NO   & IO & NO   & IO\\
\hline
 1$\sigma$  &  -7.5/8.5 & -10/   & -7.6/7.8  & -9.8/ & -7.7/7.3   & -9.5/9.0 \\
 2$\sigma$  &  -19.7/ & -25.8/   & -19.6/ & -26/ & -18.6/  & -25.3/\\
 3$\sigma$  &  -33.7/  &    / & -32.6/  & / & -32/   & 55.6/ \\
\hline
\end{tabular}
\label{table:oclinchi}
\end{table}

The expected sensitivity of ICAL@INO to the density of the OC for Case III is presented in Fig. \ref{fig:ocichiws} and \ref{table:oclinchi}. A comparison of Figs. 8, 10 and \ref{fig:ocichiws} shows that compensation due to Earth mass constraint has much less effect in Case III as compared to Case II. This is because when we change the density in the OC, compensation to preserve Earth mass happens mainly in IC for Case III, while for Case II we have compensation from both IC as well as mantle. As stated before, density changes in IC do not change the probabilities much and therefore the resulting $\chi^2$ is also lower.

\section{Conclusion \label{sec:conclusions}}

Earth tomography is an important field in science. While the best estimates of Earth's density profile comes from seismology, it is pertinent to check if complementary information and/or cross-checks can be informed elsewhere. Neutrino experiments offer a promising complementary approach to tomography. Neutrinos traveling through Earth can get affected by the ambient Earth matter in two ways. Very high energy neutrinos can undergo substantial inelastic scattering via weak interactions with the ambient particles in Earth leading to an attenuation of the neutrino flux. Neutrino telescopes can use this as a signal for determining the density through which the neutrinos travel before reaching the detector. The second way in which neutrinos can probe Earth matter density is via matter effects in their flavor oscillations. This method can be effectively used in atmospheric neutrino experiments, since atmospheric neutrinos come from all zenith angles, crossing the Earth from all directions, and also experience large matter effects. In this work we quantified, for the first time, the potential of the ICAL@INO atmospheric neutrino experiment towards Earth tomography.

In this work we used the PREM profile as the reference density structure for Earth matter density. This essentially means that we simulated the atmospheric neutrino ``data" at ICAL@INO for the PREM profile. Values of oscillation parameters compatible with the current best-fit solutions were taken. Data was generated for 25 years of running of ICAL. We then statistically analysed this data with a theory where the density was allowed to be different from the PREM profile by a given percentile. The corresponding $\chi^2$ obtained was plotted as a function of the percentile change in density. The change in density was done for either the mantle or the outer core. We checked explicitly that ICAL@INO was not sensitive to density changes in the inner core, and hence this was not presented. 

We performed this study for three different cases. We started with showing the effect of density on the survival probability for each of these three cases and then went on to show the expected sensitivity of ICAL@INO to density measurements. Case I corresponded to the situation when the density in a given layer was changed without any other constraint on the analysis. This case helped us understand how sensitive the experiment will be to change in any given layer, independent of constraints coming from density changes in other layers. We found that ICAL@INO can be sensitive to density changes within $-5.4\%/+6.3\%$ in the mantle at $1\sigma$ for $\theta_{23}=45^\circ$ and NO. For the outer core the corresponding values are $-8.3\%/+7.4\%$ at $1\sigma$ for $\theta_{23}=45^\circ$ and NO. The sensitivity was seen to depend on the value of  $\theta_{23}$ as well as the mass ordering.

Case II corresponded to the situation when the density in a given layer was changed with the constraint that the total mass of the Earth is constant. This implies that when the density, say in the mantle was changed by $x\%$, then one needs a change in density in all the other layers of the Earth by $y\%$, such that the mass of the Earth would still be the same. The sensitivity of ICAL@INO to density in both the mantle as well as outer core improved in this case as compared to Case I since in this case a given change in density in any layer was being accompanied by corresponding density changes in the the other layers in order to compensate for the constant total Earth mass. We found that at  $1\sigma$ ICAL@INO  can measure the mantle density to within $-3.9\%/+3.8\%$  for $\theta_{23}=45^\circ$ and NO. For the outer core the corresponding values are $-6.7\%/+6.4\%$ at $1\sigma$ for $\theta_{23}=45^\circ$ and NO.

Finally, we considered a softened version of Earth mass compensation in Case III, where we allowed compensatory density changes only in the inner regions of the Earth. In particular, density changes were allowed only in layers for which $d<2200$ km , where $d$ is the radial depth of the layer from the surface of the Earth. For this case the sensitivity of ICAL@INO was seen to be intermediate between Case I and Case III. In particular, we showed that the density in the mantle could be measured within $-7.8\%/+7.1\%$ at $1\sigma$ for $\theta_{23}=45^\circ$ and NO. The corresponding expected sensitivity for the outer core was shown to be $-x.x\%/+x.x\%$ at $1\sigma$ for $\theta_{23}=45^\circ$ and NO. The reason for lower sensitivity in this case as compared to Case II was discussed. 

For all the cases we studied the effect of systematic uncertainties on the expected sensitivity. We also considered the condition for hydrostatic equilibrium. 

With 25 years of data taking, ICAL@INO would be competitive with large neutrino telescopes IceCube-PINGU and ORCA. In Table \ref{tab:compare} we present the comparative $1\sigma$ expected sensitivity from ICAL@INO (this work), IceCube-PINGU \cite{pingu-tomo}and ORCA \cite{Capozzi2022}. We can see that ICAL@INO can be competitive despite its smaller size. In particular, we can see that while the expected sensitivity of both goes down significantly for IO, the sensitivity of ICAL@INO is similar for both mass orderings, with NO being only slightly better. The expected sensitivity for ORCA also seems to be comparable for both mass orderings. This feature is true for both mantle as well as outer core. Note that for outer core, PINGU has rather poor sensitivity for the IO case. However, expected sensitivity of ICAL@INO (and ORCA) are good even for this case. The main reason why ICAL@INO can perform at a level comparable to PINGU and ORCA, especially for IO, is its extremely good charge identification capability. This gives ICAL@INO very good sensitivity to Earth matter effects for both mass orderings and hence the good expected sensitivity.

\begin{table}[!h]
\centering

\begin{tabular}{|c|c|c|c|c|c|c|}
\hline
  &  \multicolumn{2}{|c|}{ICAL@INO} & \multicolumn{2}{c}{PINGU} & \multicolumn{2}{|c|}{ORCA}\\
\hline
{Layer}   & NO   & IO    & NO   & IO & NO   & IO\\
\hline
 Mantle  &  -3.7/3.9 & -4.2/4.2   & -5/5.2  & -10.5/11.6 & -4/4   & -4.7/4.8 \\
 OC  &  -6.4/5.9 & -6.9/5.9   & -7.6/8.2 & -40.2/- & -5.4/6  & -6.5/7.1\\
\hline
\end{tabular}
\caption{\label{tab:compare}Percentage uncertainty in density measurement with 1$\sigma$ confidence level for different experiments. The PINGU and ORCA results are taken from ref. \cite{winter-tomo}.} 
\end{table}

\section*{Acknowledgements}
This work is performed by the members of the INO-ICAL collaboration. We thank the members of the INO-ICAL collaboration for their valuable comments and constructive inputs. SC wishes to express her gratitude to Serguey Petcov for suggesting the problem and for detailed discussions the during initial stages on this work. The HRI cluster computing facility (http://cluster.hri.res.in) is gratefully acknowledged. This project has received funding/support from the European Union’s Horizon 2020 research and innovation programme under the Marie Skłodowska -Curie grant agreement No 860881-HIDDeN.

\bibliography{earth-tomography}
  
\end{document}